\documentclass[aps,prx,singlecolumn,showpacs,superscriptaddress,amssymb,amsfonts]{revtex4}
\usepackage{amsmath,amssymb,graphicx,braket,hyperref}

\newcommand{\Rom}[1]{\uppercase\expandafter{\romannumeral #1\relax}}
\usepackage{amsthm}
\theoremstyle{definition}
\newtheorem{definition}{Definition}[section]

\newtheorem*{remark}{Remark}
\theoremstyle{theorem}
\newtheorem{theorem}{Theorem}[section] 

\theoremstyle{corollary}
\newtheorem{corollary}{Corollary}[theorem]

\theoremstyle{lemma}

\theoremstyle{Proposition} 
\newtheorem{Proposition}[theorem]{Proposition}

\begin{document}
	\title{Characterizing topological order by the information convex}
	
	\author{Bowen Shi}\author{Yuan-Ming Lu}
	\affiliation{Department of Physics, The Ohio State University, Columbus, OH 43210, USA}
	
	\date{\today}
	
	\begin{abstract}
Motivated by previous efforts in detecting topological orders from the ground state(s) wave function, we introduce a new quantum information tool, coined the information convex, to capture the bulk and boundary topological excitations of a 2D topological order. Defined as a set of reduced density matrices that minimizes the energy in a subsystem, the information convex encodes not only the bulk anyons but also the gapped boundaries of 2D topological orders. Using untwisted gapped boundaries of non-Abelian quantum doubles as an example, we show how the information convex reveals and characterizes deconfined bulk and boundary topological excitations, and the condensation rule relating them. Interference experiments in cold atoms provide potential measurements for the invariant structure of information convex. 

\end{abstract}

	\maketitle

	\widetext

\section{Introduction}

Topological orders \cite{Wen:1989iv,PhysRevB.41.9377} represent a large class of gapped quantum phases characterized by long-range entanglement \cite{2010PhRvB..82o5138C,Haah2016}. Unlike symmetry breaking orders, they support topological excitations (anyons) created by (deformable) string operators, which obey fractional braiding statistics \cite{Arovas1984}. Topological orders have locally indistinguishable states robust to decoherence \cite{2003quant.ph..6072Z}, making them excellent candidates for quantum computation \cite{2003AnPhy.303....2K,Nayak2008}. Enormous theoretical progress has been made in understanding 2D topological orders from topological quantum field theories \cite{Witten1988,PhysRevB.41.9377,kitaev2006topological,2008JHEP...05..016D}, exactly solved models \cite{2003AnPhy.303....2K,2005PhRvB..71d5110L,2008PhRvB..78k5421B,2013PhRvB..87l5114H}, and tensor category theory \cite{2005PhRvB..71d5110L,2006AnPhy.321....2K}.

Given a topologically ordered system, one important question is how to detect its topological properties? It has been shown that using the ground state(s), one can compute many invariants that characterize the topological order, such as the topological entanglement entropy (TEE) \cite{kitaev2006topological,levin2006detecting} and modular matrices \cite{2012PhRvB..85w5151Z,Moradi2015,Haah2016}. In reality, however, any experimental measurement is performed at a finite temperature, which probes a thermal density matrix rather than the ground state(s). Can one instead extract the topological order from the density matrix?

Moreover, although many theoretical efforts study topological orders on a closed manifold such as the torus, most experiments are performed on systems with open boundaries. 2D nonchiral topological orders may have gapped boundaries, where one bulk phase can have more than one boundary types \cite{1998quant.ph.11052B,2011CMaPh.306..663B,2012CMaPh.313..351K,2012arXiv1211.4644K,Levin2013,2016arXiv160902037C,2017arXiv170603329H,2017arXiv170603611B,2017CMaPh.355..645C}. How to extract this rich structure of boundary excitations in a topologically ordered system?

In this work, we develop a quantum informational tool, the \emph{information convex} $\Sigma(\Omega)$, to characterize the topological excitations in the bulk and on the gapped boundary of a 2D topological order. Most conveniently defined for any frustration-free local Hamiltonian \cite{2013CMaPh.322..277M}, $\Sigma(\Omega)$ is the (convex) set of reduced density matrices on a region $\Omega$, obtained from states which minimize all terms overlapping with region $\Omega$ in the Hamiltonian. See Sec. \ref{Appendix_A} for detailed properties of information convex, in which a more general information convex  $\Sigma(\Omega,\Omega')$ is introduced for $\Omega\subseteq \Omega'$.  Intuitively, $\Sigma(\Omega,\Omega')$ is the set of reduced density matrices on $\Omega$ obtained from states minimizing the energy on $\Omega'$.

The element $\sigma_{\Omega}=tr_{\bar{\Omega}}\vert \varphi\rangle\langle \varphi\vert$ of the information convex  $\Sigma(\Omega)$ can be obtained from states $\vert \varphi\rangle$ with no excitations within $\Omega$, where $\bar\Omega$ is the complement of $\Omega$. The information convex can be obtained numerically by performing imaginary time evolution in region $\Omega$, or experimentally by cooling down the subsystem $\Omega$ below the finite energy gap.

First, for a 2D topological order on a closed manifold, we show that the information convex $\Sigma(\Omega)$ naturally captures previously known topological invariant characterizations like TEE \cite{kitaev2006topological,levin2006detecting} and the minimal entangled (ground) states \cite{2012PhRvB..85w5151Z}. To study bulk properties, we choose the subsystem $\Omega$ away from the boundary of the system, as illustrated by $\omega_1$ and $\Omega_1$ in Fig. \ref{Omega_Copy}. The annulus $\Omega_1$ in Fig. \ref{Omega_Copy} leads to extremal points Eq. (\ref{eq:bulk}) labeled by different bulk anyon types (or \emph{bulk superselection sectors}).

Moreover, for a 2D nonchiral topological order with open boundaries, the information convex $\Sigma(\Omega)$ allows us to reveal the structure of boundary topological excitations, which are generally different from the bulk anyons \cite{2012CMaPh.313..351K,2012arXiv1211.4644K,2017arXiv170603329H}. For this purpose, we choose the subsystem $\Omega$ containing a part of the open boundary, illustrated by $\omega_2$, $\Omega_2$, and $\Omega_3$ in Fig. \ref{Omega_Copy}. For example, choosing strip $\Omega_2$ as the subsystem leads to information convex $\Sigma(\Omega_2)$, whose extremal points correspond to distinct boundary topological excitations (or \emph{boundary superselection sectors}).

\begin{figure}[h]
		\centering
		\includegraphics[scale=0.330]{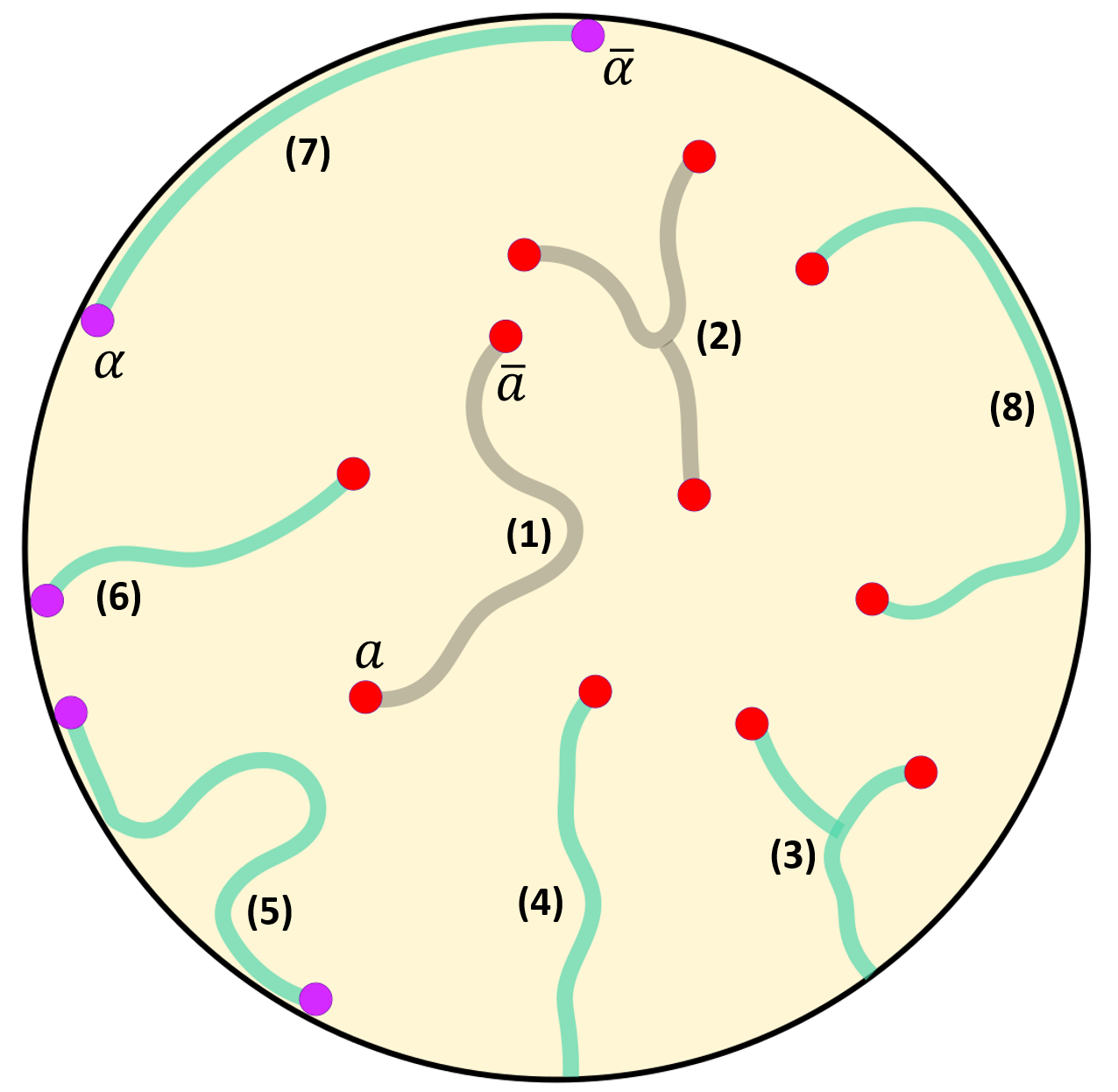}
		\caption{A system on a disk $D^2$, which has a single gapped boundary. Also shown are examples of string operators and the topological excitations they create. The grey strings are in the bulk and they could be deformed topologically. The green strings are deformable but could not detach from the boundary. Excitations inside the bulk are shown in red and excitations along the boundary are shown in purple.}\label{Boundary_excitations}
	\end{figure}

While bulk anyons (red dots in Fig. \ref{Boundary_excitations}) of a 2D topological order can be created by deformable strings within the bulk (grey lines in Fig.  \ref{Boundary_excitations}), with open boundaries there are also deformable strings that cannot detach from the boundary (green lines in Fig.  \ref{Boundary_excitations}). Some of these nondetachable strings are attached to the bulk anyons, some are attached to
 boundary topological excitations (purple dots in Fig. \ref{Boundary_excitations}). We discuss how the topological invariants of $\Sigma(\Omega)$ capture boundary superselection sectors $\{\alpha\}$, their corresponding quantum dimensions $\{d_{\alpha}\}$, and the condensing process from a bulk anyon $a$ to boundary excitations $\sum\alpha$. We also discuss an interesting peculiarity of certain non-Abelian topological orders with  condensation multiplicity greater than $1$. In this case, the information convex $\Sigma(\Omega_3)$ has an infinite number of extremal points forming a manifold whose structure reveals the nontrivial condensation  multiplicity.

This paper is organized as follows.
In Sec. \ref{Section_Main}, we introduce our main results using a simple example, decoupling the physics of information convex from the calculation from which the results are obtained. In Sec. \ref{Appendix_A}, we gives a rigorous definition of information convex $\Sigma(\Omega)$ and $\Sigma(\Omega,\Omega')$ in the context of frustration-free local Hamiltonian and explore some basic properties. In Sec. \ref{Quantum Double}, we provide a detailed study of information convex in  the quantum double model with a gapped boundary, including the calculations and a few theorems. In Sec. \ref{Sec. K={1}}, we provide a few more results for quantum double models with a specific boundary type ($K=\{1\}$).

\section{The main results}	\label{Section_Main}
While our results follow from a concrete (but involved) calculation in the  quantum double models, the main physical results are quite compact and accessible  without the relative heavy details. Therefore, we choose to convey the physical messages in this section focusing on a simple example. This section also serves as a map pointing to the more detailed calculations and theorems in later sections.
	
\subsection{The models}
	
To demonstrate our main results, we perform explicit calculations on the quantum double models (with a finite group $G$) with an untwisted boundary labeled by a subgroup $K\subseteq G$ \cite{2003AnPhy.303....2K,2008PhRvB..78k5421B,2011CMaPh.306..663B}. We focus on a simple example $G=S_3$, $K=\{1\}$, where most of the nontrivial intuitions can be seen (see Sec. \ref{Quantum Double},\ref{Sec. K={1}} for details and generalizations). For simplicity, we put the model on a disk $D^2$ with a single open boundary. The Hilbert space is defined on a lattice within the disk, and there is a unique ground state $\vert\psi\rangle$.

	$S_3=\{1,r,r^2,s, sr,sr^2\}$ with  $r^3=s^2=1$, $sr=r^2s$ is the simplest non-Abelian finite group. The $G=S_3$ quantum double has 8 bulk anyon types (bulk superselection sectors) labeled by a pair $a=(c, R)$ with   quantum dimension $d_a$. Here, $c$ is a conjugacy class of $G$ and $R$ is an irreducible representation of the centralizer group. We list only useful information for later discussions, while more details can be found in  \cite{2003AnPhy.303....2K,2008PhRvB..78k5421B,2013PhRvB..88k5133K} and Appendix \ref{Appendix E}.  Note that for the $S_3$ quantum double model, each bulk anyon is its own antiparticle.
	\begin{table}[h]
		\centering
	
		\begin{tabular}{|c|| c| c | c | c | c | c | c | c |}
			\hline
			\multicolumn{1}{|c||}{Conjugacy class $c$}& \multicolumn{3}{c|}{$c_1=\{1 \}$}& \multicolumn{3}{c|}{$c_r=\{r,r^2 \}$}& \multicolumn{2}{c|}{$c_s=\{s,sr,sr^2 \}$}\\
			\hline
			 $a$ & \,$1$\,& \,$A$\, & $J^w$ & \,$J^x$\, & \,$J^y$\, & \,$J^z$\, & \,\,\,\,\,$K^a$\,\,\,\,\, & $K^b$\\
			\hline
			 $d_a$ & 1&1 &2 &2 &2 &2 &3 &3\\
			\hline
			
		\end{tabular}\label{tab:S3}
	 \end{table}


	\begin{figure}[h]
		\centering\includegraphics[scale=0.35]{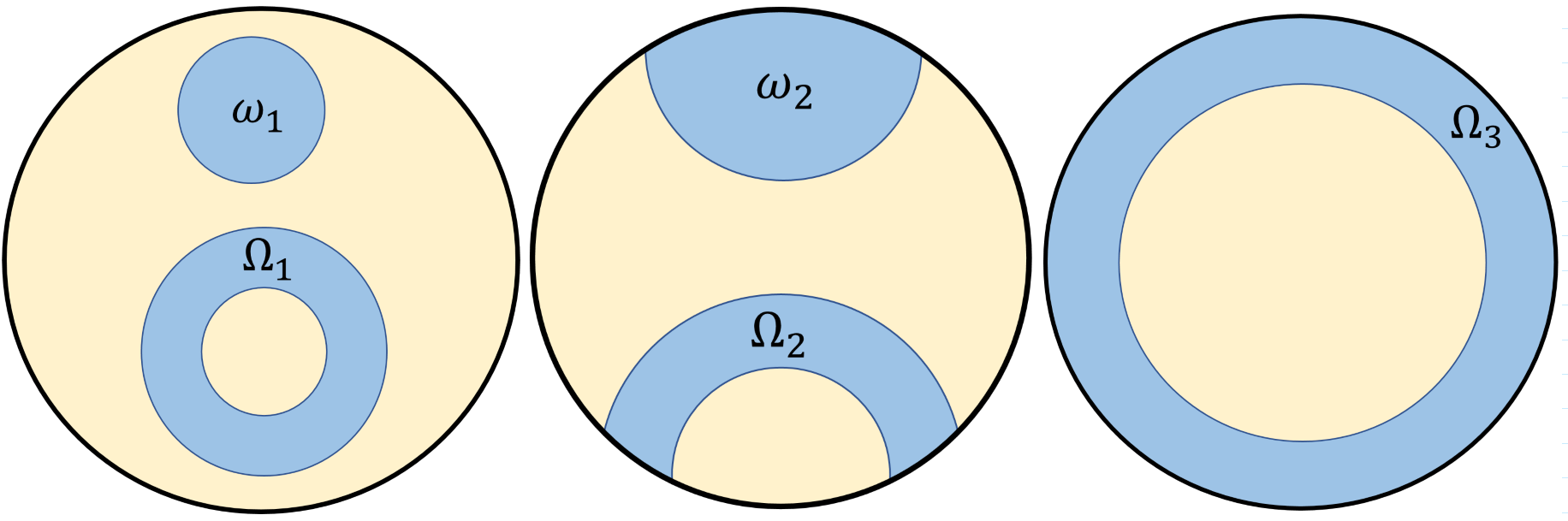}
		\caption{Subsystem choices: (left) to study the bulk; (mid)-(right) to study the gapped boundaries.}\label{Omega_Copy}
	\end{figure}

\subsection{Characterizing bulk anyons}

$\Sigma(\Omega)$  always  contains the reduced density matrix $\sigma^1_{\Omega}\equiv \textrm{tr}_{\bar{\Omega}}\vert\psi\rangle\langle\psi\vert$ of the ``global'' ground state $|\psi\rangle$. For a topologically trivial subsystem $\omega$ (e.g. $\omega_1$ and $\omega_2$ in Fig. \ref{Omega_Copy}, we have $\Sigma(\omega)=\{\sigma_\omega^1\}$. In other words, states locally minimizing the energy of subsystem $\omega$ are indistinguishable from the global ground state on the subsystem. Though simple, $\Sigma(\omega)$ determines the possible types of deformable strings in Fig. \ref{Boundary_excitations} and that the string operators can be unitary (see Sec. \ref{HJW Section} for details).

\begin{figure}[h]
    	\centering\includegraphics[scale=0.3]{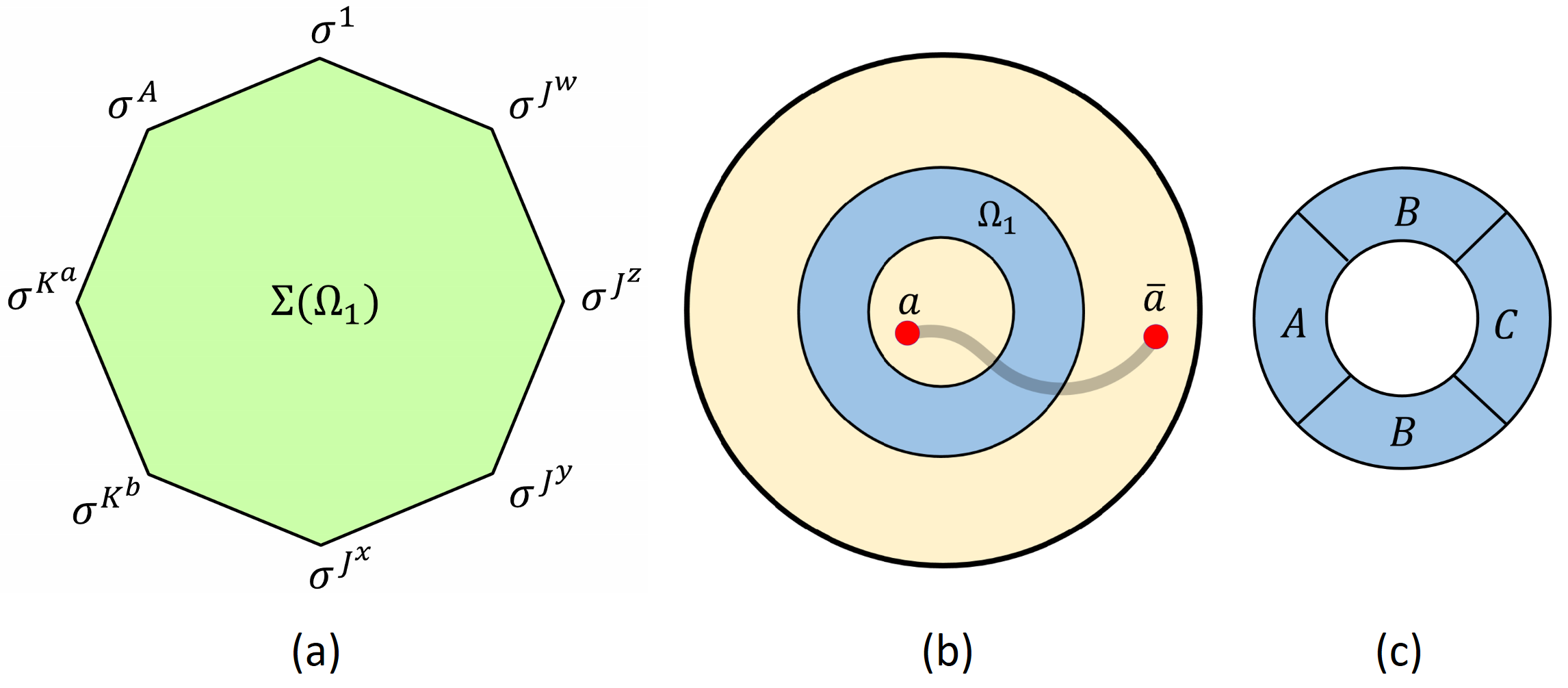}
    	\caption{(a) A 2D projection of 7D convex set $\Sigma(\Omega_1)$ which has 8 extremal points. (b) Preparing an extremal point  $\sigma_{\Omega_1}^a\in \Sigma(\Omega_1)$ by a bulk process. (c) Divide $\Omega_1$ into $A,B,C$.
    	}\label{Convex_1}
    \end{figure}

     On the other hand, the information convex of a bulk annulus $\Omega_1$ has a richer structure:
    \begin{equation}\label{eq:bulk}
    \Sigma(\Omega_1)=\{\sigma_{\Omega_1}\vert \sigma_{\Omega_1}=\sum_a p_a\sigma_{\Omega_1}^a  \} ,
    \end{equation}
    where $a$ labels bulk superselection sectors (bulk anyon types), with $a=1$ for the vacuum sector. $\{p_a|\sum_a p_a=1\}$ is a probability distribution. Clearly, $\Sigma(\Omega_1)$ is a convex set and $\sigma^a_{\Omega_1}$ are its extremal points. 
    Under continuous deformations of $\Omega_1$ and the Hamiltonian, $\Sigma({\Omega_1})$ exhibits the following topological invariant structure:
    \begin{eqnarray}
    S(\sigma^{a}_{\Omega_1}) &=& S(\sigma^{1}_{\Omega_1})+\ln d_a^2 \label{E-Omega_1} \\
    \sigma^a_{\Omega_1}\cdot\sigma^b_{\Omega_1} &=& 0\quad \textrm{for}\,\,\,\, a\ne b;\quad \label{Eq_Tr_Bulk}\\
    \frac{tr[\sigma^a_{\Omega_1}\cdot \sigma^a_{\Omega_1}]}{tr[\sigma^1_{\Omega_1}\cdot \sigma^1_{\Omega_1}]}&=&\frac{1}{d^2_a}   
    \end{eqnarray}
    where $S(\sigma)=-tr(\sigma\ln\sigma)$ is the von Neumann entropy. One can further show the extremal point $\sigma^a_{\Omega_1}=tr_{\bar\Omega_1}|\varphi^a\rangle\langle\varphi^a|$ can be obtained from an excited state $|\varphi^a\rangle$  with an anyon pair $(a,\bar{a})$ created by a bulk string shown in Fig. \ref{Convex_1}(b). When $\Omega_1$ is a noncontractible annulus on a torus $T^2$, each extremal point can be obtained from the corresponding minimal entangled state on torus \cite{2012PhRvB..85w5151Z}.

    In the example of $S_3$ quantum double, there are 8 extremal points (8 anyons), leading to a 7-dimensional information convex $\Sigma(\Omega_1)$.

\subsection{TEE as a saturated lower bound}

   The information convex $\Sigma(\Omega_1)$ naturally encodes the TEE as a saturated lower bound. Let  $\tilde{\sigma}_{\Omega_1}$ be the element located at the ``center" of $\Sigma(\Omega_1)$, written as 
   \begin{equation}
   \tilde{\sigma}_{\Omega_1}=\sum_a \frac{d_a^2}{\mathcal{D}^2}\,\sigma^a_{\Omega_1}
   \end{equation}
   It has the maximally entanglement entropy among all density matrices in  $\Sigma(\Omega_1)$. Then, let us take the partition  $\Omega_1=ABC$ as is shown in Fig.  \ref{Convex_1}(c)  and define TEE $S_{topo}\equiv(S_{AB}+S_{BC}-S_{B}-S_{ABC})\vert_{\sigma^1}$ in accordance with Levin-Wen \cite{levin2006detecting}. Then one could derive a lower bound $S_{topo}\ge S(\tilde{\sigma}_{\Omega_1})-S(\sigma^1_{\Omega_1})$ by noticing  the following two facts:
   \begin{itemize}
   	\item The form of $S_{topo}$ above is the conditional mutual information. It is an important theorem (the strong subadditivity condition) that conditional mutual information is always nonnegative, i.e. $(S_{AB}+S_{BC}-S_{B}-S_{ABC})\vert_{\sigma}\ge 0$ for any density matrix $\sigma$.
   	\item All density matrices in $\Sigma(\Omega_1)$ have the same reduced density matrix on $AB$, $BC$ and $B$ because $\Sigma({\omega_1})$ contains a single element.
   \end{itemize}
   Furthermore, it is known that under general assumptions in \cite{2016PhRvA..93b2317K}, $\tilde{\sigma}_{\Omega_1}$ saturates the strong subadditivity condition. This allows us to reformulate the celebrated TEE as a saturated lower bound:
   \begin{equation}
   S_{topo}= S(\tilde{\sigma}_{\Omega_1})-S(\sigma^1_{\Omega_1})=\ln \mathcal{D}^2
   \end{equation}
   where $\mathcal{D}\equiv\sqrt{\sum_a d_a^2}$ is the total quantum dimension. See Sec. \ref{TEE Subsection} for more details of the derivation and further discussions.

\subsection{Characterizing boundary topological excitations} 

\begin{figure}[h]
    	\centering\includegraphics[scale=0.32]{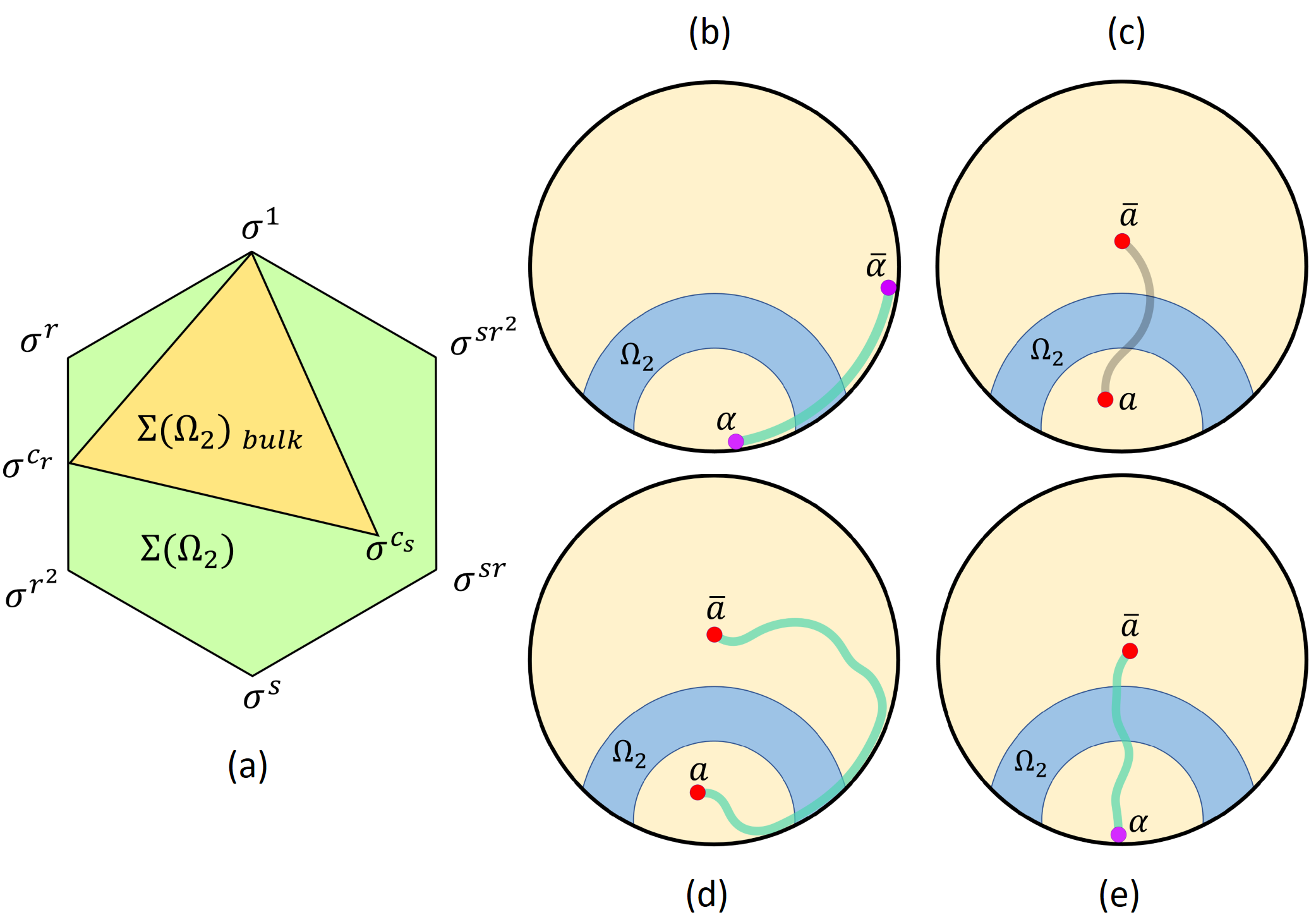}
    	\caption{(a) A 2D projection of 5D convex set $\Sigma(\Omega_2)$ for $S_3$ quantum double, which has 6 extremal points. $\Sigma(\Omega_2)_{bulk}\subseteq \Sigma(\Omega_2)$ is a  2D convex set  with 3 extremal points. (b) A boundary process that prepares an extremal point $\sigma^{\alpha}_{\Omega_2}\in\Sigma(\Omega_2)$. (c) A bulk process that prepares an extremal point of $\Sigma(\Omega_2)_{bulk}$. (d) A pair of bulk non-Abelian anyons $(a,\bar{a})$ created by a boundary process. (e) A $(\bar{a},\alpha)$ pair created by a boundary process.} \label{Convex_2}
    \end{figure}

The topological excitations on the gapped boundary can be extracted by choosing subsystem $\Omega_2$ in Fig. \ref{Omega_Copy}:
\begin{eqnarray}
    \Sigma(\Omega_2) &=& \{\sigma_{\Omega_2}\vert \sigma_{\Omega_2}=\sum_{\alpha} p_{\alpha}\sigma_{\Omega_2}^{\alpha}  \}.\\
     S(\sigma^{\alpha}_{\Omega_2}) &=& S(\sigma^1_{\Omega_2})+\ln{d_{\alpha}^2},\\
    \sigma^{\alpha}_{\Omega_2}\cdot\sigma^{\beta}_{\Omega_2} &=& 0\quad \textrm{for}\,\,\,\, \alpha\ne \beta;\quad \label{Eq_Boundary}\\
     \frac{tr[\sigma^{\alpha}_{\Omega_2}\cdot \sigma^{\alpha}_{\Omega_2}]}{tr[\sigma^1_{\Omega_2}\cdot \sigma^1_{\Omega_2}]}&=&\frac{1}{d^2_{\alpha}},
    \end{eqnarray}
    where $\{p_{\alpha}\}$ is a probability distribution.
    The structure of $\Sigma(\Omega_2)$ is very similar to that of $\Sigma(\Omega_1)$, except that the $\alpha,\beta$ label the \emph{boundary superselection sectors} instead of the bulk ones. The name comes from the fact that every extremal point $\sigma_{\Omega_2}^{\alpha}$ can be obtained from an excited state with a unitary string operator acting along the boundary, which creates a pair of boundary topological excitations $(\alpha,\bar{\alpha})$ as shown in Fig.  \ref{Convex_2}(b).

    For a $K=\{1\}$ boundary of $G$ quantum double, $\alpha\in G$ labels the ``flux" type and $d_{\alpha}=1$, $\forall~\alpha\in G$. For $G=S_3$:

    	\begin{table}[h]
    	\centering
    	
    	\begin{tabular}{|c|| c| c | c | c | c | c | }
    		\hline
    		$\alpha$ & \,$1$\,& \,$r$\, & $r^2$ & \,$s$\, & \,$sr$\, & \,$sr^2$\\
    		\hline
    		$d_{\alpha}$ & 1&1 &1 &1 &1 &1 \\
    		\hline
    		
    	\end{tabular}
    \end{table}

    For a general quantum double model with an untwisted $K\subseteq G$ boundary, we use the information convex to identify deconfined topological excitations along the boundary, in contrast to the confined boundary excitations discussed in Ref. \cite{2017CMaPh.355..645C}. However, our calculations show that they share the same algebraic structure as in \cite{2017CMaPh.355..645C}.

We use $\Sigma(\Omega_2)$ to demonstrate the relation between bulk anyons and boundary topological excitations, focusing on non-Abelian topological orders. For quantum double with Abelian group $G$ and any untwisted $K\subseteq G$ boundary, all extremal points of $\Sigma(\Omega_2)$ can be obtained by creating a bulk anyon pair $(a,\bar{a})$ with a bulk string crossing $\Omega_2$, see Fig. \ref{Convex_2}(c). On the other hand, for a non-Abelian $G$ with a $K=\{1\}$ boundary, bulk excitations in Fig. \ref{Convex_2}(c) can only explore a (convex) subset $\Sigma(\Omega_2)_{bulk}\subsetneq\Sigma(\Omega_2)$ of the information convex $\Sigma(\Omega_2)$.

For $G=S_3$, $K=\{1\}$ case, we find:
   \begin{equation}
   \Sigma(\Omega_2)_{bulk}=\{\sigma_{\Omega_2}\vert \sigma_{\Omega_2}=p_1 \sigma^1_{\Omega_2} +p_{c_r}\sigma^{c_r}_{\Omega_2}+p_{c_s}\sigma^{c_s}_{\Omega_2} \},
   \end{equation}
    where $\{p_{1}, p_{c_r}, p_{c_s} \}$ is a probability distribution, and
    \begin{equation}
   \sigma^{c_r}_{\Omega_2} \equiv \frac{1}{2}(\sigma_{\Omega_2}^r +\sigma_{\Omega_2}^{r^2}),\quad \sigma^{c_s}_{\Omega_2} \equiv \frac{1}{3}(\sigma^s_{\Omega_2} +\sigma^{sr}_{\Omega_2}+\sigma^{sr^2}_{\Omega_2}).\label{eq:condense}
    \end{equation}

    The bulk anyon pair $(a,\bar{a})$ associated with extremal points of $\Sigma(\Omega_2)_{bulk}$ is $a\in\{1,A,J^w \}$ for $\sigma_{\Omega_2}^{1}$, $a\in \{J^x,J^y,J^z\}$ for $\sigma_{\Omega_2}^{c_r}$, and $a\in \{K^a,K^b\}$ for $\sigma_{\Omega_2}^{c_s}$. In this case, while boundary topological excitations can lead to all extremal points in $\Sigma(\Omega_2)$, only one extremal point $\sigma_{\Omega_2}^1$ can be obtained by anyons connected by a bulk string.

    What are the relation and distinction between bulk and boundary topological excitations? First of all,  each boundary topological excitation as in Fig. \ref{Convex_2}(b) can be deformed into the bulk as in Fig. \ref{Convex_2}(d) by local unitary operators, although  the string may not completely detach from the boundary. However, such a bulk pair $(a,\bar a)$ connected by a boundary string in Fig. \ref{Convex_2}(d) should not be identified with a pair connected by a pure bulk string in Fig. \ref{Convex_2}(c), since they may correspond to different elements in the information convex.

     In  $G=S_3,~K=\{1\}$ quantum double, a boundary excitation pair $(\alpha=r,\bar\alpha=r^2)$ in Fig. \ref{Convex_2}(b) can be deformed into a bulk pair $(a=J^x,\bar a=J^x)$ connected by a boundary string in Fig. \ref{Convex_2}(d), and they both lead to the extremal point $\sigma^\alpha_{\Omega_2}=\sigma^r_{\Omega_2}$ in Fig. \ref{Convex_2}(a). In contrast, a pair of bulk anyons $(J^x,J^x)$ in Fig. \ref{Convex_2}(c) only gives rise to $\sigma^{c_r}_{\Omega_2}$ in Fig. \ref{Convex_2}(a).

     Secondly, each bulk anyon $a$ in Fig. \ref{Convex_2}(c), when ``moved'' to the boundary in Fig. \ref{Convex_2}(e)  by local unitary operators, is generally deformed into a superposition $\sum\alpha$ of boundary topological excitations. This process is governed by certain condensation rules $a\rightarrow\sum\alpha$ \cite{2012CMaPh.313..351K,2011CMaPh.306..663B,2017CMaPh.355..645C}, which manifests in the information convex.

     Specifically in $G=S_3,~K=\{1\}$ case,  Eq. (\ref{eq:condense}) of the information convex $\Sigma(\Omega_2)$ corresponds to the anyon condensation rules of $K=\{1\}$ boundary of $G=S_3$ quantum double, summarized below:
        	\begin{table}[h]
    	\centering
    	
    	\begin{tabular}{|c|| c| c | c | c|c | c|c | c|c | c|c|c | c|c|c |}
    		\hline
    		\multicolumn{1}{|c||}{$c$}& \multicolumn{3}{c|}{$c_1=\{1 \}$}& \multicolumn{6}{c|}{$c_r=\{r,r^2 \}$}& \multicolumn{6}{c|}{$c_s=\{s,sr,sr^2 \}$}\\
    		\hline
    		$a$ & \,$1$\,& \,$A$\, & $J^w$ &    \multicolumn{2}{c|}{\,\,\,\,$J^x$\,\,\,\,} & \multicolumn{2}{c|}{\,\,\,\,$J^y$\,\,\,\,} & \multicolumn{2}{c|}{\,\,\,\,$J^z$\,\,\,\,} & \multicolumn{3}{c|}{\,\,\,\,$K^a$\,\,\,\,}&\multicolumn{3}{c|}{\,\,\,\,$K^b$\,\,\,\,}\\
    		\hline
    	    $\alpha$ & \,$1$\,& \,$1$\, & $2\cdot 1$ & \,$r$\, & $ r^2$ & \,$r$\, & $ r^2$ & \,$r$\, & $ r^2$ & \,$s$\,& $sr$ & $sr^2$ & \,$s$\,& $sr$ & $sr^2$\\
    		\hline
    	\end{tabular}
    \end{table}

Here each of the bulk anyons $\{J^x,J^y,J^z, K^a,K^b\}$ becomes a superposition of boundary topological excitations, once moved to the boundary as in Fig. \ref{Convex_2}(e) by a local unitary operator. For instance, consider a bulk anyon pair $(a=J^x,\bar a=J^x)$ in Fig. \ref{Convex_2}(c). Local unitaries can deform the bulk string and move anyon $a=J^x$ to the boundary as in Fig. \ref{Convex_2}(e). However, this bulk anyon $a=J^x$ cannot turn into a single boundary superselection sector by any local unitary: instead, it ``condenses into'' a superposition of boundary topological excitations $J^x\rightarrow r+r^2$, in accordance with  Eq. (\ref{eq:condense}).

\subsection{Anyon condensation to the boundary}

\begin{figure}[h]
    	\centering\includegraphics[scale=0.27]{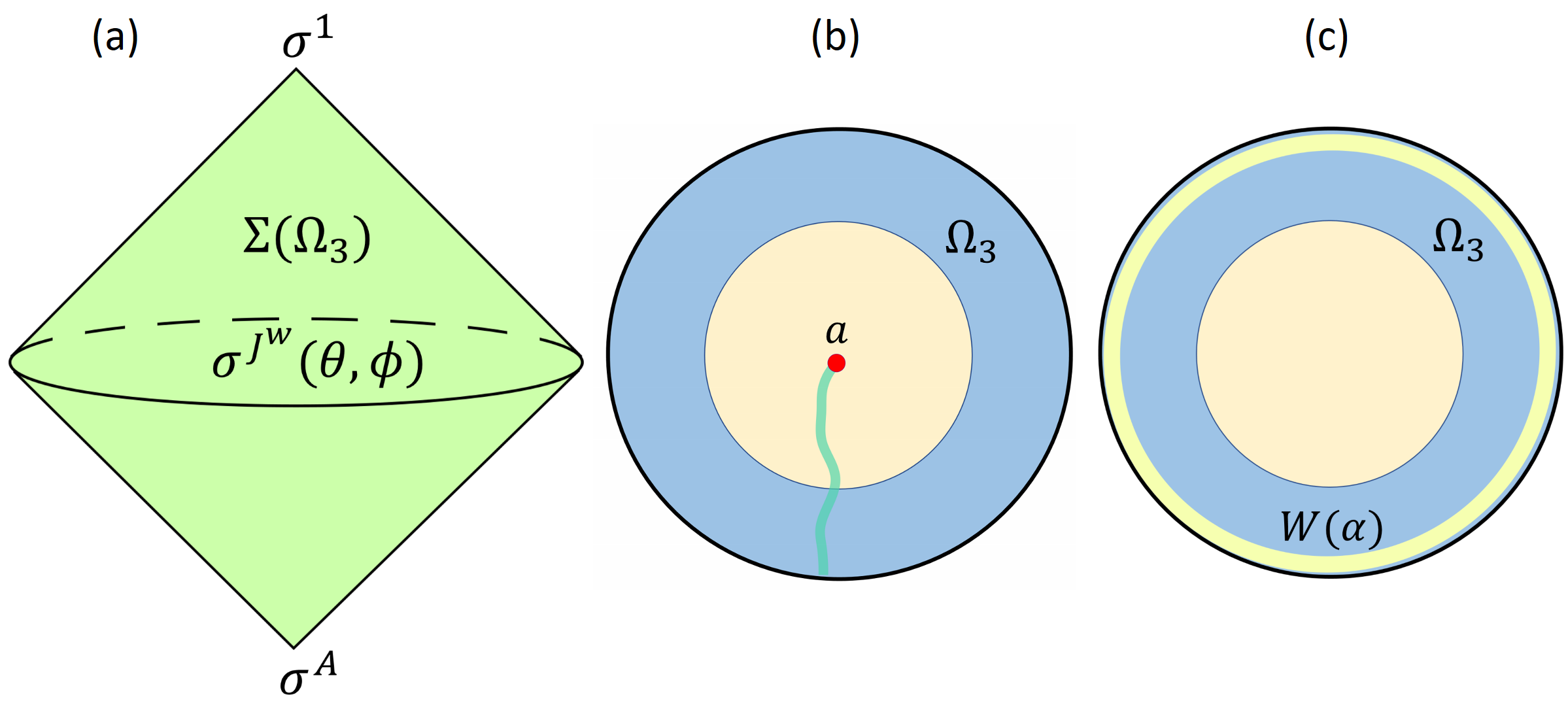}
    	\caption{(a) A 3D projection of 5D convex set $\Sigma(\Omega_3)$ which has extremal points on a sphere $S^2$ plus two isolated points.
    		(b) A boundary process that prepares an extremal point of $\Sigma(\Omega_3)$， where $a=1, A, J^w$.
    		(c) Braiding an $(\alpha,\bar{\alpha})$ pair along the boundary corresponds to a unitary operation $W(\alpha)$ supported on the yellow loop.
    	}\label{Convex_3}
    \end{figure}

Previously, $\Sigma(\Omega_2)$ already demonstrates the anyon condensation on the gapped boundary \cite{2012CMaPh.313..351K,2012arXiv1211.4644K,2017CMaPh.355..645C,Hung2015}. Here we discuss a unique consequence of boundary anyon condensation rules of non-Abelian topological orders, where the information convex $\Sigma(\Omega_3)$ for subsystem $\Omega_3$ (see Fig. \ref{Convex_3}) can have an \emph{infinite} number of extremal points.

%
%

Consider annulus $\Omega_3$  covering the boundary in Fig. \ref{Convex_3}, for $K=\{1\}$ gapped boundary of $G=S_3$ quantum double, the information convex $\Sigma(\Omega_3)$ has the following extremal points: $\sigma^1_{\Omega_3}$,
   $\sigma^{A}_{\Omega_3}$, and $\sigma^{J^w}_{\Omega_3}(\theta,\phi)$, where $(\theta, \phi)$ parametrize a sphere $S^2$ as shown in Fig. \ref{Convex_3}(a),
    \begin{eqnarray}
    	\Sigma(\Omega_3) = \{\sigma_{\Omega_3}\vert \sigma_{\Omega_3}&=& p_1 \sigma_{\Omega_3}^1 +p_{A}\sigma^A_{\Omega_3} \nonumber\\
    	&&+\int d\theta d\phi\,\, p(\theta,\phi)\, \sigma^{J^w}_{\Omega_3}(\theta,\phi) \},\qquad
    \end{eqnarray}
    with nonnegative $p_1$, $p_A$,  $p(\theta,\phi)$ satisfying $p_1 +p_A +\int d\theta d\phi \, p(\theta,\phi)=1$. Moreover, $\Sigma(\Omega_3)$ has the following structure:
    \begin{eqnarray}
    S(\sigma^A_{\Omega_3}) &=& S(\sigma^1_{\Omega_3}) +\ln d_A \label{A},\\
    S(\sigma^{J^w}_{\Omega_3}(\theta,\phi)) &=& S(\sigma^1_{\Omega_3}) + \ln d_{J^w}\label{J^w}\label{TEE Omega3};\\
    \sigma^1_{\Omega_3}\cdot \sigma^A_{\Omega_3}=\sigma^1_{\Omega_3}\cdot \sigma^{J^w}_{\Omega_3}(\theta,\phi)&=& \sigma^A_{\Omega_3}\cdot \sigma^{J^w}_{\Omega_3}(\theta,\phi) = 0;\qquad
     \\
     \frac{tr[\sigma^{A}_{\Omega_3}\cdot \sigma^{A}_{\Omega_3} ]}{tr[\sigma^1_{\Omega_3}\cdot \sigma^1_{\Omega_3} ]}&=&\frac{1}{d_{A}};\\
     \frac{tr[\sigma^{J^w}_{\Omega_3}(\theta,\phi)\cdot \sigma^{J^w}_{\Omega_3}(\theta',\phi') ]}{tr[\sigma^1_{\Omega_3}\cdot \sigma^1_{\Omega_3} ]}&=&\frac{1}{d_{J^w}}\cdot\frac{1+\hat{n}\cdot \hat{n}'}{2},\label{Interference Observable}
     \end{eqnarray}
      where the unit vector $\hat{n}=(\sin\theta\cos\phi,\sin\theta\sin\phi,\cos\theta)$, and similarly for $\hat{n}'$ in terms of $(\theta',\phi')$.

   We  notice that braiding a $(\alpha,\bar{\alpha})$ pair around the boundary before annihilation, implemented by a unitary operator $W(\alpha)$ supported on the closed loop in Fig. \ref{Convex_3}(c), do not change the energy density anywhere. Therefore this operation generates a (structure preserving) bijective map $\sigma_{\Omega_3}\in\Sigma(\Omega_3)\rightarrow W(\alpha)\sigma_{\Omega_3}W({{\alpha}})^{\dagger} \in \Sigma(\Omega_3)$. Explicitly, each $W(\alpha)$ keeps the extremal points $\sigma^1_{\Omega_3}$ and $\sigma^A_{\Omega_3}$ fixed but rotates on the sphere of $\sigma^{J^w}_{\Omega_3}(\theta,\phi)$. The set of rotations are generated by $W(r)$:   $(\theta,\phi)\rightarrow (\theta,\phi +2\pi/3)$ and $W(s)$: $(\theta,\phi)\rightarrow (\pi-\theta,-\phi)$, and they realize the $S_3$ group action on sphere $S^2$.

 While $\sigma^1_{\Omega_3}$ is obtained from the  ground state, the extremal point $\sigma^A_{\Omega_3}$ (or $\sigma^{J^w}_{\Omega_3}(\theta,\phi)$) is prepared by
    an excited state with a single anyon $A$ (or $J^w$) in the bulk, created by a string attached to the boundary as in Fig. \ref{Convex_3}(b).
    The $(\theta,\phi)$ dependence for $\sigma_{\Omega_3}^{J^w}$ comes from the  condensation multiplicity $2$ in $J^w\to 2\cdot 1$. The two ways to condense $J^w$ into the boundary lead to a two-dimensional protected Hilbert space, which result in a set of reduced density matrices $\{\sigma_{\Omega_3}^{J^w}(\theta,\phi)\}$ parametrized by $(\theta,\phi)\in S^2$. Such condensation multiplicity and infinite extremal points are unique to non-Abelian topological orders.

Though $\sigma^{J^w}_{\Omega_3}$ with different $(\theta,\phi)$ share the same entanglement entropy and entanglement spectrum \cite{2009PhRvL.103z1601F}, their ``interference pattern'' in Eq. (\ref{Interference Observable}) leaves a clear signature for the infinite extremal points $\sigma^{J^w}_{\Omega_3}(\theta,\phi)$. 

\subsection{Potential measurements of information convex structure}
It is an interesting question  whether the structure of the information convex could be observed experimentally. One challenge is the creation of anyons and another is the  measurement of properties directly related to density matrices. Recently, cold atom experiments seem to have made progress in both directions. Anyons are claimed to be created in a minimal toric code Hamiltonian \cite{2016arXiv160205709D} and the corresponding braiding properties are studied. The recent interference experiments \cite{2015arXiv150901160I,2016Sci...353..794K} allow people to measure $tr [\sigma_{\Omega}^{(1)}\cdot \sigma_{\Omega}^{(2)}]$ for any $\Omega$. In the interference experiment,  two identical copies of a cold atom system are created. Then, the authors prepare the two copies of the system in pure state $\vert \varphi^{(1)}\rangle$ and $\vert \varphi^{(2)}\rangle$ respectively. The quantum state of the two copies of the system $\vert \varphi^{(1)}\rangle$ and $\vert \varphi^{(2)}\rangle$ can be either the same or different and  
\begin{equation}
\sigma_{\Omega}^{(1)} = tr_{\bar{\Omega}} \vert \varphi^{(1)}\rangle \langle   \varphi^{(1)} \vert,\quad \sigma_{\Omega}^{(2)} = tr_{\bar{\Omega}} \vert \varphi^{(2)}\rangle \langle   \varphi^{(2)} \vert
\end{equation} 
The interference of these two copies of the system allows people to measure $tr [\sigma_{\Omega}^{(1)}\cdot \sigma_{\Omega}^{(2)}]$.
It seems possible to observe the structure of information convex in this type of cold atom experiment.

One could cool down the system except for several isolated points such that $\Omega$, a subsystem being cooled down, contains no excitations. Then the information convex gives prediction for the measurement result of $tr [\sigma_{\Omega}^{(1)}\cdot \sigma_{\Omega}^{(2)}]$ for topological orders. For example:
\begin{itemize}
	\item First, in the simplest situation, both  $\vert \varphi^{(1)}\rangle$ and  $\vert \varphi^{(2)}\rangle$ are in the ground state. Then, the interference experiment measures $tr[\sigma^1_{\Omega}\cdot \sigma^1_{\Omega}]$ for all subsystems $\Omega$. It is always a positive number, which may be used to normalize the rest of the results.
	
	\item Suppose on the state $\vert \varphi^{(1)}\rangle$ a pair of bulk anyons $(a,\bar{a})$ is created and the state  $\vert \varphi^{(2)}\rangle$ is the ground state. Here $a\ne 1$. Then, according to Eq. (\ref{Eq_Tr_Bulk}), we get $0$ on any annulus surrounding the anyon $a$, since $tr[\sigma^1_{\Omega_1}\cdot \sigma^a_{\Omega_1}]=0$. A similar result holds for any annulus surrounding the anyon $\bar{a}$.
	
	\item Suppose on the state $\vert \varphi^{(1)}\rangle$ a pair of boundary topological excitations $(\alpha,\bar{\alpha})$ is created and the state  $\vert \varphi^{(2)}\rangle$ is the ground state. Here $\alpha \ne 1$. Then, according to Eq. (\ref{Eq_Boundary}), we get $0$ on any subsystem of $\Omega_2$ topology surrounding the boundary topological excitation $\alpha$, since $tr[\sigma^1_{\Omega_2}\cdot \sigma^{\alpha}_{\Omega_2}]=0$.  A similar result holds for any subsystem of $\Omega_2$ topology surrounding the boundary topological excitation $\bar{\alpha}$.

	\item Suppose on the state $\vert \varphi^{(1)}\rangle$ a bulk anyon $J^w$ discussed above is created with a condensation channel labeled by $(\theta,\phi)$, and  on the state $\vert \varphi^{(2)}\rangle$  a bulk anyon $J^w$ discussed above is created with a condensation channel labeled by $(\theta',\phi')$. Note that, we do not require the two anyons be created at the same location. Then, according to Eq. (\ref{Interference Observable}),  for each subsystem of $\Omega_3$ topology, we get an interference result depending on the condensation channel,  i.e.
	\begin{equation}
	tr [\sigma^{J^w}_{\Omega_3}(\theta,\phi)\cdot \sigma^{J^w}_{\Omega_3}(\theta',\phi')]\sim \frac{1 + \hat{n}\cdot \hat{n}'}{2} 
	\end{equation}
    \item For a system with multiple gapped boundaries (or closed manifold like a torus) which typically give rise to multiple ground states, we could observe signatures even without excitations.
\end{itemize}
 On the other hand, no features listed above are expected for any short-range entangled phase without topological excitations.

For real experiments, a challenge is to prepare relatively large identical copies of the system. Another challenge is to  make accurate interference measurement in large systems. Typically, the number of measurements to make a prediction for  $tr [\sigma_{\Omega}^{(1)}\cdot \sigma_{\Omega}^{(2)}]$ with a given precision grows very fast as system size grows. Therefore, it would be a difficult experiment. A good news is that a single interference  simultaneously measure $tr [\sigma_{\Omega}^{(1)}\cdot \sigma_{\Omega}^{(2)}]$ for a lot of different subsystems $\Omega$. Therefore, it should be possible to obtain a good accuracy of information convex structure with a much fewer number of measurements than what is required to measure the 2nd-Renyi entropy for a single subsystem $\Omega$. We hope this type of experimental detection of the information convex structure will be possible in the future.

%

\section{The Information convex}\label{Appendix_A}
We provide a definition of information convex $\Sigma(\Omega,\Omega')$ and $\Sigma(\Omega)$ for frustration-free local Hamiltonians and discuss some basic properties. Generalizations beyond frustration-free local Hamiltonians are briefly discussed.

\subsection{Frustration-free local Hamiltonians}
We use the following definition of frustration-free local Hamiltonians in the context of lattice models. This definition of frustration-free local Hamiltonian is similar to the definition in Ref. \cite{2013CMaPh.322..277M}.
\begin{definition}[Frustration-free local Hamiltonians]
	A frustration-free local Hamiltonian is a Hamiltonian written as $H=\sum_{i}h_{i}$, which satisfies the following:
	
	1) Each $h_i$ is a Hermitian operator acting on links within a local region of finite size $R$. To simplify our notations below, we will assume the minimal eigenvalue of each $h_i$ to be $0$.
	
	2)  $h_iP_0=0,\forall i$, where $P_0$ is the projector onto the subspace of ground states of $H$. In other words, every $h_i$ obtains its minimal eigenvalue $0$ on a ground state $\vert\psi\rangle$, i.e. $h_i\vert\psi\rangle=0,\,\forall i $.\label{Frustration_free definition}
\end{definition}

Let $H_{\Omega'}$ be the Hamiltonian of subsystem $\Omega'$, i.e. keeping terms of $H$ which are supported on the subsystem $\Omega'$. 
One can easily check that the ground state $\vert\psi\rangle$ minimize the Hamiltonian $H_{\Omega'}$, i.e. $H_{\Omega'}\otimes 1_{\bar{\Omega}'}\vert\psi\rangle=0$. Here $\bar{\Omega}'$ is the complement of $\Omega'$.

\subsection{The information convex for frustration-free local Hamiltonians}

Let us define the information convex for a general frustration-free local Hamiltonian satisfying definition \ref{Frustration_free definition} and study a few basic properties. Note that frustration-free local Hamiltonians include commuting projector Hamiltonians as a subset. Therefore, the definition applies to many exactly solved models of topological orders in 2D, 3D,  exactly solved SET models and models related to these exactly solved models by a finite depth quantum circuit.

\begin{figure}[h]
	\centering\includegraphics[scale=0.20]{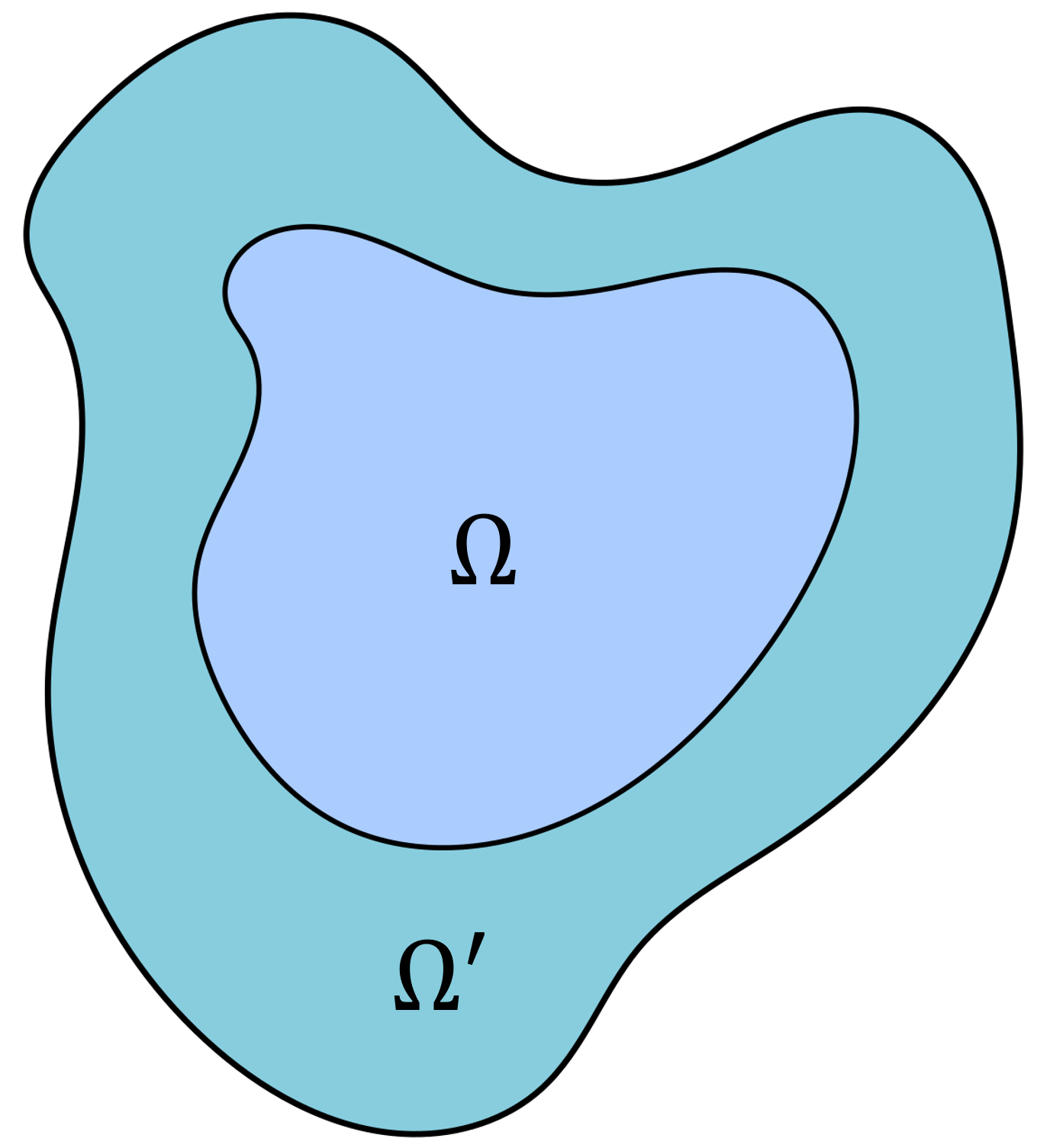}
	\caption{ An illustration of subsystem $\Omega$ and $\Omega'$. Here $\Omega\subseteq \Omega'$.
	}\label{Omega_pi}
\end{figure}

\begin{definition}[The information convex]
	For a frustration-free local Hamiltonian, define the information convex $\Sigma(\Omega,\Omega')$ to be the set of reduced density matrices on subsystem $\Omega$  obtained from reduced density matrices on a larger subsystem $\Omega'$ (see Fig.\ref{Omega_pi}, $\Omega\subseteq \Omega'\subseteq S$, $S$ is the whole system) which minimize the Hamiltonian $H_{\Omega'}$, i.e.:
	\begin{equation}
	\Sigma(\Omega,\Omega')\equiv \{\,\sigma_{\Omega}\,\vert\, \sigma_{\Omega}=tr_{[\Omega'\backslash \Omega]}\,\rho_{\Omega'} \,\,\,\,\textrm{where}\,\, \,\, tr[H_{\Omega'} \rho_{\Omega'}]=0\,\}. \label{Equation Info Convex definition}
	\end{equation}
	For the set to be interesting, we require $\Omega'$ to contain all terms in $H$ which overlap with $\Omega$. We use a simpler notation $\Sigma(\Omega)$ when we choose the minimal $\Omega'$.
	\label{Information convex definition}
\end{definition}

The definition of $\Sigma(\Omega,\Omega')$ may be motivated by the consideration that while the set of general reduced density matrices on $\Omega$  has a complicated structure due to the large number of possible excitations, the set of reduced density matrices that  minimize the energy around $\Omega$ should have a much simpler structure. Another motivation is that, as we will see, for the quantum double model (which is a zero correlation length commuting projector model) of topological orders, $\Sigma(\Omega,\Omega')$ and $\Sigma(\Omega)$ are small dimensional but nontrivial convex sets depending on subsystem topologies. Homotopically increase $\Omega'$ would not change the set $\Sigma(\Omega,\Omega')$.
Its structure contains important information about the phase.  
On the other hand, there are frustration-free Hamiltonian models with nonzero correlation length. For these models, the dimension of $\Sigma(\Omega)$ may be sensitive to the boundary length and we do not expect $\Sigma(\Omega)$ to be stable under an increase of $\Omega'$. Nevertheless, if the correlation length is finite, we expect $\Sigma(\Omega,\Omega')$ to  (approximately)  have a low dimension and simple structure when $\Omega'$ is bigger than $\Omega$ by a few correlation lengths. In this case, it seems better to consider $\Sigma(\Omega,\Omega')$ instead of $\Sigma(\Omega)$. 

In the present paper, we do not have to worry about this issue since the calculation is done in a zero correlation length commuting projector model. Nevertheless, some useful properties can be proved with the assumption of frustration-free and we will  consider this general class of Hamiltonians in the next section.

\subsection{Some general properties}

This section contains a few general properties of the information convex $\Sigma(\Omega,\Omega')$. Most importantly, it is shown that $\Sigma(\Omega,\Omega')$ is always a \emph{compact convex set}. This allows us to borrow tools from convex analysis and explore the structure of the convex set $\Sigma(\Omega,\Omega')$ in this new context.  The concept  \emph{extremal points} is introduced. Additional discussions concern some general properties of $\Sigma(\Omega,\Omega')$ under a change of $\Omega$ or $\Omega'$.

\begin{theorem}
	$\Sigma(\Omega,\Omega')$ is a convex set. \label{Thm convex}
\end{theorem}
Here, the set $\Sigma(\Omega,\Omega')$ being a convex set means the condition that for any two reduced density matrices $\sigma^{(1)}_{\Omega},\sigma^{(2)}_{\Omega}\in \Sigma(\Omega,\Omega')$ and arbitrary $p\in [0,1]$, we always have $p\,\sigma^{(1)}_{\Omega} +(1-p)\sigma^{(2)}_{\Omega} \in \Sigma(\Omega,\Omega')$.
\begin{proof}
	For $\sigma^{(1)}_{\Omega},\sigma^{(2)}_{\Omega}\in \Sigma(\Omega,\Omega')$, by definition, there exist $\rho^{(1)}_{\Omega'}$ and $\rho^{(2)}_{\Omega'}$ such that:
	\begin{equation}
	\sigma^{(1)}_{\Omega}=tr_{[\Omega'\backslash \Omega]}\rho^{(1)}_{\Omega'},\quad \sigma^{(2)}_{\Omega}=tr_{[\Omega'\backslash \Omega]}\rho^{(2)}_{\Omega'},\quad 
	tr[\rho^{(1)}_{\Omega'}H_{\Omega'}]=tr[\rho^{(2)}_{\Omega'}H_{\Omega'}]=0.
	\end{equation}
	Therefore, $p \rho^{(1)}_{\Omega'} +(1-p) \rho^{(2)}_{\Omega'}$ with $p\in [0,1]$ is also a density matrix that minimize the Hamiltonian $H_{\Omega'}$, and:
	\begin{equation}
	p \sigma^{(1)}_{\Omega} +(1-p)\sigma^{(2)}_{\Omega}=tr_{[\Omega'\backslash \Omega]} [p \rho^{(1)}_{\Omega'} +(1-p)\rho^{(2)}_{\Omega'}] \quad\Rightarrow\quad p \sigma^{(1)}_{\Omega} +(1-p)\sigma^{(2)}_{\Omega}\in \Sigma(\Omega,\Omega').
	\end{equation}
\end{proof}

\begin{theorem}
	$\Sigma(\Omega,\Omega')$ is a compact subset of $R^{{N}}$. Here ${N}$ is a finite number which could depend on the choice of $\Omega$ and $\Omega'$.\label{Thm Compact convex}
\end{theorem}
\begin{proof}
	The set of all reduced density matrices on $\Omega$ is a compact (closed and bounded) subset of $R^{{M}^2}$. Here ${M}= \dim \mathcal{H}(\Omega)$ is the dimension of the Hilbert space on subsystem $\Omega$. $M$ is finite for a lattice model with $\Omega$ containing a finite number of links (or sites) and each link (or site) has a corresponding finite-dimensional Hilbert space.
	$\Sigma(\Omega,\Omega')$ is a subset of the set of all reduced density matrices on $\Omega$ and therefore it is a bounded subset of $R^{{M}^2}$. Furthermore, 	$\Sigma(\Omega,\Omega')$ is closed. Therefore, $\Sigma(\Omega,\Omega')$ is a compact subset of $R^{{N}}$ with finite $N$.
\end{proof}
\begin{remark}
	Theorem \ref{Thm convex} and Theorem \ref{Thm Compact convex} show that $\Sigma(\Omega,\Omega')$ is a compact convex subset of $R^N$ for some finite $N$. This applies to all the convex sets we will discuss in this paper. We will call them convex sets for short and it is understood that they are compact convex subsets of $R^N$ for some finite $N$.
\end{remark}
\begin{definition}[Extremal point]
	An extremal point of $\Sigma(\Omega,\Omega')$ is a reduced density matrix $\sigma_{\Omega}\in \Sigma(\Omega,\Omega')$ such that if $\sigma_{\Omega}=p\, \sigma^{(1)}_{\Omega} +(1-p)\sigma^{(2)}_{\Omega}$ with $\sigma^{(1)}_{\Omega}, \sigma^{(2)}_{\Omega}\in \Sigma (\Omega,\Omega')$ and $p\in(0,1)$, then $\sigma^{(1)}_{\Omega}=\sigma^{(2)}_{\Omega}=\sigma_{\Omega}$.
\end{definition}
In other words, an extremal point is a point (reduced density matrix) in $\Sigma(\Omega,\Omega')$ which could not be prepared by other points in $\Sigma(\Omega,\Omega')$ with a probability distribution.
\begin{Proposition}
	$\Sigma(\Omega,\Omega')$ is uniquely determined by the set of extremal points. Furthermore, if $\Sigma(\Omega,\Omega')$ is $n$-dimensional, then any point in $\Sigma(\Omega,\Omega')$ can be written as a convex combination of at most $n+1$ extremal points. \label{Thm extremal unique}
\end{Proposition}
Here, convex combination is a combination with a probability distribution $\{p_i\}_{i=1}^{n+1}$. In other words, for any $\sigma_{\Omega}\in \Sigma(\Omega,\Omega')$  it is possible to find a (sub)set of extremal points $\{\sigma^i_{\Omega}\}_{i=1}^{n+1}$ and  a probability distribution $\{p_i\}_{i=1}^{n+1}$ such that $\sigma_{\Omega}=\sum_{i=1}^{n+1} p_{i}\sigma^i_{\Omega}$.
\begin{proof}
	First, notice that  $\Sigma(\Omega,\Omega')$ is a compact convex subset of $R^N$ for some finite $N$, i.e., the result of theorems \ref{Thm convex} and  \ref{Thm Compact convex}. Then, use the Minkowski-Caratheodory theorem, \footnote{The Minkowski-Caratheodory theorem together with its generalization to infinite dimension, i.e., the Krein-Milman theorem can be found in the following link: http://math.caltech.edu/Simon\_Chp8.pdf.}, which says that, for $\Sigma$, a compact convex subset of dimension $n$ ($n$ is finite and $\Sigma$ is a subset of $R^N$ for some finite $N$), any point in $\Sigma$ can be written as a convex combination of at most $n+1$ extremal points.
\end{proof}

\begin{Proposition}
	Every extremal point of $\Sigma(\Omega,\Omega')$ has a purification in $\Omega'$. In other words, there exists a pure state  $\vert\varphi\rangle_{\Omega'}$ such that $\sigma_{\Omega}=tr_{[\Omega'\backslash \Omega]} \vert\varphi\rangle_{\Omega'\,\Omega'}\langle\varphi\vert$, if $\sigma_{\Omega}$ is an extremal point of  $\Sigma(\Omega,\Omega')$. \label{Purify an extremal}
\end{Proposition}

In the following, we discuss a few properties of $\Sigma(\Omega,\Omega')$ when one tries to change $\Omega$ or $\Omega'$.

\begin{theorem}
	One obtains a convex subset of $\Sigma(\Omega,\Omega')$ when one replaces $\Omega'$ by a larger subsystem $\Omega''$, i.e.:
	\begin{equation}
	\Sigma(\Omega,\Omega'')\subseteq \Sigma(\Omega,\Omega') \quad \textrm{for}\quad \Omega'\subseteq \Omega''.
	\end{equation}
\end{theorem}

\begin{corollary}
	Let $\vert\psi\rangle$ be a ground state of the frustration-free local Hamiltonian $H$. Then, the corresponding reduced density matrix $\sigma_{\Omega }^{1}\equiv tr_{\bar{\Omega}}\vert\psi\rangle\langle \psi\vert$ satisfies $\sigma_{\Omega}^1\in\Sigma (\Omega,\Omega')$.
\end{corollary}

\begin{theorem}
	The mapping  $\Gamma: \Sigma(\Omega,\Omega')\rightarrow \Sigma(\omega,\Omega')$ defined by $\Gamma[\sigma_{\Omega}]=tr_{[\Omega\backslash\omega]}\sigma_{\Omega}$ is surjective and it preserves the convex structure. Here,  $\omega\subseteq \Omega$. \label{Mapping theorem}
\end{theorem}
\begin{proof} The mapping $\Gamma$ is surjective. This follows from $tr_{[\Omega'\backslash\omega]}= tr_{[\Omega\backslash\omega]} tr_{[\Omega'\backslash\Omega]}$. The following is about the convex structure.
	Let $\sigma^{(1)}_{\Omega},\sigma^{(2)}_{\Omega}\in\Sigma(\Omega,\Omega')$, and $p\in[0,1]$. From the linearity  of the $tr_{[\Omega\backslash\omega]}$  operation, we have:
	\begin{equation}
	\Gamma[p\,\sigma^{(1)}_{\Omega}+ (1-p)\sigma^{(2)}_{\Omega}]=p\,\Gamma[\sigma^{(1)}_{\Omega}] +(1-p)\Gamma[\sigma^{(2)}_{\Omega}].
	\end{equation}
	This result shows (by definition) that the mapping $\Gamma$ preserves the convex structure.
\end{proof}
Theorem \ref{Mapping theorem} gives constraints to the number of extremal points.

\begin{corollary}
	If the number of extremal points of $\Sigma(\Omega,\Omega')$ is a finite number $N_{\Omega}$, then the number of extremal points of $\Sigma(\omega,\Omega')$ is a finite number $N_\omega$ satisfying $N_\omega\le N_\Omega$. Furthermore, an extremal point of $\Sigma(\omega,\Omega')$ must be the image of some extremal point of  $\Sigma(\Omega,\Omega')$ under the mapping $\Gamma$.
	Here, $\omega\subseteq \Omega$.
\end{corollary}
\begin{proof}
	This result follows from the fact that the image of a nonextremal point of $\Sigma(\Omega,\Omega')$ cannot be an extremal point of $\Sigma(\omega,\Omega')$ unless it is also the image of an extremal point of $\Sigma(\Omega,\Omega')$. 
\end{proof}
\begin{remark}
	Constraints for the case with an infinite number of extremal points may also be deduced from theorem \ref{Mapping theorem}.
\end{remark}

\subsection{Beyond frustration-free local Hamiltonians}

In this section, we briefly discuss what we expect for generalizations of information convex to models beyond frustration-free local Hamiltonians and hope that more rigorous results will be available in the future. 

Let us first consider a frustration-free local Hamiltonian $H$ with $H_{\Omega'}$ having a finite energy gap $\Delta$ (between the ground states and the 1st excited state) and consider a generalization of $\Sigma(\Omega,\Omega')$ into $\Sigma(\Omega,\Omega'\vert \epsilon)$. Here, $0\le\epsilon\ll \Delta$,
\begin{equation}
\Sigma(\Omega,\Omega'\vert\epsilon )\equiv \{\,\sigma_{\Omega}(\epsilon)\,\vert\, \sigma_{\Omega}(\epsilon)=tr_{[\Omega'\backslash\Omega]}\,\rho_{\Omega'} \,\,\,\,\textrm{where}\,\, \,\, tr[H_{\Omega'} \rho_{\Omega'}]\in [0,\epsilon]\,\}.
\end{equation}
It is straightforward to show that $\Sigma(\Omega,\Omega'\vert\epsilon )$ is a convex set.
Comparing with Eq. (\ref{Equation Info Convex definition}), if $\epsilon>0$, then $\rho_{\Omega'}$ can have small mixture of excited states.  Due to the large number of excited states, the convex set $\Sigma(\Omega,\Omega'\vert \epsilon)$ with $\epsilon>0$ will be of a large dimension even if $\Sigma(\Omega,\Omega')$ is a small dimensional convex set. Nevertheless,  $\Sigma(\Omega,\Omega'\vert \epsilon)$ stretches out in the directions of excited states by a ``distance" suppressed by $\epsilon/\Delta$. Here, the distance could be measured by the minimal  deviation of the fidelity (between $\sigma_{\Omega}(\epsilon)\in 	\Sigma(\Omega,\Omega'\vert\epsilon )$ and $\sigma_{\Omega}\in 	\Sigma(\Omega,\Omega' )$) from 1:
\begin{equation}
l(\sigma_{\Omega}(\epsilon))\equiv  [1-F(\sigma_{\Omega}(\epsilon),\sigma_{\Omega})]_{\min},
\qquad\qquad \textrm{with}\qquad 
\sigma_{\Omega}(\epsilon)\in\Sigma(\Omega,\Omega'\vert\epsilon ) \quad\textrm{and}\quad \sigma_{\Omega}\in 	\Sigma(\Omega,\Omega' ).
\end{equation}
Here fidelity is defined as $F(\rho,\sigma)\equiv \bigg( tr\sqrt{\rho^{\frac{1}{2}}\sigma\rho^{\frac{1}{2}}} \bigg)^2$. One can show that:
\begin{equation}
l(\sigma_{\Omega}(\epsilon))\le \frac{\epsilon}{\Delta}
\quad\qquad \quad\quad \textrm{for}\quad \forall \sigma_{\Omega}(\epsilon)\in \Sigma(\Omega,\Omega'\vert\epsilon ).
\end{equation}
For $0\le\epsilon\ll \Delta$, we could still approximately treat the convex set  $\Sigma(\Omega,\Omega'\vert \epsilon)$ as small dimensional with the same structures as $\Sigma(\Omega,\Omega')$.

Now let us consider the case beyond frustration-free local Hamiltonians. We focus on the case that local perturbations are added to a gapped frustration-free local Hamiltonian (with energy gap $\Delta$), the case discussed in \cite{2010JMP....51i3512B,2013CMaPh.322..277M}. Note that we do need to generalize  $\Sigma(\Omega,\Omega')$ into $\Sigma(\Omega,\Omega'\vert \epsilon)$ with some $0\le\epsilon\ll \Delta$ in order to get meaningful structures. Consider a model with topological order and the system is defined on a torus $S=T^2$ with length $L$ and a correlation length $\xi\ll L$. For the unperturbed model, the ground state degeneracy is exact and $\Sigma(S,S)$ is a convex set with (an infinite number of) extremal points in one to one correspondence with the (pure) ground states. 
However, local  perturbations will split the ground-state energies by the order $\Delta \exp(-L/\xi)$; for a more rigorous bound  of the energy splitting see  \cite{2010JMP....51i3512B,2013CMaPh.322..277M}. 
Therefore, in order to construct a convex set with similar structure as the $\Sigma(S,S)$ of the unperturbed model, $\Sigma(S,S)$ is no longer a good choice, since it does not keep all the low energy states corresponding to the degenerate ground states of the unperturbed model. We need to choose $\Sigma(S,S \vert\epsilon)$ with $\epsilon \sim \Delta \exp(-L/\xi)$ for the perturbed theory.

More generally, we expect the $\Sigma(\Omega,\Omega')$ for the unperturbed model, with $\Omega'$ thicker than $\Omega$ by length  $L'\gg \xi$ to be generalized into $\Sigma(\Omega,\Omega'\vert\epsilon)$ with $\epsilon \sim \Delta \exp(-L'/\xi)$. Since  $\exp(-L'/\xi)\ll 1$, the convex set $\Sigma(\Omega,\Omega'\vert\epsilon)$ is approximately small dimensional and it should have very similar structure to the $\Sigma(\Omega,\Omega')$ in the unperturbed model.

\section{$G$ quantum double with $K\subseteq G$ boundary}\label{Quantum Double}
\subsection{The Hamiltonian of $G$ quantum double with $K\subseteq G$ boundary}
\begin{figure}[h]
	\centering\includegraphics[scale=0.420]{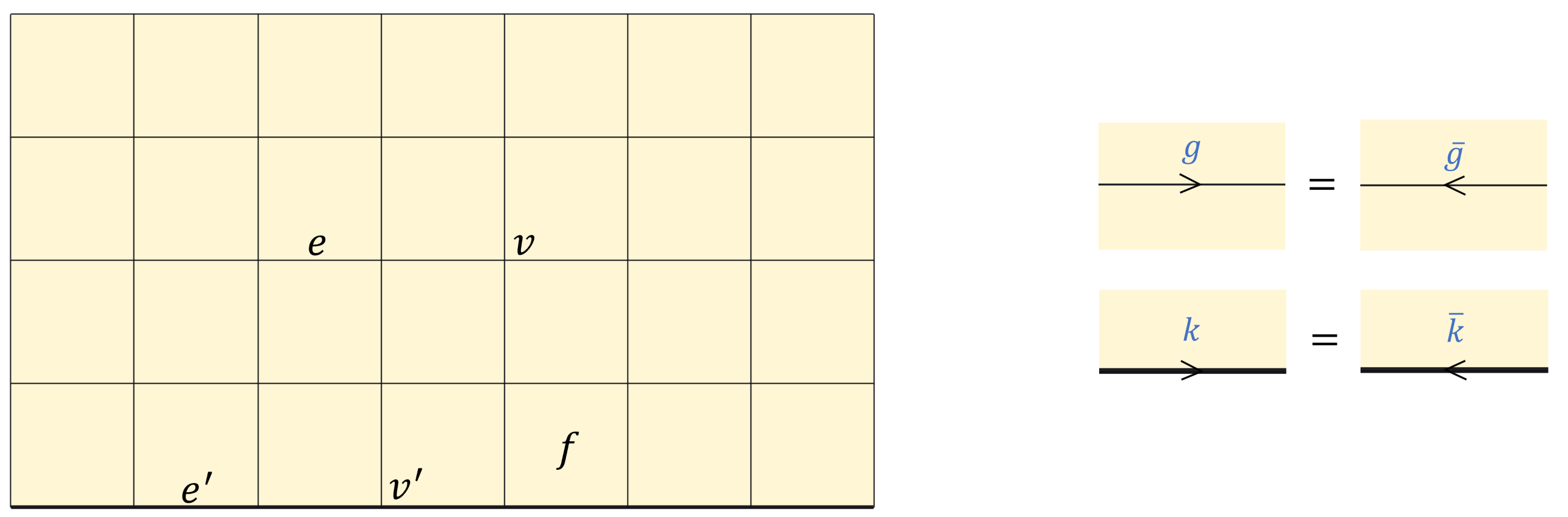}
	\caption{A square lattice with a boundary, generalization to other lattices is straightforward. Labels are as follows: bulk link $e$, boundary link $e'$, bulk vertex $v$, boundary vertex $v'$, and face $f$. Note that in our model, $e'$, $v'$ only lie in the 1D boundary lattice shown in thicker line.
	}\label{Lattice_1}
\end{figure}
A quantum double model on an orientable 2D lattice without boundary is defined for any finite group $G$ \cite{2003AnPhy.303....2K,2008PhRvB..78k5421B}. Let us consider a square lattice shown in Fig. \ref{Lattice_1}, generalization to other lattices is straightforward.
On a lattice with a boundary, a gapped  boundary can be defined for each subgroup $K\subseteq G$ \cite{2011CMaPh.306..663B,2017CMaPh.355..645C}. 
In addition to the subgroup $K$, the boundary can depend on a 2-cocycle of $K$  \cite{2011CMaPh.306..663B}. In the current work, we focus on  the untwisted boundaries, i.e. those with trivial 2-cocycles.

\begin{figure}[h]
	\centering\includegraphics[scale=0.53]{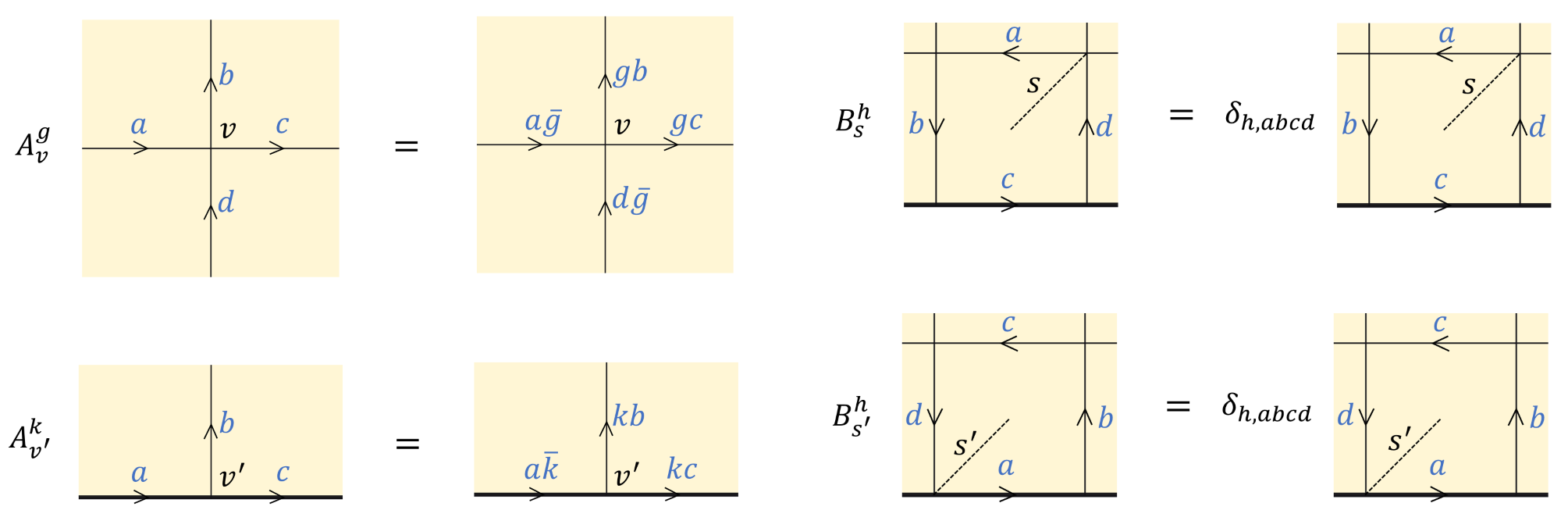}
	\caption{ The diagram shows the definitions of the operators $A_v^g$, $B_s^h$, $A_{v'}^k$, $B_{s'}^h$ with $g,h\in G$ and $k\in K$.
	}\label{Operators}
\end{figure}

The total Hilbert space is a tensor product of the Hilbert space on each link. The Hilbert space for each bulk link (labeled by $e$) is $\vert G\vert$ dimensional: $\mathcal{H}_e=span \{\,\vert g\rangle_e\,\vert\, g\in G\, \}$, where $\{\,\vert g\rangle_e\,\vert\, g\in G\, \}$ is an orthonormal basis.
The Hilbert space for each boundary link (labeled by $e'$, thicker links in Fig. \ref{Lattice_1}) is $\vert K\vert$ dimensional: $\mathcal{H}_{e'}=span \{\,\vert k\rangle_{e'} \,\vert\, k\in K \,\}$, where $\{\,\vert k\rangle_{e'} \,\vert\, k\in K \,\}$ is an orthonormal basis. 
We denote a vertex in the bulk (bulk vertex) as $v$ and a vertex on the boundary (boundary vertex) as $v'$, and denote a face as $f$. A bulk site $s=(v,f)$ is a pair containing a face $f$ and an adjacent bulk vertex $v$, a boundary site $s'=(v',f)$ is a pair containing a face $f$, and an adjacent boundary vertex $v'$. Our Hamiltonian for $G$ quantum double with a $K\subseteq G$ boundary is
\begin{equation}
H= \sum_{v}(1- A_v) +\sum_f(1- B_f) +\sum_{v'}(1- A_{v'}^{K}).    \label{The Hamiltonian}
\end{equation}
Constants are added into the Hamiltonian simply to keep the minimal eigenvalue to be zero. Here
\begin{equation}
A_v\equiv\frac{1}{\vert G\vert} \sum_{g\in G} A_v^{g};\qquad B_f\equiv B_s^1;\qquad A_{v'}^K\equiv\frac{1}{\vert K\vert} \sum_{k\in K} A_{v'}^k.
\end{equation}
Each operator $A_v^g$, $B_s^h$, $A_{v'}^k$, $B_{s'}^h$ with $g,h\in G$ and $k\in K$  acts on a few links around a vertex or a face. They are defined in Fig. \ref{Operators}.

One can easily check that all terms in the Hamiltonian commute, and that $A_v, B_f, A_{v'}^{K}$ are projectors (so do $1-A_v$, $1-B_f$, $1-A_{v'}^{K}$), i.e. $A_v^2=A_v$, $B_f^2=B_f$, $(A_{v'}^{K})^2=A_{v'}^{K}$. A state $\vert\psi\rangle$ is a ground state if and only if it satisfies
\begin{equation}
A_v\vert\psi\rangle =B_f \vert \psi\rangle =A_{v'}^{K}\vert \psi\rangle =\vert\psi\rangle,\qquad \forall\, v,f,v'.
\end{equation}
For a system  with $D^2$ topology (i.e. cover a disk $D^2$ with lattice), it can be shown that there is a unique ground state $\vert\psi\rangle$ that can be written as (up to normalization)
\begin{equation}
\vert \psi\rangle = \prod_v A_v \cdot \prod_{v'} A_{v'}^{K} \vert 1,1,\cdots,1\rangle.
\end{equation}
Here, $1$ represents the identity of  group $G$ (or $K$). This ground state is an equal weight superposition of all ``zero flux" configurations. Let us define the reduced density matrix on subsystem $\Omega$ calculated from the unique ground state $\vert\psi\rangle$ as $\sigma^1_{\Omega}$, i.e. $\sigma^1_{\Omega}\equiv tr_{\bar{\Omega}}\vert\psi\rangle\langle\psi\vert$, where $\bar{\Omega}$ is the complement of $\Omega$. It will appear many times in later sections. 
\begin{remark}
	Our Hilbert space and Hamiltonian is closely related to the ones in previous works \cite{2011CMaPh.306..663B,2017CMaPh.355..645C}, but there are differences. Our Hilbert space is ``smaller" than that in Refs. \cite{2011CMaPh.306..663B} and \cite{2017CMaPh.355..645C}. 
	It is the effective Hilbert space when certain terms in the Hamiltonian \cite{2011CMaPh.306..663B,2017CMaPh.355..645C} are not excited.  Our Hamiltonian is the effective Hamiltonian of the models in \cite{2011CMaPh.306..663B,2017CMaPh.355..645C} when no excitations of the types discussed above are present.
	The excitations that could be created in our model also appear in the models in  \cite{2011CMaPh.306..663B,2017CMaPh.355..645C}, but some of  the excitations that could be created in Refs. \cite{2011CMaPh.306..663B} or \cite{2017CMaPh.355..645C} may not be created in our  model. Especially, the confined excitations inside a boundary \cite{2017CMaPh.355..645C} do not exist in our model. On the other hand, as  will be described below, our model allows a set of deconfined topological excitations to carry boundary superselection sectors $\alpha$ and quantum dimension $d_{\alpha}$. The set of $(\alpha,d_{\alpha})$ is coincident with what has been discussed in Ref. \cite{2017CMaPh.355..645C} for confined boundary excitations. See Eq. (\ref{quantum dimension boundary}) and Secs. \ref{HJW Section},\ref{Boundary Ribbon Section},\ref{Sec. Boundary prepare Omega_2} for more details.
\end{remark}

\subsection{The calculation of the information convex for  $G$ quantum double with $K\subseteq G$ boundary}\label{The calculation of information convex}
The $G$ quantum double with $K\subseteq G$ boundary is a model with commuting projector Hamiltonian Eq. (\ref{The Hamiltonian}), and on a ground state, each projector obtains its minimal eigenvalue $0$. Therefore, it is an example of frustration-free local Hamiltonian. The information convex $\Sigma({\Omega,\Omega'})$ is a convex set uniquely determined by the set of extremal points, see Theorem \ref{Thm extremal unique}. Therefore our task here is to find the set of extremal points.

Consider a reduced density matrix $\rho_{\Omega'}$ that minimizes the Hamiltonian $H_{\Omega'}$. Let us write $\rho_{\Omega'}$ in its diagonal form, $\rho_{\Omega'}=\sum_{\alpha}\lambda_{\alpha} \vert\alpha\rangle_{\Omega'\,\Omega'}\langle\alpha\vert$ with $_{\Omega'}\langle \alpha\vert \beta\rangle_{\Omega'}=\delta_{\alpha,\beta}$. 
\begin{equation}
tr (H_{\Omega'}\rho_{\Omega'})=0 \quad\Leftrightarrow\quad  H_{\Omega'}\vert\alpha\rangle_{\Omega'}=0\quad \forall \alpha.
\end{equation}
In other words, each $\vert\alpha\rangle_{\Omega'}$ is a ground state of $H_{\Omega'}$. According to proposition \ref{Purify an extremal}, to find the extremal points of $\Sigma(\Omega,\Omega')$, it is enough to consider the set of reduced density matrices $\sigma_{\Omega}= tr_{[\Omega'\backslash\Omega]}\vert\alpha\rangle_{\Omega'\,\Omega'} \langle \alpha\vert$.

Let us take the minimal $\Omega'$, i.e. consider $\Sigma(\Omega)$. 
In this case, $H_{\Omega'}\vert\alpha\rangle_{\Omega'} =0$ is equivalent to the following conditions:\\
(a) $B_f\vert \alpha\rangle_{\Omega'}=\vert\alpha\rangle_{\Omega'}$, for $f\in \partial \Omega$.\\
(b) $B_f\vert\alpha\rangle_{\Omega'}=\vert\alpha\rangle_{\Omega'}$, for $f\in \Omega$.\\
(c) $A_{v}^{g}\vert \alpha\rangle_{\Omega'} =\vert \alpha \rangle_{\Omega'}$ and $A_{v'}^{k}\vert\alpha\rangle_{\Omega'}= \vert\alpha\rangle_{\Omega'}$, for $v,v'\in  \Omega$ and $g\in G$, $k\in K$.\\
(d) $A_{v}^{g}\vert\alpha\rangle_{\Omega'}= \vert\alpha\rangle_{\Omega'}$ and $A_{v'}^{k}\vert\alpha\rangle_{\Omega'}= \vert\alpha\rangle_{\Omega'}$, for $v,v'\in \partial \Omega$ and $g\in G$, $k\in K$.\\
Here, we say $f\in\Omega$ if $B_f$ is supported on $\Omega$; we say $f\in\partial\Omega$ if $B_f$ has support overlap with $\Omega$ but is not supported on $\Omega$; we say $v,v'\in \Omega$ if $A_v^g$ and $A_{v'}^k$ are supported on $\Omega$;  we say $v,v'\in\partial\Omega$ if $A_v^g$ and $A_{v'}^k$ have support overlap with $\Omega$ but not supported on $\Omega$.

Then, one can show it is possible to write $\vert\alpha\rangle_{\Omega'}$ in the following Schmidt basis:
\begin{equation}
\vert \alpha\rangle_{\Omega'} =\sum_{\{h^{I}_{a} \},\lambda}\,\sqrt{p^{\lambda}_{\{h^I_{a}\}}} \,\vert \{h^I_{a}\};\lambda\rangle_{\Omega}\otimes \vert\{h^{I}_{a}\};\lambda\rangle_{\Omega'\backslash\Omega}\quad \Rightarrow\quad \sigma_{\Omega}=\sum_{\{h^{I}_{a} \},\lambda}{p^{\lambda}_{\{h^I_a\}}} \vert\{h^I_a\};\lambda\rangle _{\Omega\,\,\Omega}\langle \{h^I_a\};\lambda\vert.    \label{Reduced}
\end{equation}
Here $I=1,\cdots, M$ labels the number of disconnected pieces of  $\partial\Omega\cap\partial\bar{\Omega}$ ($\bar{\Omega}$ is the complement of $\Omega$). $\{h^I_a \}$ is a set of link values $h^I_{a}\in G$, \footnote{Or $h^I_{a}\in K$ for the case involving boundary links in the $I$th piece. This will not happen in this paper.} with $a=1,\cdots N_I$ labeling the links along the $I$th piece and $h^{I}_1 h^I_{2}\cdots h^I_{N_I}=h^I$. Each $h_a^{I}$ is obtained from a product of group elements on one or more links connecting $v,v'\in \partial\Omega$, and it is important to notice that those vertices  $v,v'\in \Omega$ do not count even if they are near $\partial\Omega$.
The same set of $h^I_a$ appears in $\vert \{h^I_a\};\lambda \rangle_{\Omega}$ and  $\vert \{h^{I}_a\};\lambda\rangle_{\Omega'\backslash \Omega}$, this is due to  condition (a).

Conditions (b) and (c) are equivalent to the following:
\begin{equation}
B_f \vert \{h^I_a\};\lambda \rangle_{\Omega}= A_v\vert \{h^I_a\};\lambda\rangle_{\Omega}= A_{v'}^{K}\vert \{h^I_a\};\lambda \rangle_{\Omega}=\vert \{h^I_a\};\lambda\rangle_{\Omega}
\qquad \quad\forall f,v,v'\in \Omega.    \label{Flat}
\end{equation}
This requires  $\vert \{h^I_a\};\lambda\rangle_{\Omega}$  be an equal weight superposition of all configurations with ``zero flux," which could be related by a product of $A_v^{g}$ and $A_{v'}^k$ operators, $v,v'\in \Omega$. Note that, $A_v^g$ or $A_{v'}^k$ with $v,v'\in\Omega$ does not mix different $\{h^{I}_a\}$ sectors. Here, ``zero flux" stands for the condition that a configuration has eigenvalue $B_f=1$ for $\forall f\in\Omega$.  It may happen that two ``zero-flux" configurations could not be related by a product of $A_v^{g}$ and $A_{v'}^{k}$ operations, and the index $\lambda$ labels exactly those additional degrees of freedom. Note that the number of additional degrees of freedom $\lambda$ depends on $\{h^I_a\}$ in general. 

Finally, the operators $A_v^{g}$, $A_{v'}^k$ with $v,v'\in\partial\Omega$ mix different $\{h^I_a \}$ sectors.
Condition (d) gives constraint for the probability distribution $\{ p^{\lambda}_{\{h^I_a\}} \}$. 
Each of the $A_v^g$, $A_{v'}^k$ operators with $v,v'\in \partial \Omega$ is a unitary operator that could be written as products of unitary operators on each link. Define $A_{v}^{g}(\Omega)$, $A_{v'}^k(\Omega)$ to be the ``truncation"  of the operators $A_v^g$, $A_{v'}^k$ (with $v,v'\in \partial \Omega$) onto $\Omega$  in the fashion that $U_{\Omega'}(\Omega)= U_{\Omega}$ for $U_{\Omega'}=U_{\Omega}\otimes U_{\Omega'\backslash\Omega}$. Let us define a group of operators $A(\Omega)$ as following:

\[
A(\Omega)\equiv \{ A(g,k)\,\vert\, A(g,k)\equiv\prod_{v\in\partial \Omega}[A_{v}^{g(v)}(\Omega)]\cdot\prod_{v'\in \partial \Omega}[A_{v'}^{k(v')}(\Omega)]\textrm{ with } g(v)\in G, k(v')\in K \}.
\]
Now, we can write down the constraint on $\sigma_{\Omega}$ caused by (d): 
\begin{eqnarray}
A(g,k) \sigma_{\Omega} A(g,k)^{\dagger}&=&\sigma_{\Omega}\qquad \forall \,A(g,k)\in A(\Omega)\nonumber\\
&\Leftrightarrow& A_v^g(\Omega)\sigma_{\Omega}A^{\bar{g}}_{v}(\Omega) = A_{v'}^{k}(\Omega)\sigma_{\Omega}A_{v'}^{\bar{k}}(\Omega)\quad  \forall v,v'\in\partial\Omega,\,\,\,\, g\in G, k\in K.
\label{V_rotation}
\end{eqnarray}

\begin{figure}[h]
	\centering\includegraphics[scale=0.420]{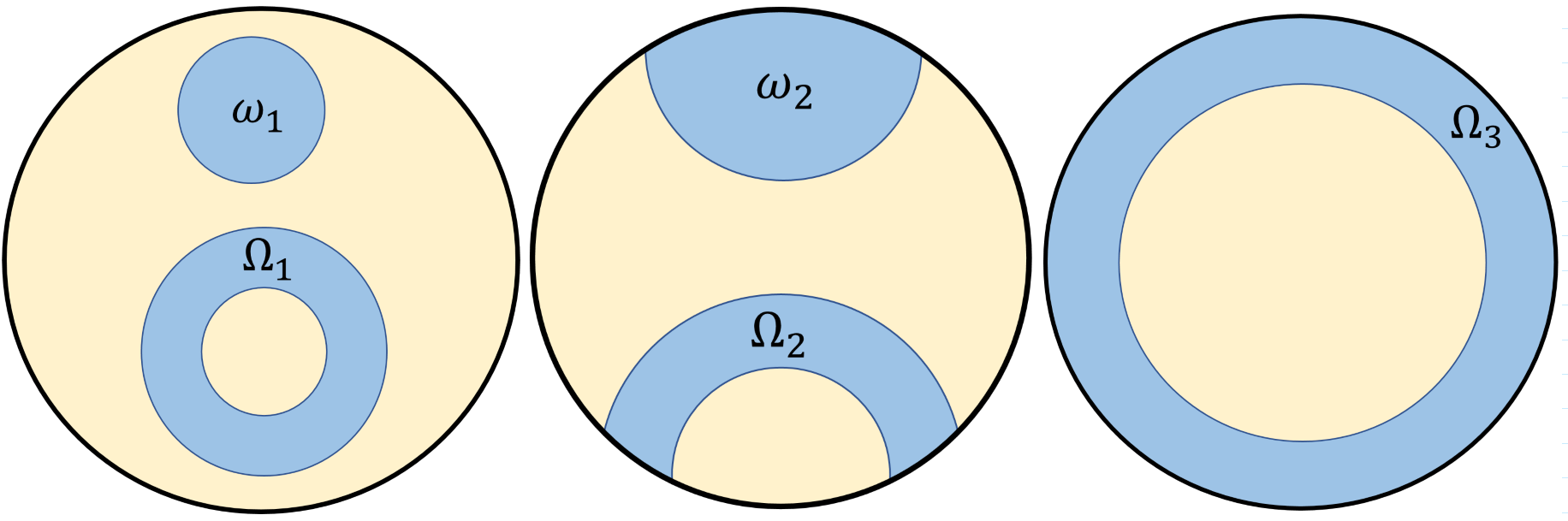}
	\caption{An illustration of topologically distinct subsystem types: $\omega_1$, $\Omega_1$, $\omega_2$, $\Omega_2$ and $\Omega_3$. The $G$ quantum double model lives   on a disk $D^2$ with a single gapped $K\subseteq G$ boundary.
		Note that, the relation to the boundary is considered as part of the topological data.}\label{Omega}
\end{figure}

Next, let us go to some examples. First, recall the subsystem choices $\omega_1,\Omega_1,\omega_2,\Omega_2,\Omega_3$ discussed in the paper, see Fig. \ref{Omega}. We will see below that the structures of the information convex of these choices of subsystem all have simple physical meaning. 

Our strategy for solving a information convex $\Sigma(\Omega)$ is to apply the method developed above, i.e. follow Eqs. (\ref{Reduced},\ref{Flat},\ref{V_rotation}). In practice we find that, the problem is reduced to a problem for some \emph{minimal diagram}, a simplified lattice with less links and a corresponding Hilbert space $\mathcal{H}^{\ast}(\Omega)$. The problem is to solve $\Sigma^{\ast}(\Omega)$, a suitably defined set of density matrices on $\mathcal{H}^{\ast}(\Omega)$, which satisfies a set of requirements very similar to Eqs. (\ref{Reduced},\ref{Flat},\ref{V_rotation}). 

$\Sigma(\Omega)$ and $\Sigma^{\ast}(\Omega)$  have identical convex structures, i.e. there is a naturally defined bijective mapping $\pi:\Sigma(\Omega)\to\Sigma^{\ast}(\Omega)$, which preserves the convex structure (maps a line segment to a line segment). The number of extremal points does not change under this mapping.
Furthermore, there are physical properties (properties of the density matrices) of $\Sigma(\Omega)$, which are invariant under continuous deformations of $\Omega$, e.g., the entanglement entropy difference between two extremal points (in the case with more than one extremal point).
We call these properties \emph{topological invariant structures}  (or \emph{structures} for short) of the information convex $\Sigma(\Omega)$.  $\Sigma^{\ast}(\Omega)$ captures all the topological invariant structures of $\Sigma(\Omega)$.

\subsubsection{The calculation of  $\Sigma(\omega_1)$}\label{omega_1}

For subsystem $\omega_1$, i.e. a disk in the bulk, $\partial\omega_1\cap \partial \bar{\omega}_1=\partial\omega_1$ contains one piece. 
Therefore, $M=1$. Let us relabel $h^1_a \rightarrow h_a$ and $N_1\rightarrow n$. It is well known that $\Sigma(\omega_1)$ contains a single element, i.e. the reduced density matrix calculated from the ground state $\vert\psi\rangle$,
\begin{equation}
\Sigma(\omega_1)=\{\sigma^1_{\omega_1} \}\qquad\textrm{with}\qquad \sigma^1_{\omega_1}=tr_{\bar{\omega}_1}\vert\psi\rangle\langle \psi\vert. \label{Sigma_omega1}
\end{equation}
This result is consistent with the fact that topological orders have locally indistinguishable ground states.  The reduced density matrix $\sigma^{1}_{\omega_1}$ can be found in a number of references, for example, Refs. \cite{2009PhRvL.103z1601F,2011PhRvB..84s5120G}. Simple as it is, this result is a powerful statement for the study of local perturbations  (known as the TQO-2 condition) \cite{2010JMP....51i3512B}. Also, it strongly constraints the possible form of operators that create excitations on a ground state once combined with the HJW theorem, see Sec. \ref{HJW Section}.
Let us briefly recall that $\sigma^{1}_{\omega_1}$ can be written as:
\begin{equation}
\sigma^1_{\omega_1}=\frac{1}{\vert G\vert^{n-1}}\sum_{\{h_a\}} \vert \{h_a\}\rangle_{\omega_1\,\,\omega_1}\langle \{h_a \}\vert. 
\label{omega_1 reduced density matrix}
\end{equation}
Here, the sum of $\{h_a \}$ is the shorthand notation for the sum of different $\{ h_1,\cdots,h_n \}$, and $\vert \{h_a \} \rangle_{\omega_1}$ is a unique state fixed by  two requirements:\\
1) The set of values  on $\partial\omega_1$ are $\{h_a \}$, $a=1,\cdots, n$ with $h_a\in G$.\\
2) The requirement in Eq. (\ref{Flat}), i.e. $ B_f\vert \{h_a\} \rangle_{\omega_1}=A_{v}\vert \{h_a\} \rangle_{\omega_1}=\vert \{h_a\} \rangle_{\omega_1}$ for $\forall f,v\in \omega_1$.

The second requirement implies $h_1h_2\cdots h_n=1$, and ends up with $\vert G\vert^{n-1}$ choices for $\{h_a\}$. Also, it guarantees $\vert \{h_a\}\rangle_{\omega_1}$ to be an equal weight superposition of all zero-flux configurations with fixed $\{h_a\}$ at $\partial \omega_1$.
\begin{figure}[h]
	\centering\includegraphics[scale=0.320]{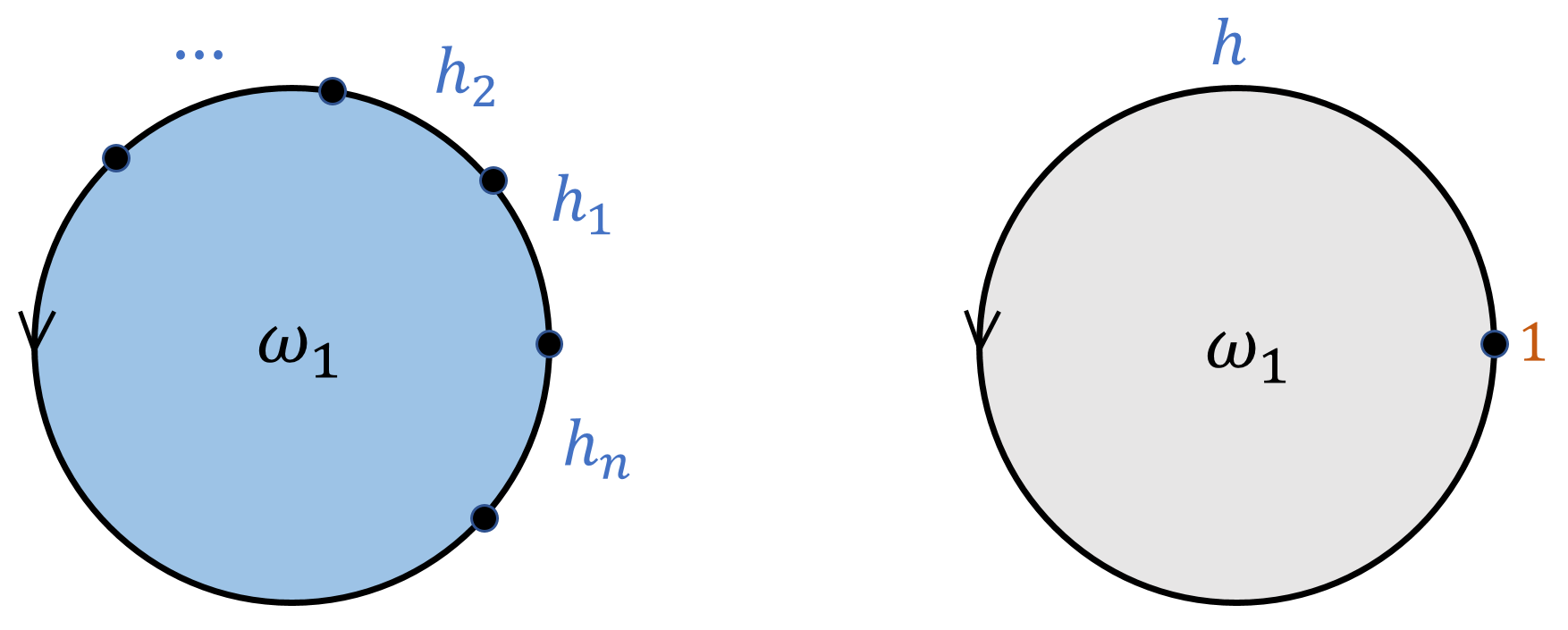}
	\caption{ An illustration of the subsystem $\omega_1$ and the corresponding minimal diagram.}\label{omega1}
\end{figure}

Using the fact that $\langle \{h_a\}\vert \{\tilde{h}_a\} \rangle=\delta_{\{h_a\},\{\tilde{h}_a \}}$, one can derive the following properties:
\begin{equation}
\sigma^{1}_{\omega_1}\cdot\sigma^1_{\omega_1}=\frac{1}{\vert G\vert^{n-1}}\sigma^{1}_{\omega_1}\quad\Rightarrow\quad
S(\sigma^{1}_{\omega_1})=(n-1)\ln\vert G\vert, \quad\qquad 
tr[{\sigma^{1}_{\omega_1}\cdot\sigma^1_{\omega_1}}]=\frac{1}{\vert G\vert^{n-1}}.    \label{TEE bulk}
\end{equation}
Here, $S(\sigma^{1}_{\omega_1})$ is the von Neumann entropy. Note that these results depends on $n$ the number of link values around $\partial \omega_1$.

On the other hand,  as is  mentioned above, there are properties invariant under topological deformations of $\omega_1$. Such as the number of extremal points and the entanglement entropy difference between two extremal points (in the case with more than one extremal points).
We call these properties \emph{topological invariant structures}  (or \emph{structures} for short) of an information convex.
Also, note that we need to be careful when talking about ``topological deformations." The relation to boundaries must be treated as topological data. $\omega_1$, $\omega_2$ and $\Omega_2$ in Fig. \ref{Omega} are all simply connected in the usual sense, but here they are treated as topologically distinct due to their different relation to the boundary.

The information convex $\Sigma(\omega_1)$ may also be calculated following Eqs. (\ref{Reduced},\ref{Flat},\ref{V_rotation}). We find that the problem is reduced to a problem for some \emph{minimal diagram} that realizes the same topological invariant structures as $\Sigma(\omega_1)$. One may solve the problem for the minimal diagram first and then go back to the original problem.

Consider the minimal diagram in Fig. \ref{omega1}. Define the corresponding Hilbert space $\mathcal{H}^{\ast}(\omega_1)=span\{\vert h\rangle\vert h\in G \}$, where $\{\vert h\rangle \}$ with $h\in G$ is an orthonormal basis. Define $\Sigma^{\ast}(\omega_1)$ as the set of density matrices $\sigma_{\omega_1}^{\ast}$ on $\mathcal{H}^{\ast}(\omega_1)$ satisfying the following  requirements:\\
1) $\sigma^{\ast}_{\omega_1}=\sum_{h\in G} p_h \vert h\rangle\langle h\vert$; here, $\{p_h \}$ is a probability distribution.\\
2) $B\,\sigma_{\omega_1}^{\ast}=\sigma_{\omega_1}^{\ast}$; here, $B\vert h\rangle =\delta_{1,h}\vert h\rangle$. \\
3) $A^g_1 \sigma^{\ast}_{\omega_1} A_1^{\bar{g}}=\sigma^{\ast}_{\omega_1}$ for $\forall g\in G$; here, $A_1^g\vert h\rangle =\vert gh\bar{g}\rangle$.

Then, it is easy to verify that 
\begin{equation}
\Sigma^{\ast}(\omega_1)=\{\sigma^{\ast 1}_{\omega_1} \},\qquad \sigma^{\ast 1}_{\omega_1}\equiv \vert 1\rangle\langle 1\vert.
\end{equation}
It has the same structures as $\Sigma(\omega_1)$. There is a naturally defined mapping $\pi: \Sigma(\omega_1)\to \Sigma^{\ast}(\omega_1)$, such that $\pi(\sigma^{1}_{\omega_1})=\sigma^{\ast 1}_{\omega_1}$. Intuitively, what the mapping $\pi$ does is to map a state  $\vert\{h_a\}\rangle_{\omega_1}$ into a configuration eigenstate $\vert h\rangle \in \mathcal{H}^{\ast}(\omega_1)$ with $h=h_1h_2\cdots h_n$. Trivial as the minimal diagram for $\omega_1$ is, similar constructions will be very useful in the  more involved examples below.

\subsubsection{The calculation of  $\Sigma(\Omega_1)$}\label{Omega_1_Calculation Section}
\begin{figure}[h]
	\centering\includegraphics[scale=0.320]{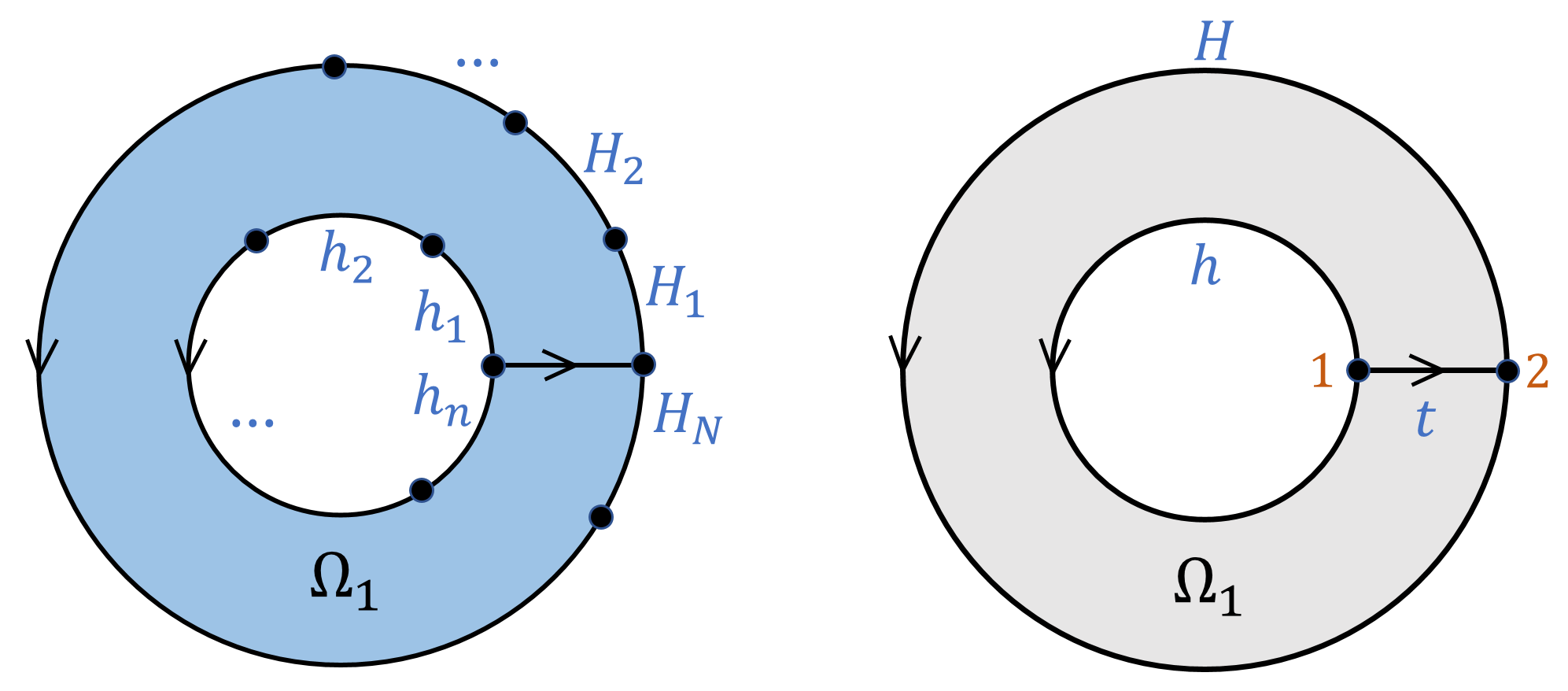}
	\caption{An illustration of the subsystem $\Omega_1$ and the corresponding minimal diagram.}\label{Omega_1}
\end{figure}

For $\Omega_1$ topology, i.e. an annulus in the bulk, $M=2$, let us relabel $h^1_a\rightarrow h_a$ with $a=1,\cdots,n$ and  $h^2_b\rightarrow H_b$ with $b=1,\cdots, N$ and  $h\equiv h_1\cdots h_n$, $H=H_1\cdots H_N$. As is discussed above, the calculation of $\Sigma(\Omega_1)$ can be done following Eqs.  (\ref{Reduced},\ref{Flat},\ref{V_rotation}), but a simpler way is to consider a minimal diagram, see Fig. \ref{Omega_1}.

Define  $\mathcal{H}^{\ast}(\Omega_1)=span\{\vert h,H,t\rangle \vert h,H,t\in G \}$ to be the Hilbert space for the minimal diagram. Here $\{\vert h,H,t\rangle \vert h,H,t\in G \}$ is an orthonormal basis. Define  $\Sigma^{\ast}(\Omega_1)$ to be the set of density matrices $\sigma^{\ast}_{\Omega_1}$ on $\mathcal{H}^{\ast}(\Omega_1)$ satisfying the following requirements:\\
1) $\sigma^{\ast}_{\Omega_1}=\sum_{h,H\in G}\sum_{\lambda} \,p^{\lambda}_{\{h, H\}}\vert \{h, H\};\lambda\rangle \langle \{h, H\}; \lambda\vert$,  where $\{ p^{\lambda}_{\{h,H\}} \}$ is a probability distribution and $\vert\{h, H\};\lambda\rangle\equiv \sum_{t\in G} c_{\lambda}(t)\vert h,H,t\rangle$ with complex coefficients $c_{\lambda}(t)$ satisfying $\sum_{t\in G} \vert c_{\lambda}(t)\vert^2=1$.\\
2) $B\,\sigma^{\ast}_{\Omega_1}= \sigma^{\ast}_{\Omega_1}$, where  $B \vert h,H,t\rangle =\delta_{h, tH\bar{t}}\vert h,H,t\rangle$.\\
3) $A^{ g}_{1} \sigma^{\ast}_{\Omega_1} A^{\bar{g}}_{1}=A^{ g}_{2} \sigma^{\ast}_{\Omega_1} A^{\bar{g}}_{2}=\sigma^{\ast}_{\Omega_1}$ for $\forall g\in G$, where $A^{g}_{1}\vert h,H,t\rangle =\vert gh\bar{g}, H,gt\rangle$ and $A^{ g}_{2}\vert h,H,t\rangle =\vert h, gH\bar{g}, t\bar{g} \rangle$.

From these requirements, on can verify:
\begin{equation}
\Sigma^{\ast}(\Omega_1)=\{\sigma^{\ast}_{\Omega_1}\vert \, \sigma^{\ast}_{\Omega_1}=\sum_{(c,R)} p_{(c,R)} \sigma^{\ast (c,R)}_{\Omega_1} \},
\qquad\quad c\in (G)_{cj},\quad R\in (E(c))_{ir}.
\end{equation}
Here, $\{p_{(c,R)} \}$ is a probability distribution and therefore $\Sigma^{\ast}(\Omega_1)$ is a convex set. $\sigma^{\ast (c,R)}_{\Omega_1}$ is an extremal point
\begin{equation}
\sigma^{\ast (c,R)}_{\Omega_1}=\frac{1}{\vert c\vert^2 \cdot n_R^2}\sum_{u}\sum_{v}\vert(c,R)(u,v)\rangle \langle (c,R)(u,v)\vert
\end{equation}
and the state $\vert(c,R)(u,v)\rangle$ is defined as (note the similarity of the following result with the results in Appendix \ref{Invariant operators})
\begin{eqnarray}
\vert (c,R)(u,v)\rangle &\equiv& A^{ p_i}_{1} A_{2}^{p_{i'}} \sum_{t\in E(c)} \sqrt{\frac{n_R}{\vert E(c)\vert}}\, \bar{\Gamma}_{R}^{jj'}(t)\vert r_c,r_c,t\rangle 
= \sum_{t\in E(c)} \sqrt{\frac{n_R}{\vert E(c)\vert}}\,\bar{\Gamma}_{R}^{jj'}(t) \vert c_i, c_{i'}, p_i t\bar{p}_{i'} \rangle\\
&\Rightarrow&\quad \langle (c,R)(u,v)\vert (c',R')(u',v')\rangle =\delta_{c,c'}\delta_{R,R'}\delta_{u,u'}\delta_{v,v'}.
\end{eqnarray}
Here, \\
1) $c\in (G)_{cj}$, i.e. $c=\{g \,r_c\,\bar{g}\,\vert \, g\in G \}$ and  $r_c$ is a representative of $c$. \\
2) $E(c)$ is the centralizer group of $c$, defined as $E(c)\equiv \{g\in G \,\vert\, g\,r_c=r_c\,g \}$.\\
3) $R\in (E(c))_{ir}$ and $n_R$ is the dimension of $R$.  $\Gamma_{R}$ is the unitary $n_R\times n_R$ matrix associated with $R$, with components  ${\Gamma}_{R}^{jj'}$. $\bar{\Gamma}_{R}^{jj'}$ is the complex conjugate of ${\Gamma}_{R}^{jj'}$.\\
4) $P(c)=\{p_i \}_{i=1}^{\vert c\vert}$ is a set of representatives of $G/E(c)$. $c=\{c_i \}_{i=1}^{\vert c\vert}$ with $c_i=p_i r_c\bar{p}_i$.\\
5) $u=(i,j), v=(i',j')$ with $i,i'=1,\cdots, \vert c\vert$ and $j,j'=1,\cdots, n_R$.\\
For more explanations of the notation, see Appendix \ref{Basic group theory}.

Now, introduce the label $a=(c,R)$ for $c\in (G)_{cj}$ and $R\in (E(c))_{ir}$, which will be identified as the label of bulk superselecton sector (bulk anyon type). $d_a\equiv \vert c\vert\cdot n_R$  is the quantum dimension for bulk anyons in quantum double models.  One can easily check
$\sum_{a} d_a^2=\vert G\vert^2=\mathcal{D}^2$ with $ \mathcal{D}\equiv \sqrt{\sum_a d_a^2}= \vert G\vert$. Here, $\mathcal{D}$ is the total quantum dimension.
We have the following results about $\Sigma^{\ast}{(\Omega_1)}$:
\begin{equation}
\sigma^{\ast a}_{\Omega_1}\cdot\sigma_{\Omega_1}^{\ast b}=\frac{\delta_{a,b}}{d^2_a}\sigma^{\ast a}_{\Omega_1}\quad\Rightarrow\quad 
S(\sigma^{\ast a}_{\Omega_1})=\ln d^2_a,
\qquad\quad  tr[\sigma^{\ast a}_{\Omega_1}\cdot\sigma_{\Omega_1}^{\ast b}]=\frac{\delta_{a,b}}{d^2_a}.
\end{equation}

Knowing the similarities between $\Sigma(\Omega_1)$ and $\Sigma^{\ast}(\Omega_1)$, we conclude that the set $\Sigma(\Omega_1)$  has extremal points $\sigma^a_{\Omega_1}$, with $a=(c,R)$:
\begin{equation}
\Sigma(\Omega_1)=\{\sigma_{\Omega_1}\vert \, \sigma_{\Omega_1}=\sum_{a} p_{a}\, \sigma^{a}_{\Omega_1} \},
\qquad\qquad \{p_a\}\textrm{ is a probability distribution.}
\end{equation}
The extremal points of $\Sigma(\Omega_1)$ have  the following properties:
\begin{equation}
\sigma^a_{\Omega_1}\cdot\sigma^b_{\Omega_1}=\frac{\delta_{a,b}}{d^2_a\cdot \vert G\vert^{n+N-2}}\sigma^{a}_{\Omega_1}\quad \Rightarrow\quad
S (\sigma^{a}_{\Omega_1})= \ln d^2_a +(n+N-2)\ln\vert G\vert,
\qquad tr[\sigma^a_{\Omega_1}\cdot\sigma^b_{\Omega_1}]=\frac{\delta_{a,b}}{d^2_a\cdot \vert G\vert^{n+N-2}}.
\end{equation} 
We will use $a=1$ as a shorthand notation for $c$ being the conjugacy class containing the identity element $1\in G$, i.e. $c=\{1\}$ with the one-dimensional identity representation $R=Id$. One could verify, $\sigma_{\Omega_1}^1$ is the reduced density matrix calculated from the ground state $\vert\psi\rangle$, i.e. $\sigma^{1}_{\Omega_1}=tr_{\bar{\Omega}_1}\vert\psi\rangle\langle\psi\vert$, and that $d_1=1$.

The following structures of $\Sigma(\Omega_1)$ are invariant under topological deformations of $\Omega_1$:
\begin{equation}
S(\sigma^{a}_{\Omega_1})=S( \sigma^{1}_{\Omega_1}) +\ln d^2_{a},
\qquad\qquad 
\sigma^{a}_{\Omega_1}\cdot \sigma^{b}_{\Omega_1}=0=tr[\sigma^{a}_{\Omega_1}\cdot \sigma^{b}_{\Omega_1}] \quad \textrm{for}\quad a\ne b,
\qquad \qquad 
\frac{tr[\sigma^{ a}_{\Omega_1}\cdot \sigma^{a}_{\Omega_1}] }{tr[\sigma^{1}_{\Omega_1}\cdot \sigma^{1}_{\Omega_1}] }=\frac{1}{d_a^2}.
\end{equation}

\subsubsection{The calculation of  $\Sigma(\omega_2)$}\label{omega_2_Calculation Section}
A subsystem with  $\omega_2$ topology attaches to the boundary at one piece, see Fig. \ref{Omega}. $\partial\omega_2\cap\partial\bar{\omega}_2$ has a single piece, $M=1$. Relabel $h^I_{a}\to h_{a}$ with $a=1,\cdots, n$. 
\begin{figure}[h]
	\centering\includegraphics[scale=0.36]{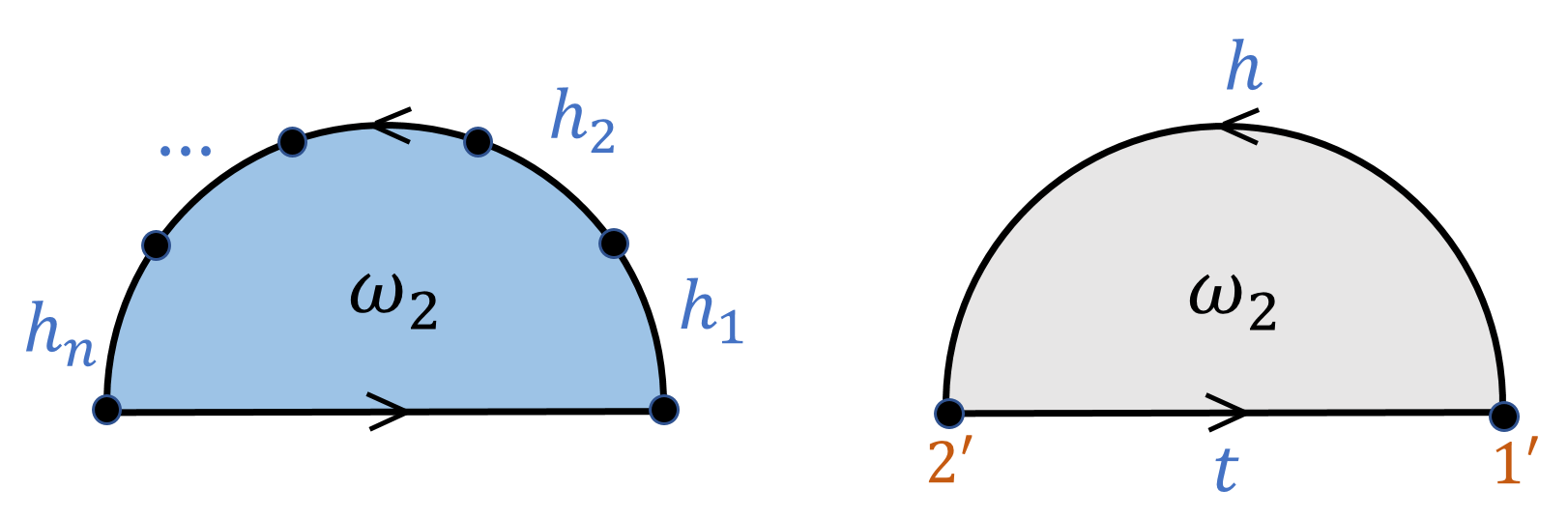}
	\caption{An illustration of the subsystem $\omega_2$ and the corresponding minimal diagram.}\label{omega2}
\end{figure}

Again, in order to find $\Sigma(\omega_2)$, we consider a corresponding minimal diagram in Fig. \ref{omega2}. Define the Hilbert space for the minimal diagram to be $\mathcal{H}^{\ast}(\omega_2)=span \{\vert h,t\rangle\vert h\in G, t\in K \}$, where $ \{\vert h,t\rangle\vert h\in G, t\in K \}$ is an orthonormal basis. Define $\Sigma^{\ast}(\omega_2)$ to be the set of density matrices $\sigma^{\ast}_{\omega_2}$ on $\mathcal{H}^{\ast}(\omega_2)$ satisfying the following requirements:\\
1) $\sigma^{\ast}_{\omega_2}=\sum_{h\in G} \sum_{\lambda}\, p^{\lambda}_{h}\vert \{h\};\lambda\rangle\langle \{h\};\lambda\vert$, where $\{p^{\lambda}_{h} \}$ is a probability distribution and $\vert \{h\};\lambda\rangle \equiv \sum_{t\in K} c_{\lambda}(t)\vert h,t\rangle$ with complex coefficients $c_{\lambda}(t)$ satisfying $\sum_{t\in K} \vert c_{\lambda}(t)\vert^2=1$.    \\
2) $B \sigma^{\ast}_{\omega_2}=\sigma^{\ast}_{\omega_2}$, where $B\vert h,t\rangle= \delta_{1,ht}\vert h,t\rangle$.\\
3) $A^{k}_{1'} \sigma^{\ast}_{\omega_2} A^{\bar{k}}_{1'}=A^{k}_{2'} \sigma^{\ast}_{\omega_2} A^{\bar{k}}_{2'}=\sigma^{\ast}_{\omega_2}$, for $\forall k\in K$; here, $A^{k}_{1'}\vert h,t\rangle =\vert kh, t\bar{k}\rangle$ and $A^{k}_{2'}\vert h,k\rangle = \vert h\bar{k}, kt\rangle$.

Then, it is easy to verify that:
\begin{equation}
\Sigma^{\ast}(\omega_2)=\{\sigma^{\ast 1}_{\omega_2} \} \quad\textrm{with}\quad \sigma^{\ast 1}_{\omega_2}=\frac{1}{\vert K\vert} \sum_{k\in K} \vert k,\bar{k}\rangle \langle k,\bar{k}\vert.
\end{equation}
with the following properties of the extremal point:
\begin{equation}
\sigma^{\ast 1}_{\omega_2}\cdot\sigma^{\ast 1}_{\omega_2}=\frac{1}{\vert K\vert}\sigma^{\ast 1}_{\omega_2}\quad\Rightarrow\quad 
S(\sigma^{\ast 1}_{\omega_2})=\ln \vert K\vert,
\qquad tr[\sigma^{\ast 1}_{\omega_2}\cdot\sigma^{\ast 1}_{\omega_2}]=\frac{1}{\vert K\vert}.
\end{equation}
From the similarity of $\Sigma^{\ast}(\omega_2)$ and $\Sigma(\omega_2)$, one can show that $\Sigma(\omega_2)$ contains a single element, i.e. the reduced density matrix calculated from the ground state $\vert\psi\rangle$,
\begin{equation}
\Sigma(\omega_2)=\{ \sigma^1_{\omega_2} \},\qquad\qquad \sigma^1_{\omega_2}=tr_{\bar{\omega}_2} \vert \psi\rangle\langle \psi\vert,
\end{equation}
with the following properties: 
\begin{equation}
\sigma^1_{\omega_2}\cdot\sigma^1_{\omega_2}=\frac{1}{\vert K\vert \cdot \vert G\vert^{n-1}} \sigma^1_{\omega_2}\quad \Rightarrow \quad S(\sigma^{1}_{\omega_2})=(n-1)\ln \vert G\vert +\ln\vert K\vert,
\qquad tr[\sigma^1_{\omega_2}\cdot\sigma^1_{\omega_2}]=\frac{1}{\vert K\vert \cdot \vert G\vert^{n-1}} .    \label{TEE boundary}
\end{equation}

\subsubsection{The calculation of  $\Sigma(\Omega_2)$}\label{Omega_2_Calculation Section}
Now consider a subsystem with $\Omega_2$ topology, see Fig. \ref{Omega}. It attaches to the boundary at two pieces. $\partial\Omega_2\cap \partial \bar{\Omega}_2$ contain two pieces, $M=2$. Relabel $h^1_{a}\rightarrow h_{a}$ with $a=1,\cdots, n$ and $h^2_a\rightarrow H_{b}$ with $b=1,\cdots,N$.
\begin{figure}[h]
	\centering\includegraphics[scale=0.32]{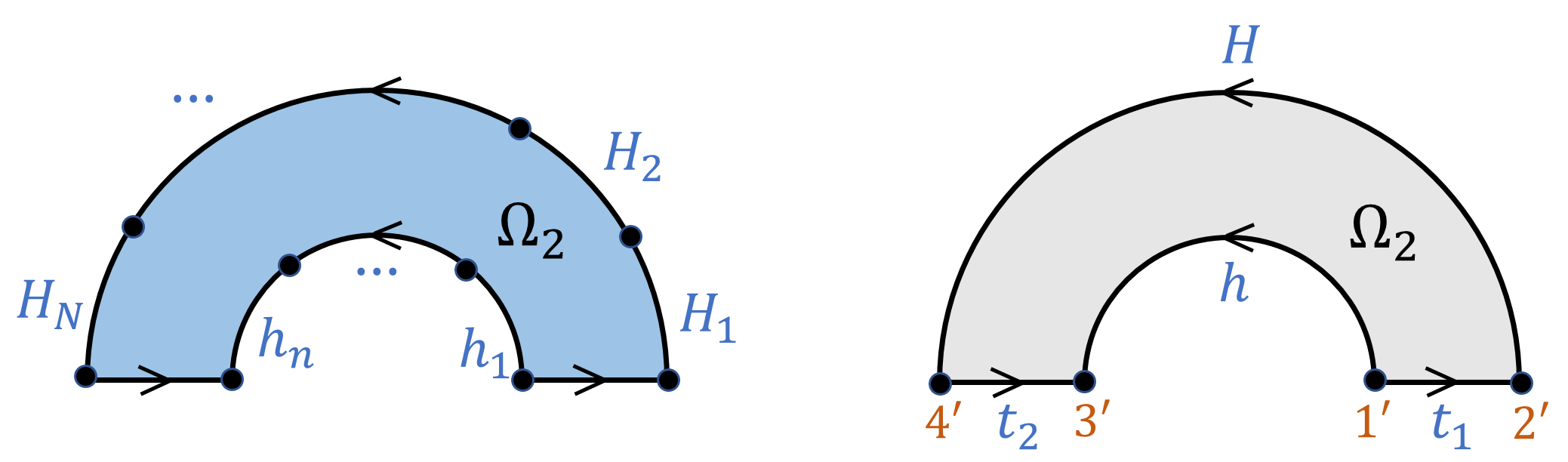}
	\caption{An illustration of the subsystem $\Omega_2$ and the corresponding minimal diagram.}\label{Omega_2}
\end{figure}

Again, to calculate $\Sigma(\Omega_2)$, we consider a minimal diagram in Fig. \ref{Omega_2}. Define the Hilbert space for the minimal diagram $\mathcal{H}^{\ast}(\Omega_2)=span\{\vert h, H,t_1,t_2\rangle \vert h, H\in G, t_1,t_2\in K \}$. Define $\Sigma^{\ast}(\Omega_2)$ to be the set of  density matrices $\sigma^{\ast}_{\Omega_2}$ on $\mathcal{H}^{\ast}(\Omega_2)$ satisfying the following requirements:\\
1) $\sigma^{\ast}_{\Omega_2}=\sum_{h,H\in G} \sum_{\lambda}\, p^{\lambda}_{\{h,H\}}\vert \{h,H\};\lambda\rangle \langle \{ h,H\};\lambda\vert$,  where $\{p^{\lambda}_{\{h,H\}} \}$ is a probability distribution and the state $\vert \{h,H\};\lambda\rangle\equiv \sum_{t_1,t_2\in K}\, c_{\lambda}(t_1,t_2)\vert h, H, t_1, t_2\rangle$ with complex coefficients $c_{\lambda}(t_1,t_2)$ satisfying $\sum_{t_1,t_2\in K}\vert c_{\lambda}(t_1,t_2)\vert^2=1$.\\
2) $B\sigma^{\ast}_{\Omega_2}=\sigma^{\ast}_{\Omega_2}$, where  $B \vert h,H,t_1,t_2\rangle = \delta_{h,t_1 Ht_2}\vert h, H,t_1,t_2\rangle$.\\
3) $A^{ k}_{1'}\sigma^{\ast}_{\Omega_2}A^{\bar{k}}_{1'}=A^{k}_{2'}\sigma^{\ast}_{\Omega_2}A^{\bar{k}}_{2'}=A^{ k}_{3'}\sigma^{\ast}_{\Omega_2}A^{\bar{k}}_{3'}=A^{k}_{4'}\sigma^{\ast}_{\Omega_2}A^{\bar{k}}_{4'}=\sigma^{\ast}_{\Omega_2}$, for $\forall k\in K$;  here,
\begin{eqnarray}
A^{k}_{1'}\vert h, H, t_1, t_2\rangle &=& \vert kh, H, kt_1,t_2\rangle, \qquad \qquad
A^{k}_{2'}\vert h, H, t_1,t_2\rangle \,\,=\,\, \vert h,kH, t_1\bar{k}, t_2\rangle,
\nonumber\\
A^{k}_{3'}\vert h, H, t_1, t_2\rangle &=& \vert h\bar{k}, H, t_1, t_2\bar{k}\rangle, \qquad\qquad
A^{k}_{4'}\vert h, H, t_1, t_2\rangle \,\,=\,\, \vert h,H\bar{k}, t_1,k t_2\rangle. \nonumber
\end{eqnarray}
From these requirements, on can verify:
\begin{equation}
\Sigma^{\ast}(\Omega_2)=\{\sigma^{\ast}_{\Omega_2}\vert \, \sigma^{\ast}_{\Omega_2}=\sum_{(T,R)} p_{(T,R)} \sigma^{\ast (T,R)}_{\Omega_2} \},
\qquad\quad T\in K\backslash G/K,\quad R\in (K^{r_T})_{ir}.
\end{equation}
Here $\{p_{(T,R)} \}$ is a probability distribution and $\sigma^{\ast (T,R)}_{\Omega_2}$ is an extremal point,
\begin{equation}
\sigma^{\ast (T,R)}_{\Omega_2}=\frac{1}{\vert T\vert^2\cdot n_R^2}\sum_{u,v}\sum_{k_3,k_4\in K} \vert (T,R)(u,v)(k_3,k_4)\rangle \langle(T,R)(u,v)(k_3,k_4)\vert
\end{equation}
with
\begin{eqnarray}
\vert (T,R)(u,v)(k_3,k_4)\rangle &\equiv& A^{q_i}_{1'}A^{q_i'}_{2'} A_{3'}^{k_3} A_{4'}^{k_4}\sqrt{\frac{n_R}{\vert K^{r_T}\vert}}\sum_{t_1\in K^{r_T}} \bar{\Gamma}^{jj'}_{R} (t_1)\vert r_T, r_T, t_1, \bar{r}_T\bar{t}_1 r_T\rangle\\
&=& \sqrt{\frac{n_R}{\vert K^{r_T}\vert}}\sum_{t_1\in K^{r_T}} \bar{\Gamma}^{jj'}_{R} (t_1)\vert q_i r_T \bar{k}_3,\, q_{i'} r_T \bar{k}_4 ,\,q_i t_1 \bar{q}_{i'}, \,k_4 \bar{r}_T\bar{t}_1 r_T \bar{k}_3\rangle
\end{eqnarray}
with $k_3, k_4\in K$. One can verify that:
\begin{equation}
\langle (T,R)(u,v)(k_3,k_4)\vert (T',R')(u',v')(k'_3,k'_4)\rangle=\delta_{T,T'}\,\delta_{R,R'}\,\delta_{u,u'}\,\delta_{v,v'}\,\delta_{k_3,k'_3}\,\delta_{k_4,k'_4}.
\end{equation}
Here:\\
1) $T\in K\backslash G/K$ is a double coset, i.e. $T=\{k_1 r_T k_2\,\vert\, k_1,k_2\in K \}$. $r_{T}\in G$ is a representative of $T$.\\
2) $K^{r_T}\equiv K\bigcap r_{T}K \bar{r}_{T}$ is a subgroup of $K$, and it depends on the choice of $r_T$ in general.\\
3) $R\in (K^{r_T})_{ir}$ and $n_R$ is the dimension of $R$.  $\Gamma_{R}$ is the unitary $n_R\times n_R$ matrix associated with $R$, with components  ${\Gamma}_{R}^{jj'}$. $\bar{\Gamma}_{R}^{jj'}$ is the complex conjugate of ${\Gamma}_{R}^{jj'}$.\\
4) $Q=\{q_i \}$, $i=1,\cdots, \vert Q\vert$ is a set of representatives of $K/K^{r_T}$. $\vert Q\vert =\vert K\vert/\vert K^{r_T}\vert= \vert T\vert/\vert K\vert$.  $s_i\equiv q_i r_T \bar{q}_i$.\\
5) $u=(i,j)$, $v=(i',j')$ with $i,i'=1,\cdots,\vert Q\vert$ and $j,j'=1,\cdots, n_R$.\\
For more explanations of the notation, see Appendix \ref{Basic group theory}.

Now let us introduce label $\alpha=(T,R)$ for $T\in K\backslash G/K$ and $R\in (K^{r_T})_{ir}$, which will be identified with the label of boundary superselection sector and the corresponding quantum dimension:
\begin{equation}
d_{\alpha}=\frac{\vert T\vert\cdot n_R}{\vert K\vert}=\frac{\vert K\vert\cdot n_R}{\vert K^{r_T}\vert}\qquad \qquad \alpha=(T,R).
\label{quantum dimension boundary}
\end{equation}
One can easily check that $\sum_{\alpha} d_{\alpha}^2= \sqrt{\sum_a d_a^2}=\vert G\vert =\mathcal{D}$.
We note that the quantum dimension $d_{\alpha}$ has been discovered algebraically in \cite{2017CMaPh.355..645C} in a different physical context.
One could verify the following properties of $\Sigma^{\ast}(\Omega_2)$:
\begin{equation}
\sigma_{\Omega_2}^{\ast\alpha}\cdot\sigma_{\Omega_2}^{\ast\beta}=\frac{\delta_{\alpha,\beta}}{d_{\alpha}^2 \cdot \vert K\vert^2} \sigma_{\Omega_2}^{\ast\alpha} \quad\Rightarrow\quad
S(\sigma^{\ast\alpha}_{\Omega_2})=\ln d^2_{\alpha} +2\ln\vert K\vert,
\qquad tr[\sigma^{\ast\alpha}_{\Omega_2}\cdot \sigma^{\ast\beta}_{\Omega_2}]=\frac{\delta_{\alpha,\beta}}{d^2_{\alpha}\cdot \vert K\vert^2} .
\end{equation}
One can obtain $\Sigma(\Omega_2)$ from its similarity to $\Sigma^{\ast}(\Omega_2)$:
\begin{equation}
\Sigma(\Omega_2)=\{\sigma_{\Omega_2}\vert \sigma_{\Omega_2}=\sum_{\alpha} p_{\alpha}\, \sigma^{ \alpha }_{\Omega_2} \}, \qquad\qquad \{p_{\alpha}\}\textrm{ is a probability distribution.}
\end{equation}
The extremal points have the following properties:
\begin{eqnarray}
\sigma^{\alpha}_{\Omega_2}\cdot\sigma^{\beta}_{\Omega_2}=\frac{\delta_{\alpha,\beta}}{d_{\alpha}^2\cdot \vert K\vert^2\cdot\vert G\vert^{n+N-2}} \sigma^{\alpha}_{\Omega_2}\quad&\Rightarrow&\quad S(\sigma^{\alpha}_{\Omega_2})=\ln d_{\alpha}^2 + 2\ln\vert K\vert + (n+N-2)\ln \vert G\vert, \\
&&\quad tr[\sigma^{\alpha}_{\Omega_2}\cdot\sigma^{\beta}_{\Omega_2}] = \frac{\delta_{\alpha,\beta}}{d_{\alpha}^2\cdot \vert K\vert^2\cdot\vert G\vert^{n+N-2}}.
\end{eqnarray}
Let us use the notation $\alpha=1$ for $T=K$ and $R=Id$  the one dimensional identity representation of $K^{r_T}$ (in fact $K^{r_T}=K$ for $T=K$). $\sigma^{1}_{\Omega_2}$ is the reduced density matrix calculated from the ground state $\vert\psi\rangle$, i.e. $\sigma^{1}_{\Omega_2}=tr_{\bar{\Omega}_2}\vert \psi\rangle\langle \psi\vert$ and the quantum dimension $d_{1}=1$.

The information convex $\Sigma(\Omega_2)$ has the following topological invariant structures:
\begin{equation}
S(\sigma^{\alpha}_{\Omega_2})=S(\sigma^1_{\Omega_2}) +\ln d^2_{\alpha},
\qquad \sigma^{\alpha}_{\Omega_2}\cdot \sigma^{\beta}_{\Omega_2}=0\quad \textrm{for}\quad \alpha\ne \beta,
\qquad \frac{tr[\sigma^{\alpha}_{\Omega_2}\cdot\sigma^{\alpha}_{\Omega_2}]}{tr[\sigma^{1}_{\Omega_2}\cdot\sigma^{1}_{\Omega_2}]}=\frac{1}{d^2_{\alpha}}.
\end{equation}

\subsection{Topological excitations, unitary string operators and superselection sectors} \label{HJW Section}
Perhaps, the most well-known examples of \emph{topological excitations} are anyons in 2D topological orders on a system without boundaries. They  could not be created by local unitary operators supported around the excitations but could be created  (usually need to create more than one) by unitary operators supported on a deformable string. Different excitations that could be related by a local unitary operation (acting around the excitations) are in the same \emph{superselection sector}. Superselection sector is the label of anyon type (let us denote the vacuum superselection sector as $a=1$).

In this section, we discuss a way to establish possible deformable unitary string operator types  for 2D topological orders with a gapped boundary (for both excitations inside the bulk and excitations along the boundary). The method makes use of the structure of $\Sigma(\omega_1)$, $\Sigma(\omega_2)$ and the HJW theorem. Then, we give a definition of bulk superselection sectors and boundary superselection sectors using the results of $\Sigma(\Omega_1)$ and $\Sigma(\Omega_2)$ and discuss what type of unitary string operators could realize topological excitations of each bulk/boundary superselection sector.
\subsubsection{Deformable unitary string operators from the HJW theorem} \label{Deformable HJW}

For Abelian models it is usually straightforward to construct the unitary operators creating $(a,\bar{a})$ for each anyon type $a$. The unitary operators have stringlike support and  the strings are deformable. 
For non-Abelian models, like a non-Abelian quantum double model, the proof of the existence of such unitary operators is less well-known but conceptually important  \cite{2015PhRvB..92k5139K}. Things that make the story complicated for non-Abelian models are (1) the ribbon operators (see Secs. \ref{Bulk Ribbon Section} and \ref{Boundary Ribbon Section}) though deformable are not unitary in general; (2) the support of the unitary operators can be slightly ``fatter" than the ribbon operators.

Here, we provide a proof of the existence of the  unitary string operators for both quantum double model on a sphere $S^2$ and quantum double model on $D^2$ with a single boundary making use of the result of $\Sigma(\omega_1)$ and $\Sigma(\omega_2)$ in Sec. \ref{omega_1}, Sec. \ref{omega_2_Calculation Section} and the HJW theorem. The proof is quite general and it is generalizable to systems on other manifold topologies. Given the suitable structure of information convex, the proof can be generalized to other topological orders in 2D and topological orders in higher dimensions.

First, let us review the HJW theorem \cite{1993PhLA..183...14H}. Consider the Hilbert space of system $AB$, which can be written as a tensor product of Hilbert spaces of subsystems $A$ and $B$, i.e. $\mathcal{H}_{AB}=\mathcal{H}_{A}\otimes \mathcal{H}_B$. For $\vert \psi\rangle,\vert \varphi\rangle \in \mathcal{H}_{AB}$, the HJW theorem implies (which could be verified easily using Schmidt decomposition):
\begin{equation}
tr_{A} \vert \varphi\rangle\langle \varphi\vert = tr_{A} \vert \psi\rangle\langle \psi\vert
\quad\Leftrightarrow \quad  
\vert \varphi\rangle = U_{A}\otimes 1_{B}\vert \psi\rangle.      \label{HJW}
\end{equation} 
Here $U_{A}$ is a unitary operator acting on $\mathcal{H}_A$ and $1_B$ is the identity operator acting on $\mathcal{H}_{B}$. 

\begin{figure}[h]
	\centering\includegraphics[scale=0.30]{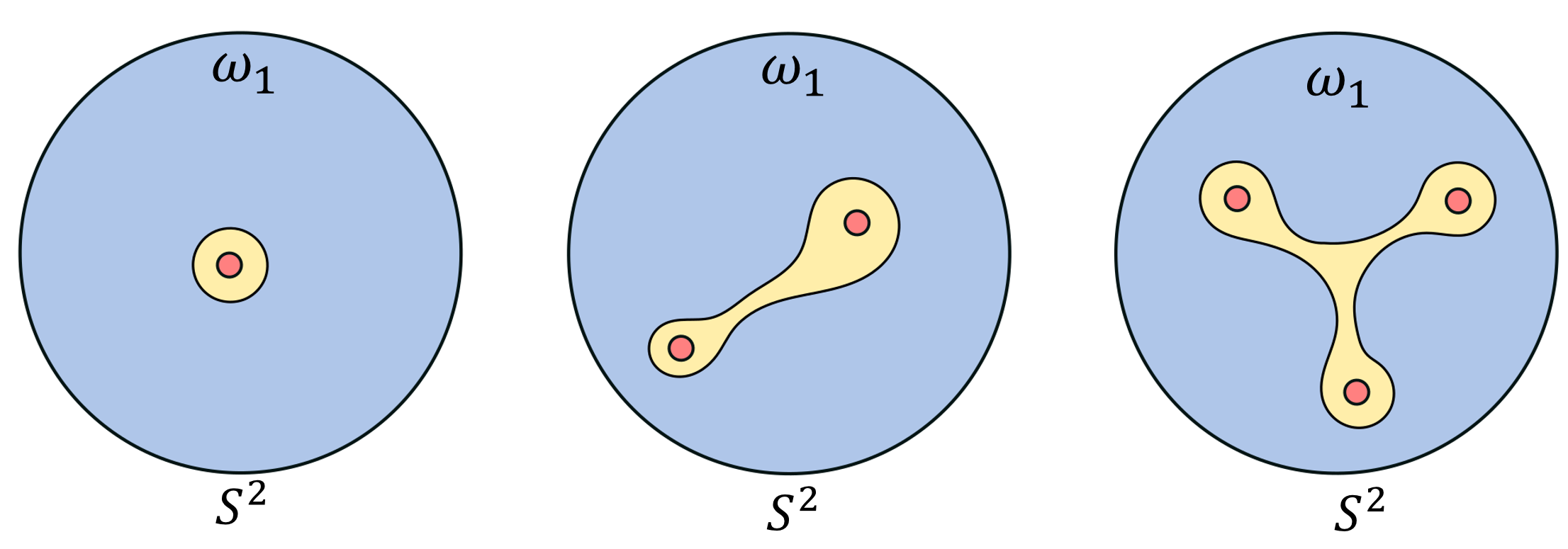}
	\caption{A system on a sphere $S^2$. Excitations are in the red area, which is a subset of the yellow region $\bar{\omega}_1$. Using the HJW theorem, one can show there must be a unitary operator supported on $\bar{\omega}_1$, which creates the excitations.}\label{Unitary_String_S2}
\end{figure}

Now let us consider a system defined on a sphere $S^2$. We have $\Sigma(\omega_1)=\{\sigma^{1}_{{\omega}_1}\equiv tr_{\bar{\omega}_1}\vert \psi\rangle\langle \psi\vert \}$. Here $\vert\psi\rangle $ is the unique ground state on $S^2$ and $\omega_1$ is any simply connected subsystem in the bulk (need to look at a scale bigger than a few lattice spacing for ``topology" to make sense).  Consider a few examples of excitations in the red areas of  Fig. \ref{Unitary_String_S2}, and let us call the corresponding excited state $\vert\varphi\rangle$. Since only the topology of $\omega_1$ matters, we may choose $\omega_1$ as large as possible, but it does not overlap with the excitations. 
From the HJW theorem:
\begin{equation}
tr_{\bar{\omega}_1} \vert \varphi\rangle\langle \varphi\vert\in \Sigma(\omega_1) \quad \Rightarrow\quad  tr_{\bar{\omega}_1} \vert \varphi\rangle\langle \varphi\vert = tr_{\bar{\omega}_1} \vert \psi\rangle\langle \psi\vert
\quad \Rightarrow\quad \vert \varphi\rangle = U_{\bar{\omega}_1}\otimes 1_{\omega_1}\vert \psi\rangle.
\end{equation}
In other words, there exists a unitary operator supported on the yellow region $\bar{\omega}_1$ which could create the excitations (when acting on the ground state). Since $\omega_1$ can be topologically deformed, the yellow region and therefore the support of the string operators can be topologically deformed also.\\
Explicitly:

A single excitation on $S^2$ can always be created using a local unitary operator acting  around the excitation. Therefore, it carries the trivial superselection sector. The method we used is an alternative way to prove a statement in \cite{2008PhRvB..78k5421B}. 
Be aware that, on torus $T^2$ it is possible (for non-Abelian models) to have a single excitation carry a nontrivial superselection sector \cite{2008PhRvB..78k5421B}. This result is also suggested by our method. A pair of excitations separated by an arbitrary distance on $S^2$ can be created using a  unitary operator supported on a deformable string connecting the pair. The thickness of the string does not grow with the distance between the excitations, and for the exactly solved quantum double model, it is just a few lattice spacings. Three excitations on $S^2$ can always be created by a unitary operator supported on a deformable treelike string.

\begin{figure}[h]
	\centering\includegraphics[scale=0.330]{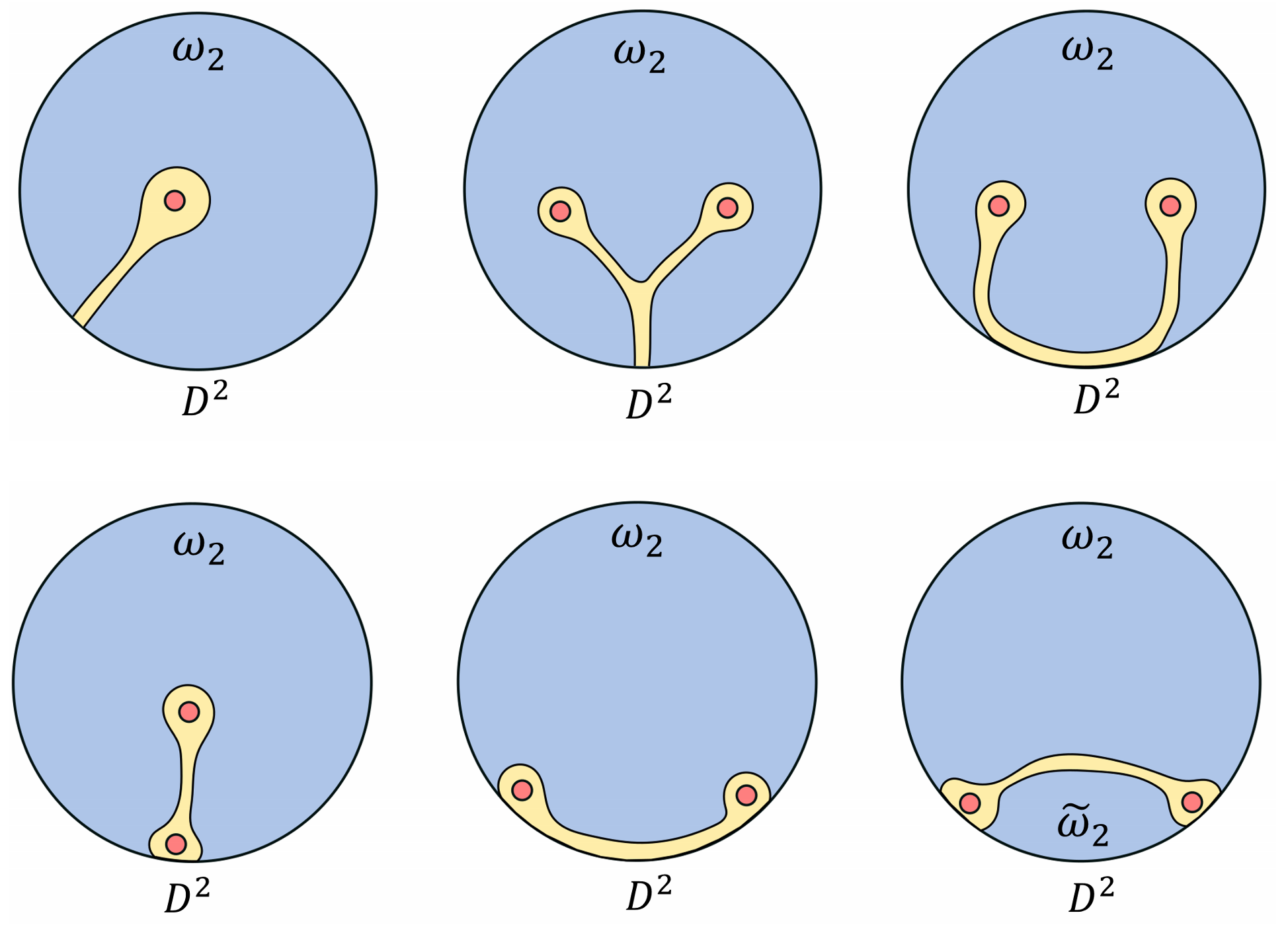}
	\caption{A system on a disk $D^2$. Excitations are in the red area, which is a subset of the yellow region $\bar{\omega}_2$ (or $\bar{\omega}_2\cap \bar{\tilde{\omega}}_2$ for the last diagram). Using the HJW theorem, one can show there must be a unitary operator supported on the yellow region $\bar{\omega}_2$ (or $\bar{\omega}_2\cap \bar{\tilde{\omega}}_2$) which creates the excitations.}\label{Unitary_String_D2}
\end{figure}

Now let us consider a system on a disk $D^2$. The results are illustrated in Fig. \ref{Unitary_String_D2}. The proof is quite similar to that discussed above, the only difference is that we now need the result $\Sigma(\omega_2)=\{\sigma^1_{\omega_2}\equiv tr_{\bar{\omega}_2} \vert \psi\rangle\langle \psi\vert \}$. Here $\omega_2$ is a simply connected subsystem attached to the boundary at one piece, see Fig. \ref{Omega} and Sec. \ref{omega_2_Calculation Section}.  $\vert \psi\rangle$ now represents the unique ground state on $D^2$. (For the last diagram, $\Sigma(\omega_2 \cup \tilde{\omega}_2)=\{\sigma^1_{\omega_2\cup \tilde{\omega}_2}\equiv tr_{\bar{\omega}_2\cap \bar{\tilde{\omega}}_2} \vert \psi\rangle\langle \psi\vert \}$ is needed, where  $\tilde{\omega}_2$ has the same topology as $\omega_2$.)

A new feature is that now a generic excited state with localized excitations need to be created using a process that involves the boundary, i.e. $\vert \varphi\rangle\ne U_{bulk}\vert \psi\rangle$ in general. Here $U_{bulk}$ is a unitary operator supported on a bulk subsystem.

Explicitly, single excitation away from the boundary of $D^2$ can be created by a unitary operator supported on a string attached to the boundary. The string could be deformed, and the string can end at anywhere of the boundary. Unlike a single excitation on $S^2$, this single excitation on $D^2$ may carry nontrivial bulk superselection sector. On the other hand, a single excitation located along the boundary can always be created by a local unitary operator acting around the excitation and therefore it carries a trivial boundary superselection sector.
A pair of excitations away from the boundary of $D^2$ could be created by a unitary operator supported on a stringlike region attached to the boundary. Note that there is no guarantee that the pair of excitations could be created by a unitary operator supported on a bulk string. Indeed, for some non-Abelian models,  there can be an excited state $\vert \varphi\rangle$ with a bulk anyon pair $(a,\bar{a})$ located away from the boundary, but $\vert \varphi\rangle\ne U_{bulk}\vert \psi\rangle$, see Sec. \ref{Condensing Section} for more details.
A pair of excitations  around the boundary could be created  by a unitary operator supported on a string along the boundary. The middle part of the string can be deformed into the bulk.

\subsubsection{Unitary string operators which create excitations of all possible superselection  sectors }
We have identified possible string types that connect pointlike excitations. Here, we discuss what type of string operators could guarantee the excitations to have all possible superselection sectors. One way  is to do the explicit calculation using ribbon operators, see Sec. \ref{Bulk Ribbon Section} and Sec. \ref{Boundary Ribbon Section}. Here, we discuss an alternative way which is applicable to models whose ribbon operators are hard (if not impossible) to be written down.

\begin{figure}[h]
	\centering\includegraphics[scale=0.250]{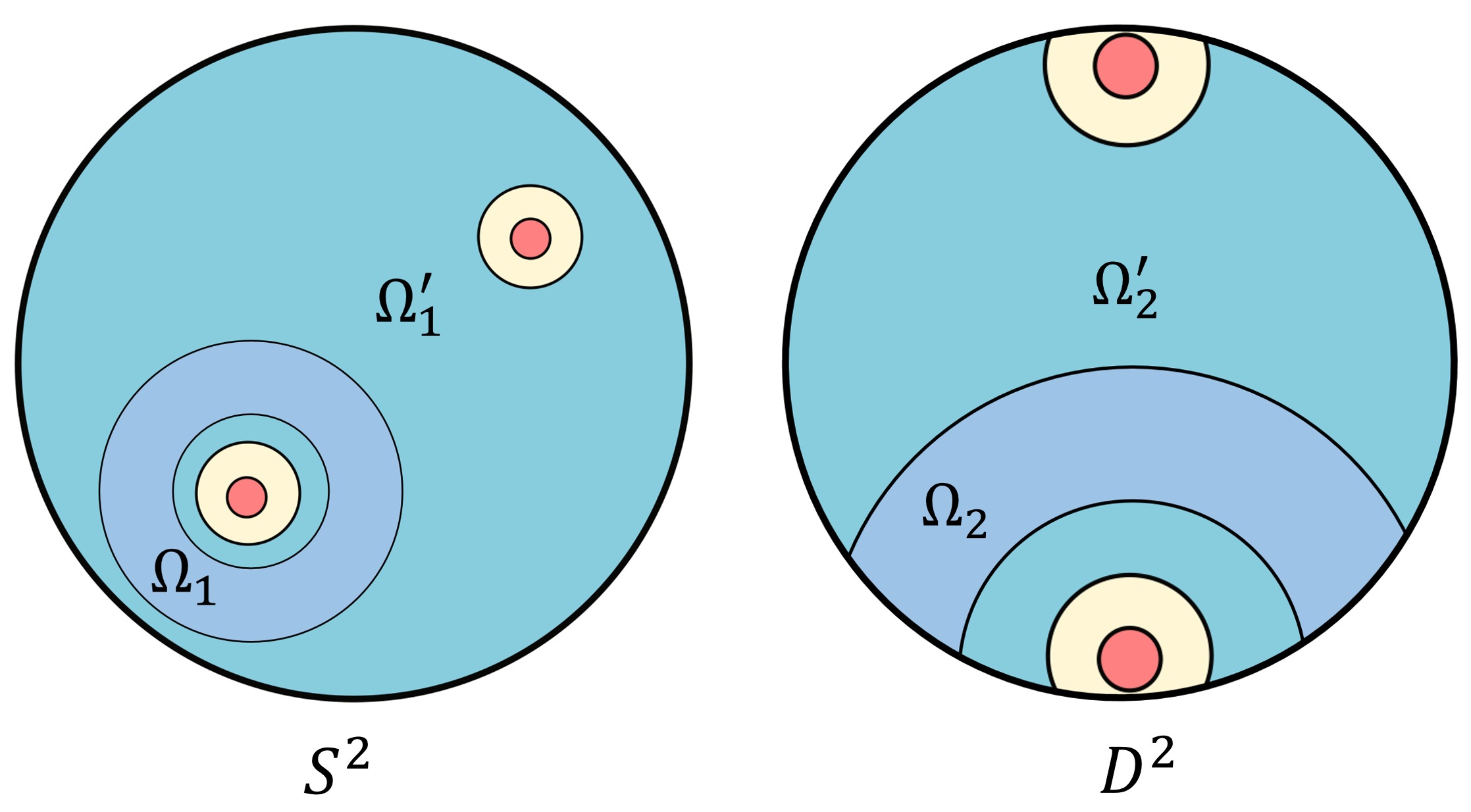}
	\caption{An illustration of subsystems used in the proof, $\Omega_1\subseteq \Omega_1'$, $\Omega_2\subseteq \Omega_2'$ and the red areas are excitations.}\label{Thicker}
\end{figure}

\begin{theorem}
	For a quantum double on $S^2$ and an annulus subsystem $\Omega_1$,
	every extremal point of $\Sigma(\Omega_1)$ can be written as $tr_{\bar{\Omega}_1}\vert\varphi\rangle\langle\varphi\vert$. Here,  $\vert\varphi\rangle$ is an excited state with a pair of pointlike excitations, created by a unitary operator supported on a deformable bulk string crossing $\Omega_1$. \label{Theorem S2}
\end{theorem}
\begin{proof}
	We have shown that an excited state with a pair of excitations could be written as $\vert\varphi\rangle=U\vert\psi\rangle$. Here,  $\vert\psi\rangle$ is the unique ground state on $S^2$ and $U$ is  supported on a deformable string. Also, one can always make a ``thicker" annulus $\Omega_1'$ on $S^2$ which  covers the entire $S^2$ except for two localized regions, see Fig. \ref{Thicker}. Then, one can show that\\
	1) $\Sigma(\Omega_1,\Omega_1')=\Sigma(\Omega_1)$ using the explicit reduced density matrix in Sec. \ref{Omega_1_Calculation Section} and tracing out some suitable subsystems. \\
	2) Each extremal point of $\Sigma(\Omega_1,\Omega_1')$ has purification on the system. This can be deduced from proposition \ref{Purify an extremal}.
	
	Therefore, there exists an excited state $\vert\varphi^a\rangle$ with a pair of pointlike excitations in $\bar{\Omega}_1'$ which could be created using a unitary string operator and it prepares an extremal point $\sigma^a_{\Omega_1}\in \Sigma(\Omega_1)$, i.e. $\sigma^a_{\Omega_1}=tr_{\bar{\Omega}_1}\vert\varphi^a\rangle\langle \varphi^a\vert $.	
\end{proof}
This also gives a natural way to define the superselection sector of pointlike excitations, i.e. by looking at what reduced density matrix of $\Sigma(\Omega_1)$ it prepares. If it prepares an extremal point $\sigma^a_{\Omega_1}$, then the excitation circled by the annulus is in the  $a$ superselection sector. If it prepares a nonextremal point, then it carries a superposition (or mixture) of superselection sectors.

\begin{theorem} 
	For a quantum double on $D^2$ and a subsystem $\Omega_2$,
	every extremal point of $\Sigma(\Omega_2)$ can be written as $tr_{\bar{\Omega}_2}\vert\varphi\rangle\langle\varphi\vert$. Here $\vert\varphi\rangle$ is an excited state with a pair of pointlike excitations along the boundary, created by a unitary operator supported on a string along the boundary crossing $\Omega_2$. The middle part of the string can be deformed into the bulk.  \label{Theorem Boundary String}
\end{theorem}
\begin{proof}
	We have shown that an excited state with a pair of excitations along the boundary  could be written as $\vert\varphi\rangle=U\vert\psi\rangle$. Here $\vert\psi\rangle$ is the unique ground state on $D^2$ and $U$ is  supported on a string along the boundary, the middle part of which could be deformed into the bulk. Also, one can always make a ``thicker" annulus $\Omega_2'$ on $D^2$ which  covers the entire $D^2$ except for two localized regions along the boundary, see Fig. \ref{Thicker}. Then, one can show:\\
	1) $\Sigma(\Omega_2,\Omega_2')=\Sigma(\Omega_2)$ using the explicit reduced density matrix in Sec. \ref{Omega_2_Calculation Section} and tracing out some suitable subsystems. \\
	2) Each extremal point of $\Sigma(\Omega_2,\Omega_2')$ has purification on the system. This can be deduced from Proposition \ref{Purify an extremal}.
	
	Therefore, there exists an excited state $\vert\varphi^{\alpha}\rangle$ with a pair of pointlike excitations in $\bar{\Omega}_2'$ which could be created using a unitary string operator along the boundary and it prepare an extremal point $\sigma^{\alpha}_{\Omega_2}\in \Sigma(\Omega_2)$, i.e. $\sigma^{\alpha}_{\Omega_2}=tr_{\bar{\Omega}_2}\vert\varphi^{\alpha}\rangle\langle \varphi^{\alpha}\vert $.
\end{proof}
Similar to the bulk case, one could define a boundary superselection sector of the excitations using the element of $\Sigma(\Omega_2)$ they prepare.

\begin{theorem}
	For a quantum double on $D^2$ and a bulk annulus   $\Omega_1$,
	every extremal point of $\Sigma(\Omega_1)$ can be written as $tr_{\bar{\Omega}_1}\vert\varphi\rangle\langle\varphi\vert$. Here $\vert\varphi\rangle$ is an excited state with a pair of pointlike excitations in the bulk, created by a unitary operator supported on a bulk string crossing $\Omega_1$. \label{Theorem Disk bulk}
\end{theorem}
\begin{proof}
	First, note the fact that the ground state of $S^2$ or $D^2$ has the same reduced density matrix on a disklike subsystem in the bulk. In other words, a disk in the bulk could not tell whether it lives on $S^2$ or $D^2$. One can choose a disklike subsystem containing the annulus $\Omega_1$ and apply a bulk string operator inside the disklike subsystem, then, use Theorem \ref{Theorem S2} to finish the proof.
\end{proof}

\subsection{Topological entanglement entropy from topological invariant structures of information convex}\label{TEE Subsection}
Topological entanglement entropy (TEE) \cite{kitaev2006topological,levin2006detecting} is an important topological invariant characterization of the ground state. In the middle steps of our derivation of $\Sigma(\omega_1)$, one could see the topological contribution explicitly, e.g. the $-\ln \vert G\vert$ in Eq. (\ref{TEE bulk}) for the bulk \footnote{The $-\ln \vert G\vert+\ln\vert K\vert$ in Eq. (\ref{TEE boundary}) may also be regarded as a topological contribution for a system attached to a boundary, but in order to extract this contribution using a linear combination canceling out local contributions of entanglement entropy, one  needs more than one boundary type.}. However, in the final step, when we keep only topological invariant structures of $\Sigma(\omega_1)$, the important constant is lost.

Nevertheless, we show that TEE can be recovered as a  lower bound (which appears to be saturated), given the topological invariant structures of $\Sigma(\omega_1)$ and $\Sigma(\Omega_1)$. In fact,  the lower bound  always saturates given a few simple assumptions  \cite{2016PhRvA..93b2317K}. In this sense, TEE is retained in the topological invariant structures of information convex.

Below is the derivation of the lower bound. First, recall some properties of $\Sigma(\Omega_1)$:
\begin{equation}
\Sigma({\Omega_1})=\{\sigma_{\Omega_1}\vert \sigma_{\Omega_1}=\sum_{a} p_a\,\sigma^a_{\Omega_1} \}\quad \textrm{with}\quad S(\sigma^a_{\Omega_1})=S(\sigma^1_{\Omega_1}) +\ln d^2_a,\qquad \sigma^{a}_{\Omega_1}\cdot \sigma^b_{\Omega_1}=0\quad \textrm{for}\quad a\ne b.
\end{equation}
From these properties, one could calculate the entanglement entropy for any $\sigma_{\Omega_1}\in \Sigma(\Omega_1)$ and find the element $\tilde{\sigma}_{\Omega_1}\in \Sigma(\Omega_1)$ with the maximal entanglement entropy:
\begin{eqnarray}
S(\sum_{a}p_a\,\sigma^a_{\Omega_1}) &=&\sum_a \big(p_a\,S(\sigma^a_{\Omega_1}) -p_a\ln p_a \big)\nonumber\\
&=& S(\sigma^1_{\Omega_1}) + \sum_{a} p_a \ln \bigg(\frac{d_a^2}{p_a}\bigg)\nonumber\\
&\le & S(\sigma^1_{\Omega_1}) +\ln \mathcal{D}^2 \qquad\qquad\qquad \textrm{``=" iff } p_a=\frac{d_a^2}{\mathcal{D}^2}\\
\Rightarrow \quad \tilde{\sigma}_{\Omega_1} &=& \sum_a \frac{d_a^2}{\mathcal{D}^2} \,\sigma^a_{\Omega_1}.
\end{eqnarray}
Here $\mathcal{D}=\sqrt{\sum_a d_a^2} $ is the total quantum dimension. It is obvious that $S(\tilde{\sigma}_{\Omega_1})-S(\sigma^1_{\Omega_1})=\ln \mathcal{D}^2$ is a topological invariance. Similar constructions apply to other subsystem topologies.

\begin{figure}[h]
	\centering\includegraphics[scale=0.330]{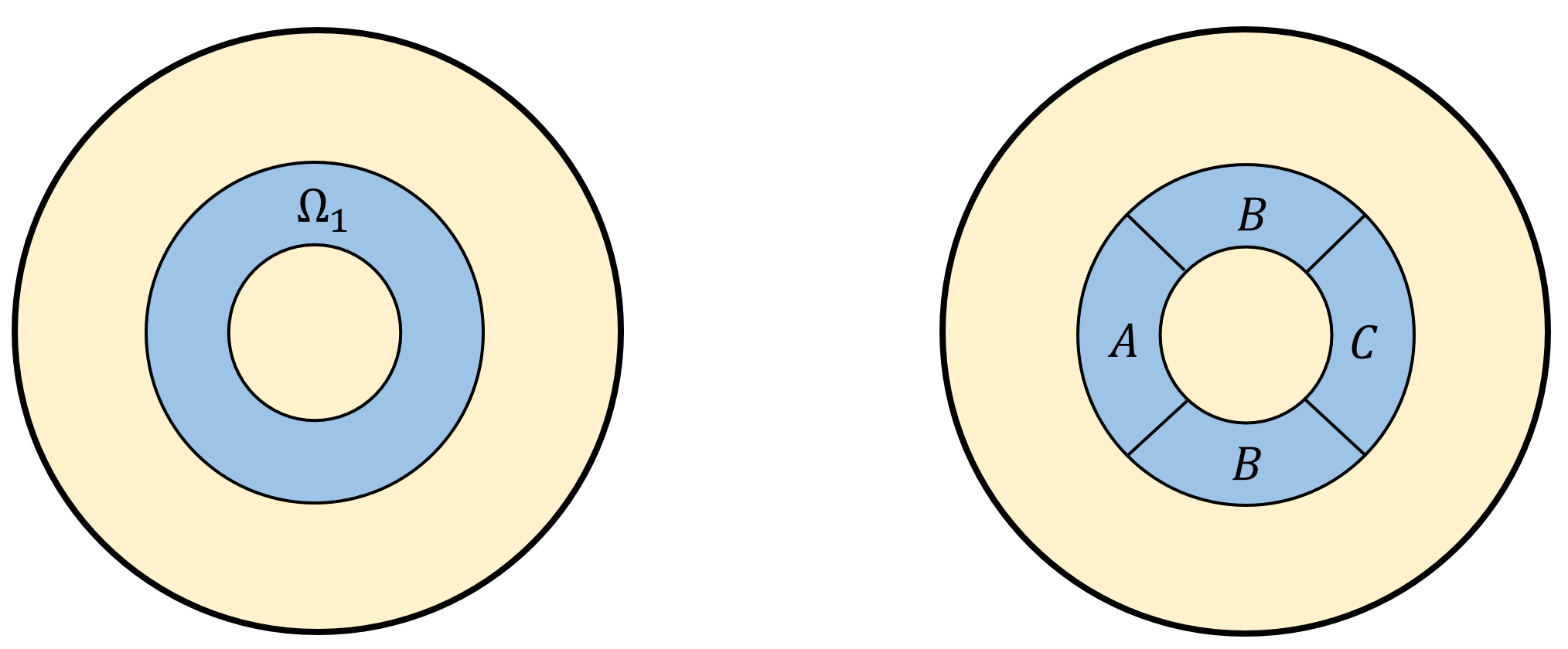}
	\caption{Divide $\Omega_1$ into $A,B,C$.}\label{TEE}
\end{figure}
Now, let us divide $\Omega_1$ into subsystems $A,B,C$ shown in Fig. \ref{TEE} and take the Levin-Wen definition of topological entanglement entropy  \cite{levin2006detecting} (an overall minus sign is added):
\begin{equation}
S_{topo}= (S_{AB} +S_{BC} -S_B -S_{ABC})\vert_{\sigma^1}= I(A:C\vert B) \vert_{\sigma^1}.
\end{equation}
Here $\sigma^1=\vert \psi\rangle \langle \psi\vert$ is the ground-state density matrix and $I(A:C\vert B)\equiv S_{AB} +S_{BC} -S_B -S_{ABC}$ is the conditional mutual information. According to the strong subadditivity $I(A:C\vert B)\ge 0$ is true in general. Also, all reduced density matrices in $\Sigma(\Omega_1)$ have the same reduced density matrix on $AB$, $BC$ [for a proof, use the structure of $\Sigma(\omega_1)$, and that $AB$ and $BC$ are of the same topology type as $\omega_1$]. Therefore,
\begin{eqnarray}
\left\{\begin{array}{l}
\tilde{\sigma}_{AB}=\sigma^1_{AB}\\
\tilde{\sigma}_{BC}=\sigma^1_{BC}
\end{array}\right. \quad\Rightarrow\quad I(A:C\vert B)\vert_{\sigma^1}&=& I(A:C\vert B)\vert_{\tilde{\sigma}} + S(\tilde{\sigma}_{\Omega_1}) -S(\sigma^1_{\Omega_1})\nonumber\\
&=& I(A:C\vert B)\vert_{\tilde{\sigma}} + \ln \mathcal{D}^2 \nonumber\\
&\ge& \ln \mathcal{D}^2 \nonumber\\
\Rightarrow\quad S_{topo}&\ge& \ln \mathcal{D}^2 \qquad\qquad\qquad \textrm{``=" iff } I(A:C\vert B)\vert_{\tilde{\sigma}}=0.
\end{eqnarray}
Comparing with the knowledge of TEE, the lower bound appears to be saturated and therefore $I(A:C\vert B)\vert_{\tilde{\sigma}}=0$. This result may be regarded as a generalization of  our previous lower bound \cite{2017arXiv170509300S} into the non-Abelian case.
We are aware that the reduced density matrix with maximal entanglement entropy $\tilde{\sigma}_{\Omega_1} \in \Sigma(\Omega_1)$ has zero conditional mutual information if the simple assumptions (I) and (II) in Ref. \cite{2016PhRvA..93b2317K} are satisfied. The assumptions are indeed satisfied for the ground states of exactly solved models for topological orders. On the other hand, given arbitrary $\rho_{ABC}$, it is in general not possible \cite{2008CMaPh.277..289I}  to find a $\sigma_{ABC}$ such that: (1) $\sigma_{AB}=\rho_{AB}$, (2) $\sigma_{BC}= \rho_{BC}$, and (3) $I(A:C\vert B)\vert_{\sigma}=0$.

To summarize,
\begin{equation}
S_{topo} = S(\tilde{\sigma}_{\Omega_1})-S(\sigma^1_{\Omega_1}) =\ln \mathcal{D}^2.
\end{equation}
This is true for exactly solved models satisfying (I) and (II) assumptions in Ref. \cite{2016PhRvA..93b2317K}. There are evidences and beliefs that $S_{topo}$ is robust  under generic local perturbations (which could be treated as finite depth quantum circuits), although very special examples like the Bravyi's counterexample \cite{2016PhRvB..94g5151Z} could change $S_{topo}$. It might be interesting to study the stability of $S({\sigma^a_{\Omega_1}})-S(\sigma^1_{\Omega_1})$, $S({\tilde{\sigma}}_{\Omega_1})-S(\sigma^1_{\Omega_1})$ and their generalizations.

\subsection{Bulk ribbon operators}\label{Bulk Ribbon Section}
Let us review bulk ribbon operators and the bulk topological excitations (bulk anyons) it creates. The review is brief and focusing on properties useful in our calculations. For more details, we refer to Refs. \cite{2003AnPhy.303....2K,2008PhRvB..78k5421B,2011CMaPh.306..663B}. 

\begin{figure}[h]
	\centering\includegraphics[scale=0.50]{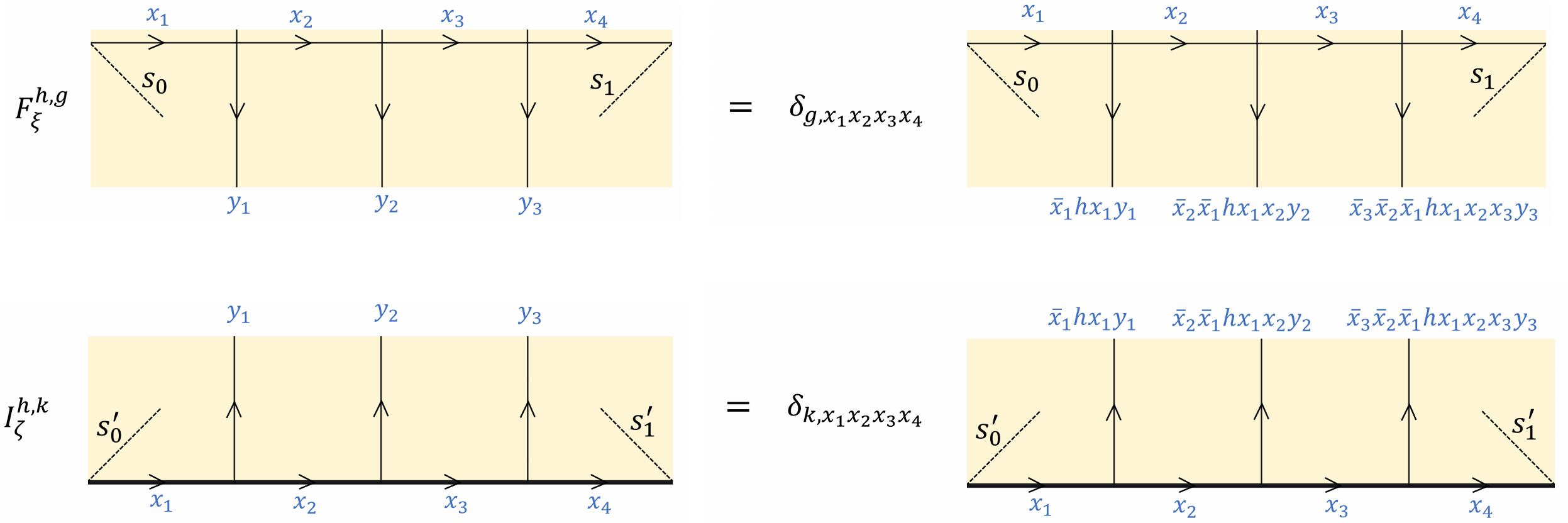}
	\caption{ Bulk ribbon operators and boundary ribbon operators. (a) $F^{h,g}_{\xi}$, $h,g\in G$ is a bulk ribbon operator. $\xi$ is a bulk ribbon consisting of bulk links and $x_1,x_2,x_3,x_4,y_1,y_2,y_3\in G$. (b) $I^{h,k}_{\zeta}$, $h\in G$ and $k\in K$ is a boundary ribbon operator. $\zeta$ is a boundary ribbon (for $K$ boundary) consisting of both boundary links with $x_1,x_2,x_3,x_4\in K$ and bulk links with $y_1,y_2,y_3\in G$.
	}\label{Ribbon_Operators}
\end{figure}

A bulk ribbon operator $F_{\xi}^{h,g}$ is defined for an open ribbon $\xi$ in the bulk (bulk ribbon) and $h,g\in G$. See Fig. \ref{Ribbon_Operators} for a bulk ribbon connecting bulk sites $s_0$ and $s_1$. Let $\vert\varphi^{h,g}_{\xi}\rangle\equiv F^{h,g}_{\xi}\vert\psi\rangle$ (not normalized). One can show $\vert\varphi^{h,g}_{\xi}\rangle$ has eigenvalues $B_f=1$, $A_v=1$, $A_{v'}^{K}=1$ for all $f,v$ that are not contained in $s_0,s_1$ and for all $v'$. In other words, $F_{\xi}^{h,g}$ (when acting on a ground state or a state locally minimizing energy) could create excitations only on $s_0$ and $s_1$. It is known that $\xi$ can be topologically deformed, in the sense that two ribbons $\xi$ and $\xi'$ both connecting $s_0$ and $s_1$, we have corresponding ribbon operators $F_{\xi}^{h,g}$ and $F_{\xi'}^{h,g}$, which are different operators, but $F^{h,g}_{\xi}\vert\psi\rangle=F^{h,g}_{\xi'}\vert\psi\rangle$. 

The bulk ribbon operators have basic properties
\begin{equation}
(F^{h,g}_{\xi})^{\dagger}= F^{\bar{h},g}_{\xi},\qquad\qquad F^{h,g}_{\xi}F^{h',g'}_{\xi}= \delta_{g,g'} F^{hh',g}_{\xi}.
\end{equation}
For $\xi=\xi_1\xi_2$, we have the ``gluing relation:"
\begin{equation}
F^{h,g}_{\xi_1\xi_2}= \sum_{l\in G} F_{\xi_1}^{h,l}F_{\xi_2}^{\bar{l}hl, \bar{l}g}.
\end{equation}
Change of basis: $\{F^{h,g}_{\xi} \}\rightarrow \{F_{\xi}^{(c,R)(u,v)} \}$:
\begin{equation}
F^{(c,R)(u,v)}_{\xi}=\sum_{t\in E(c)} \bar{\Gamma}_{R}^{jj'}(t) F_{\xi}^{\bar{c}_i, p_i t \bar{p}_{i'}}.
\end{equation}
Here, \\
1) $c\in (G)_{cj}$, i.e. $c=\{g \,r_c\,\bar{g}\,\vert \, g\in G \}$ and  $r_c$ is a representative of $c$. \\
2) $E(c)$ is the centralizer group of $c$, defined as $E(c)\equiv \{g\in G \,\vert\, g\,r_c=r_c\,g \}$.\\
3) $R\in (E(c))_{ir}$ and $n_R$ is the dimension of $R$.  $\Gamma_{R}$ is the unitary $n_R\times n_R$ matrix associated with $R$, with components  ${\Gamma}_{R}^{jj'}$. $\bar{\Gamma}_{R}^{jj'}$ is the complex conjugate of ${\Gamma}_{R}^{jj'}$.\\
4) $P(c)=\{p_i \}_{i=1}^{\vert c\vert}$ is a set of representatives of $G/E(c)$. $c=\{c_i \}_{i=1}^{\vert c\vert}$ with $c_i=p_i r_c\bar{p}_i$.\\
5) $u=(i,j), v=(i',j')$ with $i,i'=1,\cdots, \vert c\vert$ and $j,j'=1,\cdots, n_R$.\\
For more explanations of the notation, see Appendix \ref{Basic group theory}.

Note that we picked a normalization such that the ``gluing relation" in this basis looks simple:
\begin{equation}
F_{\xi_1\xi_2}^{(c,R)(u,v)}=\sum_{w} F_{\xi_1}^{(c,R)(u,w)}F_{\xi_2}^{(c,R)(w,v)}, \label{gluing_bulk}
\end{equation}
where $w=(i'',j'')$. For a (open) bulk ribbon $\xi$:
\begin{equation}
tr\bigg(F^{h,g}_{\xi}\sigma^{1}_{\xi} \bigg)=\frac{1}{\vert G\vert} \delta_{1,h}\quad \Rightarrow\quad
tr\bigg( F_{\xi}^{(c,R)(u,v)\dagger} F_{\xi}^{(c',R')(u',v')} \sigma^{1}_{\xi} \bigg)= \delta_{c,c'}\delta_{R,R'}\delta_{u,u'}\delta_{v,v'}\,\frac{1}{ \vert c\vert \cdot n_R}. \label{trace_bulk}
\end{equation}
Here $\sigma^1_{\xi}$ is the ground-state reduced density matrix on the ribbon $\xi$. $\sigma^1_{\xi}$  is proportional to the identity unless there are $A^g_v$ or $B_f$ supported on $\xi$, Eq. (\ref{trace_bulk}) is true for both cases.
This formula  will be useful in the calculation of the reduced density matrix.

For non-Abelian $G$, the ribbon operators $F^{(c,R)(u,v)}_{\xi}$ are not unitary in general, but there are corresponding unitary operators. In addition to the theorems in Sec. \ref{HJW Section}, explicit ribbon calculations can be done.
From Eq. (\ref{trace_bulk}), using the explicit wavefunction of the ground state $\vert\psi\rangle$  and the HJW theorem, one can show
\begin{equation}
\vert \varphi_{\xi}^{(c,R)(u,v)}\rangle\equiv \sqrt{\vert c\vert \cdot n_R}\, F_{\xi}^{(c,R)(u,v)}\vert\psi\rangle= U(\xi)\vert\psi\rangle.
\end{equation}
Here, $U(\xi)$ is a unitary operator supported on a stringlike region within a few lattice spacings to $\xi$. Note that the support of $U(\xi)$ can be slightly fatter than $\xi$. This result is independent from whether the system has boundaries or not, the only requirement is that $\xi$ can be contained in a disklike subsystem in the bulk.

\subsection{Boundary ribbon operators}\label{Boundary Ribbon Section}
Let us consider a new type of ribbon operator $I_{\zeta}^{h,k}$ with $h\in G$, $k\in K$ defined for open ribbon $\zeta$ that lies along the boundary (boundary ribbon), see Fig. \ref{Ribbon_Operators} for a boundary ribbon connecting boundary sites $s'_0$ and $s'_1$. Very similar to bulk ribbon operators, $I_{\zeta}^{h,k}$ can create excitations only at $s'_0$ and $s'_1$ (when acting on a ground state or a state locally minimizing energy).

The boundary ribbon operators have the basic properties:
\begin{equation}
(I^{h,k}_{\zeta})^{\dagger}= I_{\zeta}^{\bar{h}, k},\qquad\qquad I^{h,k}_{\zeta}I^{h',k'}_{\zeta}= \delta_{k,k'}\, I^{hh', k}_{\zeta}.
\end{equation}
The ``gluing relation" can be written as:
\begin{equation}
I_{\zeta_1\zeta_2}^{h,k}=\sum_{l\in K} I_{\zeta_1}^{h,l} I_{\zeta_2}^{\bar{l}hl, \bar{l}k}. \label{boundary gluing}
\end{equation}
Let us consider a linear combination: $\{I^{h,k}_{\zeta} \}\rightarrow \{I_{\zeta}^{(T,R)(u,v)} \}$:
\begin{equation}
I^{(T,R)(u,v)}_{\zeta}\equiv \sum_{t\in K^{r_T}}\bar{\Gamma}^{jj'}_{R}(t) I_{\zeta}^{{s}_i, q_i t \bar{q}_{i'}}. \label{boundary basis change}
\end{equation}
Here,\\
1) $T\in K\backslash G/K$ is a double coset, i.e. $T=\{k_1 r_T k_2\,\vert\, k_1,k_2\in K \}$. $r_{T}\in G$ is a representative of $T$.\\
2) $K^{r_T}\equiv K\bigcap r_{T}K \bar{r}_{T}$ is a subgroup of $K$, and it depends on the choice of $r_T$ in general.\\
3) $R\in (K^{r_T})_{ir}$ and $n_R$ is the dimension of $R$.  $\Gamma_{R}$ is the unitary $n_R\times n_R$ matrix associated with $R$, with components  ${\Gamma}_{R}^{jj'}$. $\bar{\Gamma}_{R}^{jj'}$ is the complex conjugate of ${\Gamma}_{R}^{jj'}$.\\
4) $Q=\{q_i \}$, $i=1,\cdots, \vert Q\vert$ is a set of representatives of $K/K^{r_T}$. $\vert Q\vert =\vert K\vert/\vert K^{r_T}\vert= \vert T\vert/\vert K\vert$.  $s_i\equiv q_i r_T \bar{q}_i$.\\
5) $u=(i,j)$, $v=(i',j')$ with $i,i'=1,\cdots,\vert Q\vert$ and $j,j'=1,\cdots, n_R$.\\
For more explanations of the notation, see Appendix \ref{Basic group theory}.

Note that, for a set of chosen $r_T$, the set $ \{I_{\zeta}^{(T,R)(u,v)} \}$ may contain a smaller number of elements than $\{I^{h,k}_{\zeta} \}$. We need to be careful to say it is a change of basis. For $K=\{1\}$, it is a change of basis.

\begin{remark}
	Our boundary ribbon operators $I_{\zeta}^{(T,R)(u,v)}$ are fundamentally different from the $Y_{\rho}^{(T,R)(u,v)}$ operators considered in Ref.  \cite{2017CMaPh.355..645C}.
	The operator $I_{\zeta}^{(T,R)(u,v)}$ is defined for a ribbon $\zeta$ that lies along the boundary, while the operator $Y_{\rho}^{(T,R)(u,v)}$ is defined for ribbon $\rho$ inside the boundary.
	Unlike the model in Ref. \cite{2017CMaPh.355..645C}, our model does not have Hilbert space for the interior of a boundary. Therefore, the excitations created by $Y_{\rho}^{(T,R)(u,v)}$ operator are not defined in our model. On the other hand, the excitation created by $I_{\zeta}^{(T,R)(u,v)}$  can be defined  not only in our model but also in the models of Ref. \cite{2011CMaPh.306..663B,2017CMaPh.355..645C}.
	The excitations created by $I_{\zeta}^{(T,R)(u,v)}$  are deconfined topological excitations (parallel to the topological excitations created by $F_{\xi}^{(c,R)(u,v)}$ in the bulk). The middle part of the boundary ribbon $\zeta$ can be  deformed into the bulk. The excitations created by $Y_{\rho}^{(T,R)(u,v)}$ are confined, and the energy cost is proportional to the length of the ribbon $\rho$. The ribbon operator $Y_{\rho}^{(T,R)(u,v)}$ cannot be deformed. 
\end{remark}

One could verify that
\begin{equation}
tr\bigg(I_{\zeta}^{h,k}\sigma^1_{\zeta} \bigg)=\frac{1}{\vert K\vert}\delta_{1,h}\quad\Rightarrow\quad
tr\bigg( I_{\zeta}^{(T,R)(u,v)\dagger} I_{\zeta}^{(T',R')(u',v')} \sigma^1_{\zeta} \bigg)=\frac{\vert K^{r_T}\vert}{n_R\cdot\vert K\vert}\delta_{T,T'}\delta_{R,R'}\delta_{u,u'}\delta_{v,v'} . \label{trace_boundary}
\end{equation}
Here, $\sigma^1_{\zeta}$ is the ground-state reduced density matrix on $\zeta$.  $\sigma^1_{\zeta}$ is generally not proportional to the identity unless $K=\{1\}$ since for $K\ne \{1\}$, and for a $\zeta$ not too short, it must contain some $A^{k}_{v}$ with $k\ne 1$.

It turns out that the nice ``gluing relation" parallel to Eq. (\ref{gluing_bulk}) only appears for boundary ribbon operators with one additional constraint of  $T$ and $r_T$, i.e. if there is a choice of $r_T\in T$ such that $r_T m=m r_T$ for $\forall m\in K^{r_T}$. When choosing this $r_T$, using Eqs. (\ref{boundary gluing},\ref{boundary basis change}), one derives
\begin{equation}
I^{(T,R)(u,v)}_{\zeta_1\zeta_2}=\sum_{w} I^{(T,R)(u,w)}_{\zeta_1} I^{(T,R)(w,v)}_{\zeta_2}  
\quad \qquad \textrm{if}\quad r_T m=m r_T \quad \forall m\in K^{r_T},     
\label{gluing_boundary}
\end{equation}
where $w=(i'',j'')$. This condition (and therefore Eq. (\ref{gluing_boundary})) holds for a large class of boundaries including\\
1) $K=\{1\}$ boundary for a general $G$ quantum double.\\
2) $K=G$ boundary for a general $G$ quantum double.

The operators $I_{\zeta}^{(T,R)(u,v)}$ are  not unitary in general, but there exist corresponding unitary operators:
\begin{equation}
\vert\varphi^{(T,R)(u,v)}_{\zeta}\rangle\equiv \sqrt{\frac{n_R\cdot \vert K\vert}{\vert K^{r_T}\vert}}\, I_{\zeta}^{(T,R)(u,v)}\vert\psi\rangle=U(\zeta)\vert\psi\rangle .
\end{equation}
Here, $U(\zeta)$ is a unitary operator supported on a stringlike region within a few lattice spacings to $\zeta$. Note that the support of $U(\zeta)$ can be slightly fatter than $\zeta$. This result has been proved (up to normalization) in Sec. \ref{Deformable HJW}, and it can be shown by explicit calculation using Eq.  (\ref{trace_boundary}), the explicit wave function of the ground state $\vert\psi\rangle$ and the HJW theorem. This calculation also determines the overall normalization.

\subsection{Bulk ribbon operators and the extremal points of $\Sigma(\Omega_1)$}\label{bulk_ribbon_extremal}
Let us consider the $\Omega_1$ subsystem discussed earlier in Sec. \ref{The calculation of information convex}. We show (with explicit calculation) that a pair of topological excitations (here are bulk anyons) $(a,\bar{a})$ created by a bulk process  prepare the extremal point $\sigma^a_{\Omega_1}\in \Sigma(\Omega_1)$. This construction confirms that the extremal points of $\Sigma(\Omega_1)$ are in one to one correspondence with bulk superselection sectors (bulk anyon types). Also, see theorem \ref{Theorem Disk bulk}, for a powerful but less explicit way.
\begin{figure}[h]
	\centering\includegraphics[scale=0.40]{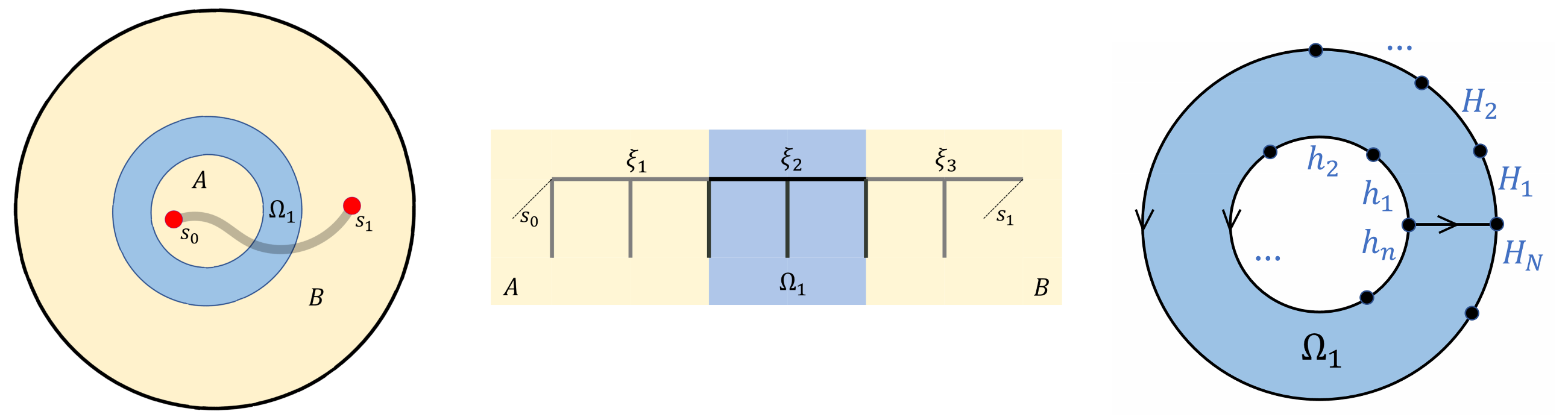}
	\caption{The whole system is a disk $D^2$, and it is divided into subsystems $A,\Omega_1,B$. The bulk ribbon $\xi$ connects bulk sites $s_0$ and $s_1$. $s_0$ is in $A$ and $s_1$ is in $C$. The ribbon $\xi=\xi_1\xi_2\xi_3$ with $\xi_1\subseteq A$, $\xi_2\subseteq \Omega_1$, and $\xi_3\subseteq B$. }\label{Omega_1_Ribbon}
\end{figure}

Consider the process and bulk ribbon shown in Fig. \ref{Omega_1_Ribbon}, the bulk ribbon $\xi=\xi_1\xi_2\xi_3$ connects bulk sites $s_0$ and $s_1$ that are separated by $\Omega_1$. Define the excited state (normalized) $\vert\varphi^{(c,R)(u,v)}_{\xi}\rangle \equiv \sqrt{\vert c\vert\cdot n_R}\, F_{\xi}^{(c,R)(u,v)}\vert \psi\rangle$. From our knowledge about bulk ribbon operators, $\vert\varphi^{(c,R)(u,v)}_{\xi}\rangle$ has excitations only at the two ends of $\xi$, i.e. $s_0$ and $s_1$. Since $s_0$ and $s_1$ are  away from $\Omega_1$, it is clear that $tr_{\bar{\Omega}_1}\vert \varphi^{(c,R)(u,v)}_{\xi}\rangle \langle \varphi^{(c,R)(u,v)}_{\xi}\vert \in \Sigma(\Omega_1)$. 

Now let us calculate $tr_{\bar{\Omega}_1}\vert \varphi^{(c,R)(u,v)}_{\xi}\rangle \langle \varphi^{(c,R)(u,v)}_{\xi}\vert$. From the ``gluing relation" (\ref{gluing_bulk}), one obtains
\begin{equation}
F_{\xi_1\xi_2\xi_3}^{(c,R)(u,v)}=\sum_{w_1,w_2} F_{\xi_1}^{(c,R)(u,w_1)}F_{\xi_2}^{(c,R)(w_1,w_2)}F_{\xi_3}^{(c,R)(w_2,v)}.
\label{gluing bulk 2}
\end{equation}
Let us write the ground state $\vert \psi\rangle$ as:
\begin{equation}
\vert \psi\rangle =\sum_{h_a,H_b} \vert\{ h_a \}\rangle_{A}\vert \{h_a\},\{H_b\}\rangle_{\Omega_1}\vert \{ H_b\}\rangle_{B}.
\end{equation}
Here,  $\{h_a\}$ with $a=1,\cdots, n$ is a set of link configurations, $h_a\in G$. Similarly, $\{H_b\}$ with $b=1,\cdots, N$ is  a set of link configurations, $H_b\in G$. The state $\vert \{h_a\}\rangle_A$ is the unique equal weight superposition of all zero flux configurations
determined by a set of $\{h_a\}$ on $\partial A$. The zero flux requirement tells us that $h_1h_2\cdots h_n=1$. Similarly, the states $\vert \{h_a\},\{H_b\}\rangle_{\Omega_1}$ and $\vert \{H_b\}\rangle_{B}$ are the unique equal weight superposition of all zero flux configurations with fixed link values $h_a$ and $H_b$ satisfying $h_1h_2\cdots h_n=H_1H_2\cdots H_N=1$.

Then, one can calculate $\sigma_{\Omega_1}(c,R)\equiv tr_{\bar{\Omega}_1} \vert \varphi_{\xi}^{(c,R)(u,v)} \rangle \langle \varphi_{\xi}^{(c,R)(u,v)}\vert$, and find (up to normalization):
\begin{equation}
\sigma_{\Omega_1}(c,R)=\sum_{w_1,w_2}\, F_{\xi_2}^{(c,R)(w_1,w_2)}  \sigma^1_{\Omega_1} F_{\xi_2}^{(c,R)(w_1,w_2){\dagger} }.
\label{Reduced_Omega1}
\end{equation}
Here, $\sigma^1_{\Omega_1}=tr_{\bar{\Omega}_1}\vert\psi\rangle\langle\psi\vert$ is the reduced density matrix of the ground state.
To get the result above, we have used the fact that 
\begin{equation}
\,_A\langle \{h_a\}\vert  F_{\xi_1}^{(c,R)(u,w_1){\dagger}} F_{\xi_1}^{(c,R)(u,\tilde{w}_1)}\vert \{\tilde{h}_a\}\rangle_{A}
=\delta_{h,\tilde{h}}\, tr_{\xi_1}\bigg( F_{\xi_1}^{(c,R)(u,w_1){\dagger}} F_{\xi_1}^{(c,R)(u,\tilde{w}_1)} \sigma^1_{\xi_1}\bigg)
= \delta_{h,\tilde{h}}\delta_{w_1,\tilde{w}_1} \frac{1}{n_R\cdot \vert c\vert}.   \label{Ortho_Bulk}
\end{equation}
Here, $\sigma^1_{\xi_1}$ is the reduced density matrix of the ground state on $\xi_1$.
$\delta_{h,\tilde{h}}$ is short for $\prod_{a}\delta_{h_a,\tilde{h}_a}$.
There is a similar equation as  Eq. (\ref{Ortho_Bulk}) when replacing $\vert \{h_a\}\rangle_A$ by $\vert \{H_b\}\rangle_B$ and replacing $\xi_1$ by $\xi_3$. 
One could  verify that the reduced density matrix $\sigma_{\Omega_1}(c,R)$ in Eq. (\ref{Reduced_Omega1}) is identical to the extremal point $\sigma_{\Omega_1}^a$, with $a=(c,R)$. In other words, every  extremal point of $\Sigma(\Omega_1)$ can be obtained by an anyon process in the bulk shown in Fig. \ref{Omega_1_Ribbon}.

This further shows that $\Sigma(\Omega_1)=\Sigma(\Omega_1,\Omega_1')$, as long as $\bar{\Omega}'_{1}$ contains  $s_0$ and $s_1$. Using Eqs. (\ref{gluing_bulk},\ref{trace_bulk}), the ground-state wave function and the HJW theorem one can show that each of the anyons $a,\bar{a}$ at the  endpoints $s_0,s_1$ can be moved by local unitary operators acting around $s_0$ and $s_1$, respectively.

\subsection{Boundary ribbon operators and the extremal points of $\Sigma(\Omega_2)$}\label{Sec. Boundary prepare Omega_2}
Now consider the $\Omega_2$ subsystem discussed in Sec. \ref{The calculation of information convex}. As is shown in theorem \ref{Theorem Boundary String}, it is possible  to create a pair of excitations along the boundary by a unitary string operator along the boundary and the excitations could carry any boundary superselection sector. On the other hand, it is nice to have explicit constructions. In practice, we find that explicit constructions are challenging beyond models with the additional requirement: \emph{ for every $T\in K\backslash G/K$, there exists $r_T \in T$ such that $r_T m =m r_T$ for $\forall m \in K^{r_T}$}, the same requirement for the ``gluing relation" (\ref{gluing_boundary}) to apply. The following constructions are restricted to models satisfying this requirement.

We show that a pair of topological excitations $(\alpha,\bar{\alpha})$ created by a process involve the boundary  prepare the extremal point $\sigma^{\alpha}_{\Omega_2}\in \Sigma(\Omega_2)$, with $\alpha=(T,R)$. This construction confirms the fact that the extremal points of $\Sigma(\Omega_2)$ have the same label as the boundary superselection sector. 
\begin{figure}[h]
	\centering\includegraphics[scale=0.40]{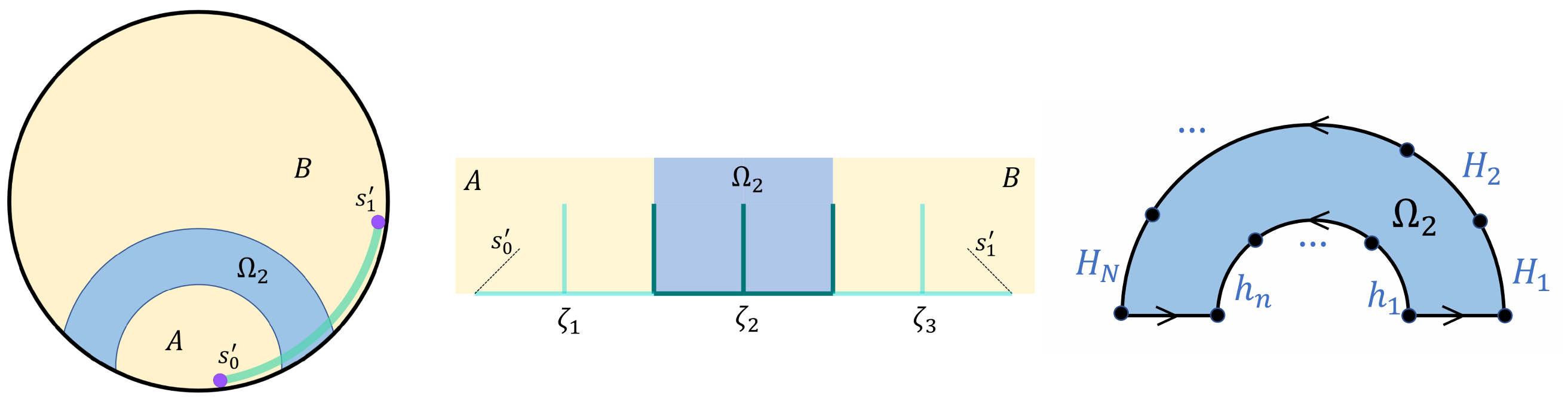}
	\caption{The whole system is a disk $D^2$ and it is divided into subsystems $A,\Omega_2,B$. The bulk ribbon $\zeta$ connects bulk sites $s'_0$ and $s'_1$. $s'_0$ is in $A$ and $s'_1$ is in $C$. The ribbon $\zeta=\zeta_1\zeta_2\zeta_3$ with $\zeta_1\in A$, $\zeta_2\in \Omega_1$, and $\zeta_3\in B$. }\label{Omega_2_Ribbon}
\end{figure}
The discussion here is very similar to that in Sec. \ref{bulk_ribbon_extremal}, and therefore will be brief. 
We have the ``gluing relation"
\begin{equation}
I_{\zeta_1\zeta_2\zeta_3}^{(T,R)(u,v)}=\sum_{w_1,w_2} I_{\zeta_1}^{(T,R)(u,w_1)} I_{\zeta_2}^{(T,R)(w_1,w_2)} I_{\zeta_3}^{(T,R)(w_2,v)}.
\end{equation}
Define a state (normalized) with excitations created by a boundary ribbon operator  and the corresponding reduced density matrix: 
\begin{equation}
\vert \varphi_{\zeta}^{(T,R)(u,v)}\rangle \equiv\sqrt{\frac{n_R \cdot \vert K\vert}{\vert K^{r_T}\vert}}\, I_{\zeta}^{(T,R)(u,v)}\vert\psi\rangle,
\qquad\qquad 
\sigma_{\Omega_2}{(T,R)}\equiv tr_{\bar{\Omega}_2} \vert\varphi_{\zeta}^{(T,R)(u,v)}\rangle \langle \varphi_{\zeta}^{(T,R)(u,v)}\vert.
\end{equation}
One can show (up to normalization):
\begin{equation}
\sigma_{\Omega_2}(T,R)= \sum_{w_1,w_2} I_{\zeta_2}^{(T,R)(w_1,w_2)} \sigma^1_{\Omega_2} I_{\zeta_2}^{(T,R)(w_1,w_2)\dagger},
\qquad\qquad \sigma^1_{\Omega_2}=tr_{\bar{\Omega}_2}\vert\psi\rangle\langle\psi\vert.
\end{equation}
From this expression, one can show $\sigma_{\Omega_2}(T,R)$ is identical to the extremal point $\sigma_{\Omega_2}^{\alpha}\in \Sigma(\Omega_2)$, with $\alpha=(T,R)$. Therefore,  $\alpha=(T,R)$ labels  both the boundary superselection sector and the extremal points of $\Sigma(\Omega_2)$.
This further shows that $\Sigma(\Omega_2)=\Sigma(\Omega_2,\Omega'_{2})$ as long as $\bar{\Omega}'_2$ contains $s_0'$ and $s'_1$. Using Eqs. (\ref{trace_boundary},\ref{gluing_boundary}), the ground-state wave function and the HJW theorem one can show that the boundary topological excitations $(\alpha,\bar{\alpha})$ at $s'_0$ and $s'_1$ can be moves by local unitary operators acting around $s'_0$ and $s'_1$, respectively.

\section{$G$ quantum double with $K=\{1\}$ boundary}\label{Sec. K={1}} 
The $K=\{1\}$ boundary ($K=\{1\}$ is the subgroup of $G$ that contains only the identity element) is particularly simple, but it already has many nontrivial features. We take the opportunity to discuss $K=\{1\}$ in some details, and also discuss some additional things like $\Sigma(\Omega_2)_{bulk}$, $\Sigma({\Omega_3})$,  etc.

\subsection{Boundary superselection sectors for a  $K=\{1\}$ boundary}
For a $K=\{1\}$ boundary, each double coset $T$ contains just one group element $T=\{g\}, g\in G$, so $r_T=g$. $K^{r_T}=\{1\}$ and there is a unique irreducible representation of $K^{r_T}$, i.e. the one-dimensional identity representation $Id$. $(T,R)\rightarrow (\{g\},Id)$, and the label $i,j$ can only take one possible value $i,j=1$. Therefore, we will drop the $i,j$ indices. 

The boundary superselection sectors $\alpha=(T,R)=(\{g\},Id)$ are in one to one correspondence with the group elements. Because of this, we will use the simplified notation $\alpha\in G$. The quantum dimension of each boundary topological excitation  is $d_{\alpha}=1$ for $\forall \alpha\in G$.

\subsection{The information convex $\Sigma(\Omega_2)$ for a $K=\{1\}$ boundary}
Here, we repeat some calculation and result of Sec. \ref{Omega_2_Calculation Section} in the simple example $K=\{1\}$. Start with the minimal diagram. For a $K=\{1\}$ boundary, the Hilbert space for the minimal diagram is $\mathcal{H}^{\ast}(\Omega_2)=span\{\vert h,H\rangle\vert h, H\in G \}$. Here $\{\vert h,H\rangle\vert h, H\in G \}$ is an orthonormal basis.
The set $\Sigma^{\ast}(\Omega_2)$ is the set of  density matrices $\sigma^{\ast}_{\Omega_2}$ on $\mathcal{H}^{\ast}({\Omega_2})$ satisfying the following requirements.\\
1) $\sigma^\ast_{\Omega_2}=\sum_{h,H\in G} \, p_{\{h,H\}}\vert h,H\rangle\langle h,H\vert$, where $\{p_{\{h,H\}} \}$ is a probability distribution.\\
2) $B\sigma^{\ast}_{\Omega_2}=\sigma^{\ast}_{\Omega_2}$, where $B\vert h,H\rangle=\delta_{h, H}\vert h,H\rangle$.

From these requirements, it is straightforward to write down a general density matrix $\sigma^{\ast}_{\Omega_2}\in \Sigma^{\ast}(\Omega_2)$:
\begin{equation}
\sigma^{\ast}_{\Omega_2}=\sum_{h\in G}\, p_h\vert h,h\rangle \langle h, h\vert,
\qquad\qquad \{p_h\}\textrm{ is a probability distribution.}
\end{equation}
From this expression, it is obvious that each  extremal point of $\Sigma^{\ast}({\Omega_2})$ is labeled by a group element:
\begin{equation}
\sigma^{\ast \alpha}_{\Omega_2}\equiv \vert \alpha,\alpha\rangle\langle \alpha,\alpha\vert \qquad \textrm{with}\quad \alpha\in G  .
\end{equation}
The quantum dimensions $d_{\alpha}=1$ for $\forall \alpha \in G$. The following properties of the extremal points can be easily checked:
\begin{equation}
\sigma^{\ast \alpha}_{\Omega_2}\cdot\sigma^{\ast \beta}_{\Omega_2}={\delta_{\alpha,\beta}}\,\sigma^{\ast\alpha}_{\Omega_2}
\quad\Rightarrow\quad S(\sigma^{\ast\alpha}_{\Omega_2})=0,\qquad  tr[\sigma^{\ast\alpha}_{\Omega_2}\cdot\sigma^{\ast\beta}_{\Omega_2}]={\delta_{\alpha,\beta}}.
\end{equation}
Now go back to $\Sigma({\Omega_2})$. It has extremal points $\sigma_{\Omega_2}^{\alpha}$ for  $\alpha\in G$ with $\pi(\sigma^{\alpha}_{\Omega_2})=\sigma^{\ast\alpha}_{\Omega_2}$ and the following properties:
\begin{equation}
\sigma^{\alpha}_{\Omega_2}\cdot\sigma^{\beta}_{\Omega_2}=\frac{\delta_{\alpha,\beta}}{\vert G\vert^{n+N-2}}\sigma^{\alpha}_{\Omega_2}\quad\Rightarrow\quad S(\sigma^{\alpha}_{\Omega_2})=(n+N-2) \ln \vert G\vert,
\qquad  tr[\sigma^{\alpha}_{\Omega_2}\cdot\sigma^{\beta}_{\Omega_2}]=\frac{\delta_{\alpha,\beta}}{\vert G\vert^{n+N-2}}.
\end{equation}
From the properties of the extremal points one  verifies the following topological invariant structures of $\Sigma(\Omega_2)$:
\begin{equation}
S(\sigma^{\alpha}_{\Omega_2})=S(\sigma^{1}_{\Omega_2}),
\qquad \sigma^{\alpha}_{\Omega_2}\cdot\sigma^{\beta}_{\Omega_2}=0 \quad \textrm{for}\quad \alpha\ne \beta,
\qquad tr[\sigma^{\alpha}_{\Omega_2}\cdot\sigma^{\alpha}_{\Omega_2}]=tr[\sigma^{1}_{\Omega_2}\cdot \sigma^1_{\Omega_2}].
\end{equation}

\subsection{Boundary strings and the extremal points of $\Sigma(\Omega_2)$}
For  a $K=\{1\}$ boundary, the boundary operators in the basis $\{I_{\zeta}^{h,k}\}$ with $h\in G$ and $k\in K$ now become $\{I_{\zeta}^{\alpha,1}\}$, with $\alpha\in G$. The Hilbert space on each boundary link $e'$ is one dimensional, and therefore any state in the total Hilbert space has a direct product on all boundary links $e'$. We could neglect the boundary links and get an effective theory with a ``rough boundary". We will not do so in order to keep it  similar to $K\ne \{1\}$ cases.

\begin{figure}[h]
	\centering\includegraphics[scale=0.25]{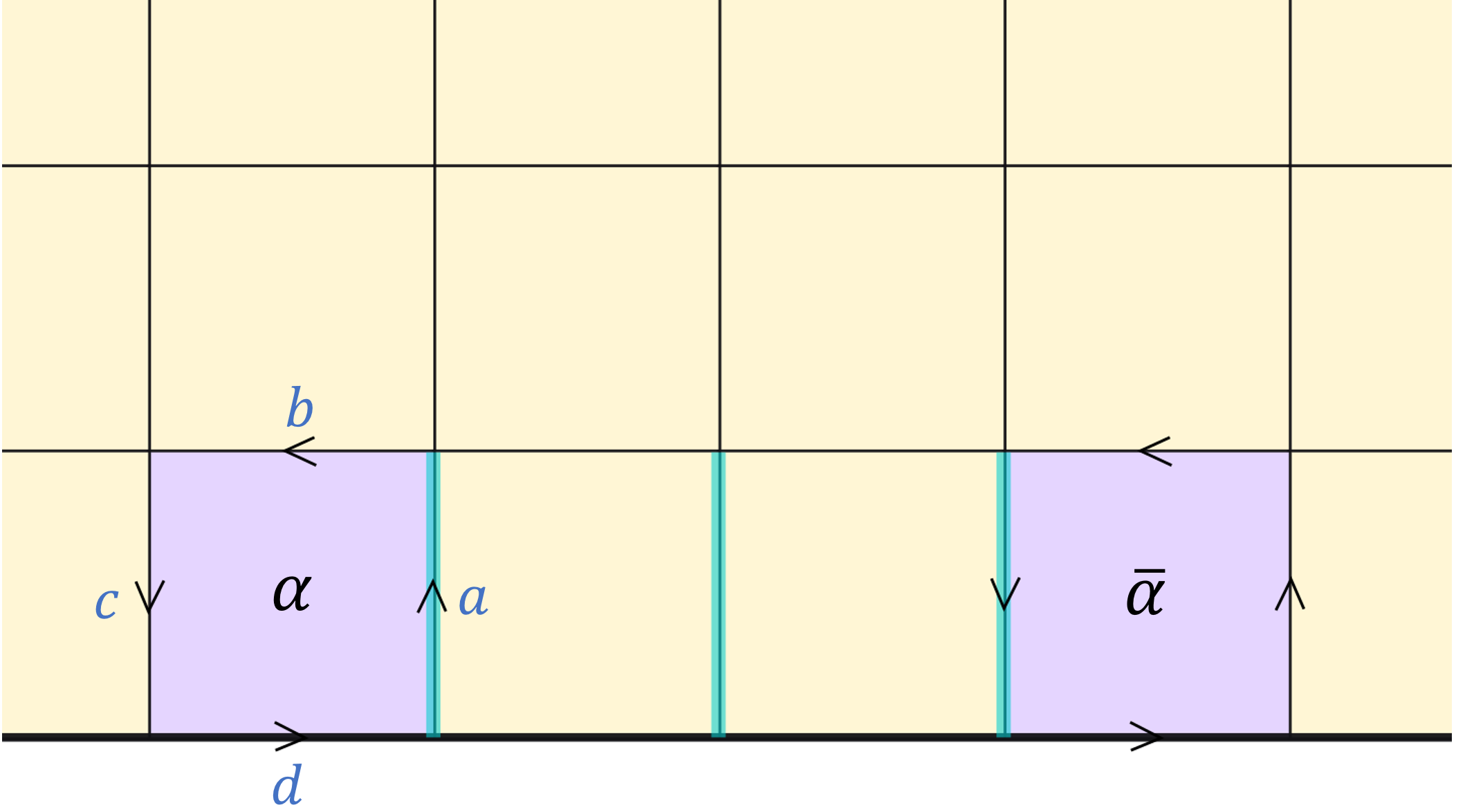}
	\caption{Along the $K=\{1\}$ boundary, a $(\alpha,\bar{\alpha})$ pair is created by a unitary operator $I^{\alpha}_{\zeta}$ which acts nontrivially on the green links. In this case, the $\alpha\in G$ labels the flux type of a single face, $abc=\alpha$. The link configurations $a,b,c\in G$ and $d=1\in K$. }\label{Alpha_flux}
\end{figure}

The basis $\{I_{\zeta}^{(T,R)(u,v)}\}$ now becomes $\{I_{\zeta}^{\alpha} \}$ with $\alpha\in G$ since  for $K=\{1\}$ we could neglect the $u,v$ labels and that $R=Id$. One could verify the following change of basis $\{I^{\alpha}_{\zeta} \}\to \{I^{\alpha,1}_{\zeta} \}$:
\begin{equation}
I^{\alpha}_{\zeta}=  I_{\zeta}^{\alpha,1},\qquad\qquad 
\forall\alpha\in G.
\end{equation}
Therefore,  each $I_{\zeta}^{\alpha}$ is a product of local unitary operators each acting on a bulk link $e\in \zeta$.  It is easy to verify the following properties:
\begin{equation}
(I_{\xi}^{\alpha})^{\dagger}= I_{\xi}^{\bar{\alpha}},\qquad\qquad
I_{\zeta}^{\alpha}I_{\zeta}^{\beta}=I_{\zeta}^{\alpha\beta}, \qquad\qquad 
tr(I_{\zeta}^{\bar{\alpha}}I_{\zeta}^{\beta})=\delta_{\alpha,\beta}\, tr(1),\qquad\qquad I_{\zeta_1\zeta_2}^{\alpha}=I_{\zeta_1}^{\alpha}I_{\zeta_2}^{\alpha},\qquad\qquad I_{\zeta}^{\alpha}I_{\zeta}^{\beta}=I_{\zeta}^{\alpha\beta}.
\end{equation}
Define $\vert\varphi^{\alpha}_{\zeta}\rangle\equiv I_{\zeta}^{\alpha}\vert\psi\rangle$. The middle part of the operator $I^{\alpha}_{\zeta}$ can be deformed into the bulk. For $K=\{1\}$, the excitation type $\alpha\in G$ has a simple interpretation as ``flux" type, see Fig. \ref{Alpha_flux}.  One may also consider the ``fusion" of two fluxes $\alpha$ and $\beta$. The ``fusion" result depends on the ordering: one obtains a flux $\alpha\beta$ if $\alpha$ was on the right of $\beta$ and one obtains a flux $\beta\alpha$ if $\alpha$ was on the left of $\beta$. $\alpha\beta \ne \beta \alpha$ unless $\alpha,\beta$ commute, even though $d_{\alpha}=1$ for $\forall \alpha\in G$. This process is more intricate than the fusion of two Abelian anyons in the bulk.

Let us reconsider the process in Fig. \ref{Omega_2_Ribbon}. Let $\zeta=\zeta_1\zeta_2\zeta_3$, then one can show that $\vert \psi^{\alpha}_{\zeta} \rangle$ prepares an extremal point  $\sigma^{\alpha}_{\Omega_2}\in \Sigma(\Omega_2)$:
\begin{equation}
tr_{\bar{\Omega}_2}\vert \varphi_{\zeta}^{\alpha}\rangle \langle \varphi_{\zeta}^{\alpha}\vert =I^{\alpha}_{\zeta_2}\sigma^1_{\Omega_2}I^{\bar{\alpha}}_{\zeta_2} =\sigma^{\alpha}_{\Omega_2}\qquad\quad\textrm{with}\quad \pi(\sigma^{\alpha}_{\Omega_2})= \vert \alpha,\alpha\rangle\langle \alpha,\alpha\vert.
\end{equation}
Therefore the boundary operators $\{I^{\alpha}_{\zeta}\}$ could prepare all the extremal points of $\Sigma(\Omega_2)$. In this case, it is straightforward to verify that excitations $(\alpha,\bar{\alpha})$ at $s'_0$ and $s'_1$ can be moved by  unitary operators acting around $s'_0$ and $s'_1$ respectively, since $I^{\alpha}_{\zeta}$ itself is a product of local unitary operators each acting on a link.

\subsection{Bulk processes and $\Sigma(\Omega_2)_{bulk}$}\label{Sec. Omega_2_bulk}
Let us consider what element of $\Sigma(\Omega_2)$ could be produced by a bulk process.
Define $\Sigma(\Omega_2)_{bulk}$ to be a subset of $\Sigma(\Omega_2)$ which could be explored by bulk processes. Explicitly:
\begin{equation}
\Sigma(\Omega_2)_{bulk}\equiv\{ \sigma_{\Omega_2}\in \Sigma(\Omega_2)\vert \sigma_{\Omega_2}=tr_{\bar{\Omega}_2}\vert\varphi_{bulk}\rangle\langle \varphi_{bulk}\vert \}\qquad \textrm{with}\quad \vert \varphi_{bulk}\rangle =U_{bulk}\vert\psi\rangle.
\end{equation}
Here $U_{bulk}$ is a unitary operator supported on a bulk subsystem (a subsystem away from the boundary). For the quantum double model, it is enough to have $U_{bulk}=U_{\bar{\Omega}_3}\otimes 1_{\Omega_3}$. Here $\Omega_3$ is an annulus covering a few layers of lattice around the boundary, see Fig. \ref{Omega_3 diagram}.

\begin{figure}[h]
	\centering\includegraphics[scale=0.30]{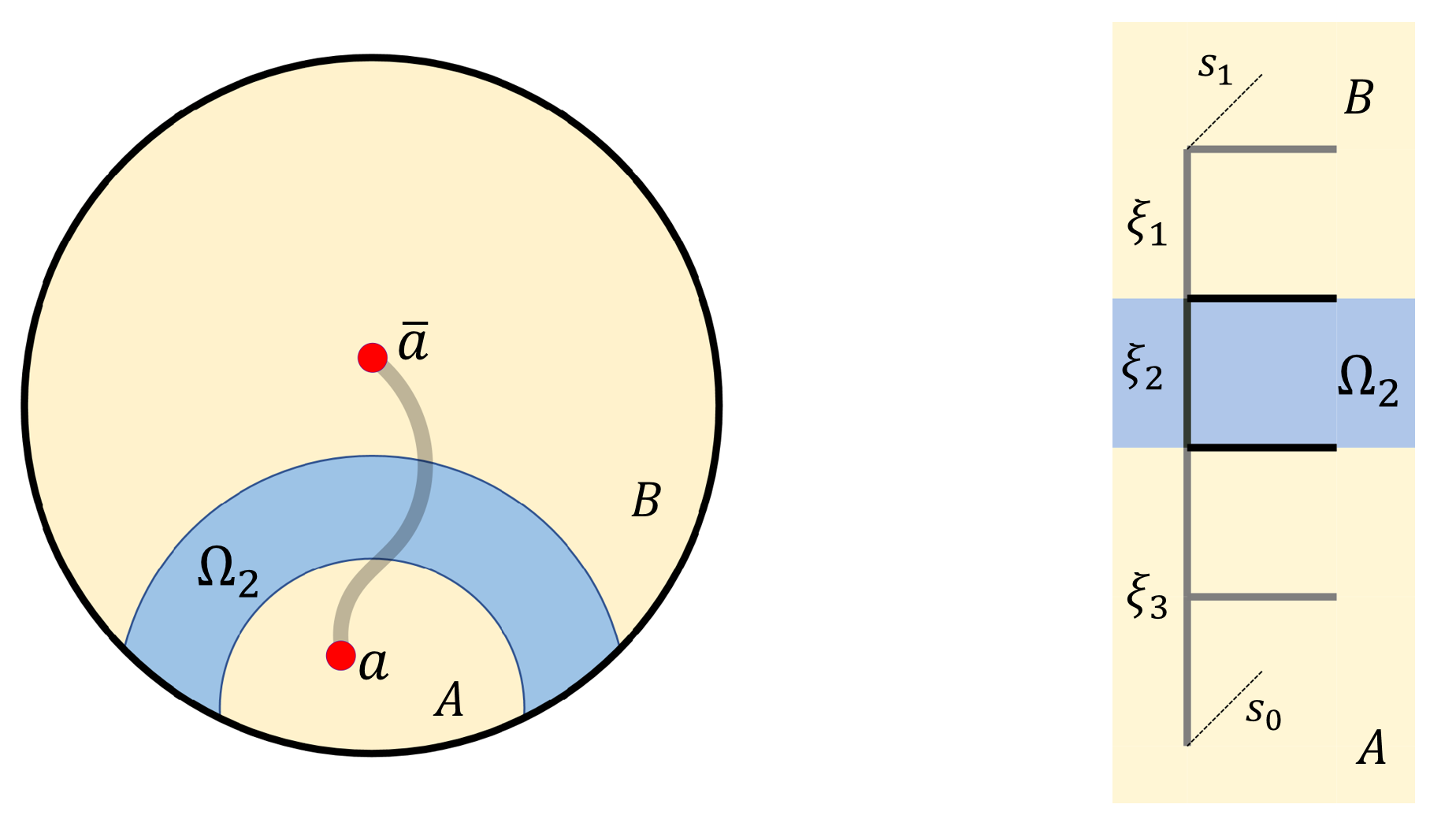}
	\caption{A bulk unitary process creating a $(a,\bar{a})$ pair which can be explicitly constructed using a bulk ribbon operator.}\label{Omega_2 bulk}
\end{figure}

For a $K=\{1\}$ boundary, consider a process involving ribbon operators in the bulk, where bulk anyon pairs $(a,\bar{a})$ are separated by $\Omega_2$, see Fig. \ref{Omega_2 bulk}. According to the discussion in Sec. \ref{Bulk Ribbon Section}, it is a unitary process in the bulk. Calculations using a similar method as  the one in Sec. \ref{bulk_ribbon_extremal} show that
\begin{equation}
\vert\varphi_{\xi}^{(c,R)(u,v)}\rangle=\sqrt{n_R\cdot \vert c\vert}\, F^{(c,R)(u,v)}_{\xi}\vert\psi\rangle\quad\Rightarrow\quad 
tr_{\bar{\Omega}_2} \vert \varphi_{\xi}^{(c,R)(u,v)}\rangle\langle \varphi^{(c,R)(u,v)}_{\xi}\vert =\frac{1}{\vert c\vert} \sum_{i=1}^{\vert c\vert} \sigma^{c_i}_{\Omega_2}. \label{Mixed boundary sector from bulk}
\end{equation}
Recall, $c_i\in c$  is a group element in conjugacy class $c$ and $\sigma^{c_i}_{\Omega_2}$ is an extremal point of $\Sigma(\Omega_2)$.

Observe that for a bulk process creating a $(a,\bar{a})$ pair, with $a=(c,R)$, if $\vert c\vert>1$, it does not prepare an extremal point of $\Sigma(\Omega_2)$. On the other hand, it can be shown (see Sec. \ref{Sec. Omega_3}) that the bulk processes, which create $(a,\bar{a})$ pairs, do prepare all the extremal points of $\Sigma(\Omega_2)_{bulk}$. Therefore $\Sigma(\Omega_2)_{bulk}$ is
\begin{equation}
\Sigma(\Omega_2)_{bulk}=\{\sigma_{\Omega_2}\vert \sigma_{\Omega_2}=\sum_{c} p_c \,\sigma^c_{\Omega_2} \}\qquad\textrm{with}\qquad
\sigma^c_{\Omega_2}\equiv\frac{1}{\vert c\vert} \sum_{i=1}^{\vert c\vert} \sigma_{\Omega_2}^{c_i}, \label{claim}
\end{equation} 
where $\{p_c\}$ is a probability distribution and $c\in (G)_{cj}$. \\
In other words, for a $K=\{1\}$ boundary:\\
1) When $G$ is Abelian we always have $\Sigma(\Omega_2)_{bulk}=\Sigma(\Omega_2)$. \\
2) When $G$ is non-Abelian, we always have $\Sigma(\Omega_2)_{bulk}\subsetneq \Sigma(\Omega_2)$.

Therefore, for non-Abelian models, the boundary superselection sectors could not be identified as a subset of bulk superselection sectors and they need to be treated as fundamental.

\subsection{Some other boundary processes}\label{Condensing Section}
\begin{figure}[h]
	\centering\includegraphics[scale=0.440]{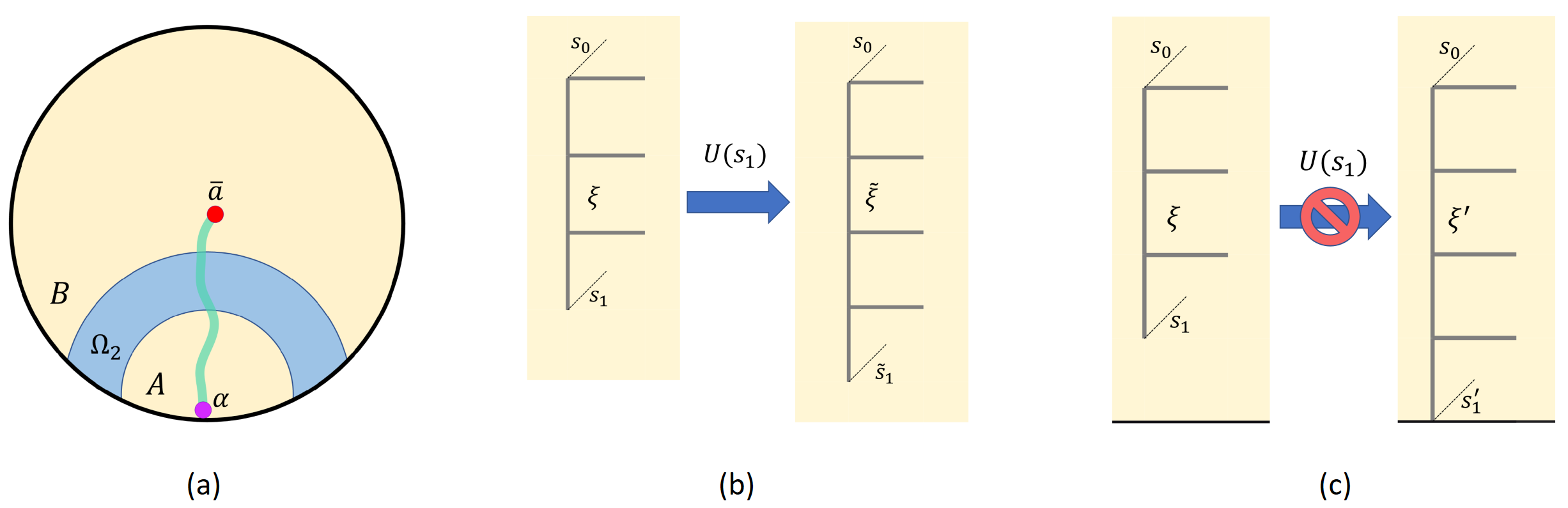}
	\caption{(a) A unitary string operator creating a $(\bar{a},\alpha)$ pair. (b) Extending a bulk ribbon in the bulk $\xi\to \tilde{\xi}$ can  be done by a local unitary operator $U(s_1)$ acting around $s_1$. (c) Extending a bulk ribbon to the boundary $\xi\to \xi'$ may not be achievable using a local unitary operator $U(s_1)$ around $s_1$.}\label{Omega_2 condense}
\end{figure}

\begin{figure}[h]
	\centering\includegraphics[scale=0.30]{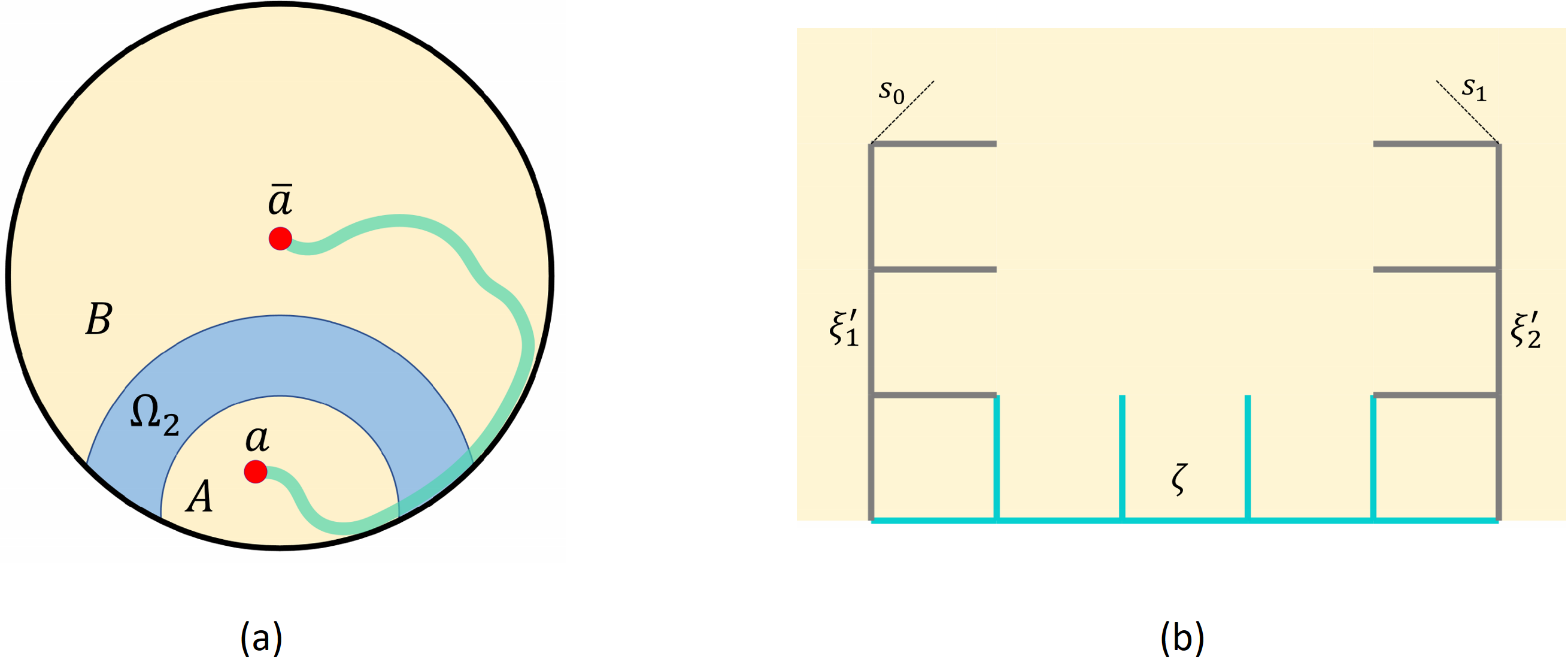}
	\caption{(a) A pair of bulk anyon $(a,\bar{a})$ with $a=(c,R)$ and $\vert c\vert >1$ created by a unitary string operator attaches to the boundary. (b) An explicit ribbon operator could be constructed that realize this  process.}\label{xizetaxi}
\end{figure}

In this section, we discuss a few more unitary processes which involve the boundary, see Figs. \ref{Omega_2 condense} and \ref{xizetaxi}.

The unitary process in Fig. \ref{Omega_2 condense}(a) creates a  $(\bar{a},\alpha)$ pair with $\bar{a}=(\bar{c},\bar{R})$ [so that $a=(c,R)$] and $\alpha=c_i$. Here, $\bar{c}$ is the conjugacy class containing $\bar{r}_c$ and $\bar{R}$ is the complex conjugate of $R$ and $\bar{R} \in (E(\bar{c}))_{ir}$ [note that $E(c)=E(\bar{c})$].

It is possible to write down an explicit ribbon operator $F_{\xi'}^{(\bar{c},\bar{R})(u,v)}$ that realizes this process. Here, the ribbon $\xi'$ connects a bulk site $s_0$ and a boundary site $s'_1$, see Fig. \ref{Omega_2 condense}(c).
The corresponding excited state (normalized) is
\begin{equation}
\vert \varphi^{(\bar{c},{\bar{R}})(u,v)}_{\xi'}\rangle \equiv \sqrt{\vert c\vert \cdot n_R} \, F^{(\bar{c},\bar{R})(u,v)}_{\xi'}\vert\psi\rangle= U(\xi')\vert\psi\rangle.
\end{equation}
Here, $u=(i,j)$ and $v=(i',j')$, where $i,i'=1,\cdots, \vert c\vert$ and $j,j'=1,\cdots, n_R$.
It is straightforward to check that:
\begin{equation}
tr_{\bar{\Omega}_2} \vert \varphi^{(\bar{c},\bar{R})(u,v)}_{\xi'}\rangle \langle  \varphi^{(\bar{c},\bar{R})(u,v)}_{\xi'}\vert =\sigma_{\Omega_2}^{\alpha}\qquad\quad\textrm{with}\quad \alpha=c_{i'}. \label{alpha from condense}
\end{equation}
It prepares an extremal point of $\Sigma(\Omega_2)$. The result  depends only on the flux type $\alpha=c_{i'}$.

One may  interpret this diagram as  condensing a bulk anyon $a=(c,{R})$ into a boundary topological excitation:
$\alpha\in c$ i.e. $a\to n_R\sum_{i=1}^{\vert c\vert}\cdot c_i$ for $a=(c,R)$. The condensation multiplicity equals $n_R$  and it matches the possible values of $j'$  (unlike the case for a bulk site, different $j'$ could not be changed by a local unitary process for $s_1'$ being a boundary site) \footnote{These condensation multiplicity can be seen from $\Sigma(\Omega)$ of some suitable $\Omega$.}.

The ribbon $\xi'$ is not a bulk ribbon since we require a bulk ribbon to be away from the boundary.  However, it is not difficult to do an extension $\xi\to\xi'$ at the level of the ribbon operator, in the same manner as extending a bulk ribbon into a longer bulk ribbon $\xi\to\tilde{\xi}$. However,  one important difference  one should be aware is seen Figs. \ref{Omega_2 condense}(b)(c). 

As is already discussed in Sec. \ref{bulk_ribbon_extremal}, the extension in Fig. \ref{Omega_2 condense}(b), which corresponds to a move of a bulk anyon $a$ from $s_1$ to $\tilde{s}_1$, can be done by a local unitary operation around $s_1$:
\begin{equation}
F_{\tilde{\xi}}^{(c,R)(u,\tilde{v})}\vert \psi\rangle = U(s_1) F_{\xi}^{(c,R)(u,v)}\vert\psi\rangle.  \quad\qquad\qquad
\end{equation}

Now consider the extension $\xi\to\xi'$, i.e. go from the state $F_{\xi}^{(c,R)(u,v)}\vert\psi\rangle$ to $F_{{\xi'}}^{(c,R)(u,{v'})}\vert \psi\rangle$.
 For $\vert c\vert=1$, we have
\begin{equation}
F_{{\xi'}}^{(c,R)(u,{v'})}\vert \psi\rangle = U_{AB}\otimes 1_{\Omega_2}\,\, F_{\xi}^{(c,R)(u,v)}\vert\psi\rangle 
\qquad \textrm{for}\quad \vert c\vert >1.    \label{U_AB equal}
\end{equation}
This result follows from Eqs. (\ref{Mixed boundary sector from bulk},\ref{alpha from condense}) and the HJW theorem. For the case $\vert c\vert=1$, $n_R=1$ explicit construction shows $F_{{\xi'}}^{(c,R)(u,{v'})}\vert \psi\rangle =U(s_1) F_{\xi}^{(c,R)(u,v)}\vert\psi\rangle$. On the other hand, even for the simple case, $c=\{1\}$, $n_R>1$, i.e. it is condensed into the vacuum $\alpha=1$, Eq. (\ref{U_AB equal}) holds only for $U_{AB}\ne U_A\otimes U_B$. Therefore $F_{{\xi'}}^{(c,R)(u,{v'})}\vert \psi\rangle\ne U(s_1) F_{\xi}^{(c,R)(u,v)}\vert\psi\rangle$.

For $\vert c\vert >1$,
\begin{equation}
F_{{\xi'}}^{(c,R)(u,{v'})}\vert \psi\rangle \ne U_{AB}\otimes 1_{\Omega_2}\,\, F_{\xi}^{(c,R)(u,v)}\vert\psi\rangle
\quad \Rightarrow\quad
F_{{\xi'}}^{(c,R)(u,{v'})}\vert \psi\rangle \ne U(s_1) F_{\xi}^{(c,R)(u,v)}\vert\psi\rangle.
\end{equation}
This result follows from Eqs. (\ref{Mixed boundary sector from bulk},\ref{alpha from condense}) and the HJW theorem.
For  $\vert c\vert>1$, $n_R=1$, one can use $U(s_1)$ to push the bulk anyon $a=(c,R)$ into an equal weight superpositon of boundary topological excitations with $\alpha=c_i$, $i=1,\cdots,\vert c\vert$.  Only after a measurement of boundary topological excitation type can we obtain a state with fixed $\alpha$.
For $\vert c\vert>1$, $n_R>1$, one need to use $U_{AB}\otimes 1_{\Omega_2}$ instead of $U(s_1)$, to push $a=(c,R)$ into an equal weight superpositon of boundary topological excitations with $\alpha=c_i$, $i=1,\cdots,\vert c\vert$. Here,  $U_{AB}\ne U_A\otimes U_B$.

In comparison, the following can always be done by local unitary operations.\\
1) Creating a $(\bar{a},\alpha)$ pair separated by a small distance. Here $a=(c,R)$ and $\alpha=c_{i'}$ are fixed.  Then one may also move $\bar{a}$ away from $\alpha$ step by step using a sequence of local unitary operations. The support of the local unitary operators in the sequence may overlap with each other.\\
2) Start from an excited state with a $(\bar{a},\alpha)$ pair, where  $a=(c,R)$ and $\alpha=c_{i'}$, see Fig. \ref{Omega_2 condense}(a).  Push $\bar{a}$ towards the boundary, and then condense $\bar{a}$ into a boundary topological excitation. In this case,  $\bar{a}$ will  condense into $\bar{\alpha}$ instead of a superposition.

Another intriguing process is to have a pair of bulk anyons $(a,\bar{a})$ created using a string attached to the boundary (which could not be deformed into the bulk completely), see Fig. \ref{xizetaxi}. This could not happen for a quantum double model with Abelian $G$. On the other hand, this type of boundary process exists for all quantum double models with non-Abelian $G$ and $K=\{1\}$ boundary.

One can write down explicit ribbon operators with support shown in Fig. \ref{xizetaxi}(b). For this unitary process:
\begin{equation}
\vert \varphi^{(a,\bar{a})}_{\xi'_1\zeta\xi'_2}\rangle \equiv \vert c\vert \cdot n_R\,\, F_{\xi'_1}^{(c,R)(u_1,v_1)} I_{\zeta}^{\alpha}F_{\xi'_2}^{(c,R)(u_2,v_2)}\vert\psi\rangle =U(\xi'_1\zeta\xi'_2)\vert\psi\rangle. 
\end{equation}
Here, $u_1=(i_1,j_1)$, $v_1=(i'_1,j'_1)$, $u_2=(i_2,j_2)$, and $v_2=(i'_2,j'_2)$, with the requirement $c_{i'_1}=c_{i_2}=\alpha$. 
$U(\xi'_1\zeta\xi'_2)$ is a unitary operator supported on a stringlike region within a few lattice spacing to $\xi'_1\zeta\xi'_2$.
Explicit calculations show
\begin{equation}
tr_{\bar{\Omega}_2} \vert \varphi^{(a,\bar{a})}_{\xi'_1\zeta\xi'_2}\rangle \langle  \varphi^{(a,\bar{a})}_{\xi'_1\zeta\xi'_2}\vert =\sigma^{\alpha}_{\Omega_2}.
\end{equation}
For $\vert c\vert>1$,  $ \vert \varphi^{(a,\bar{a})}_{\xi'_1\zeta\xi'_2}\rangle\ne U_{bulk}\vert\psi\rangle$ since $\sigma^{\alpha}_{\Omega_2}\notin \Sigma(\Omega_2)_{bulk}$ for $\alpha=c_{i'_1}$ with $\vert c\vert >1$ and therefore the string could not be deformed into the bulk \footnote{The type of process in Fig. \ref{xizetaxi}, which could not be deformed into the bulk completely actually beyond the $\vert c\vert>1$ case. }. 
Furthermore, it can be shown that this process is related to the process in Fig. \ref{Omega_2_Ribbon} and \ref{Omega_2 condense} by unitary operations in $A$ and $B$, i.e.,
\begin{equation}
\vert \varphi^{(a,\bar{a})}_{\xi'_1\zeta\xi'_2}\rangle
= U_{A} \otimes U_{B} \otimes 1_{\Omega_2} \vert \varphi^{\alpha}_{\zeta}\rangle 
=\tilde{U}_{A}\otimes 1_B\otimes 1_{\Omega_2} \vert \varphi_{\xi'}^{(\bar{c},\bar{R})(u,v)}\rangle.
\end{equation}
Therefore, it is possible to pull the pair $(\alpha$, $\bar{\alpha})$ into the bulk  using a local unitary process around $\alpha$ and $\bar{\alpha}$ to get a state with an $(a,\bar{a})$ pair.

\subsection{A new subsystem $\Omega_3$:  infinite extremal points of $\Sigma(\Omega_3)$ from condensation multiplicity}\label{Sec. Omega_3}
\begin{figure}[h]
	\centering\includegraphics[scale=0.320]{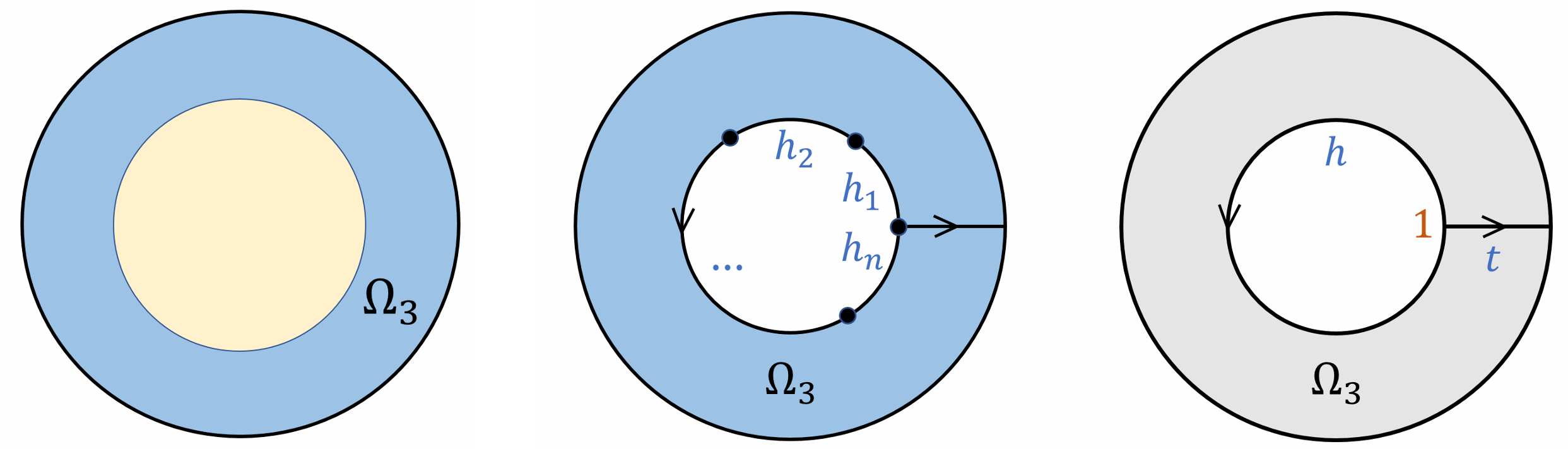}
	\caption{An illustration of the subsystem $\Omega_3$ and the corresponding minimal diagram.}\label{Omega_3 diagram}
\end{figure}

\begin{figure}[h]
	\centering\includegraphics[scale=0.280]{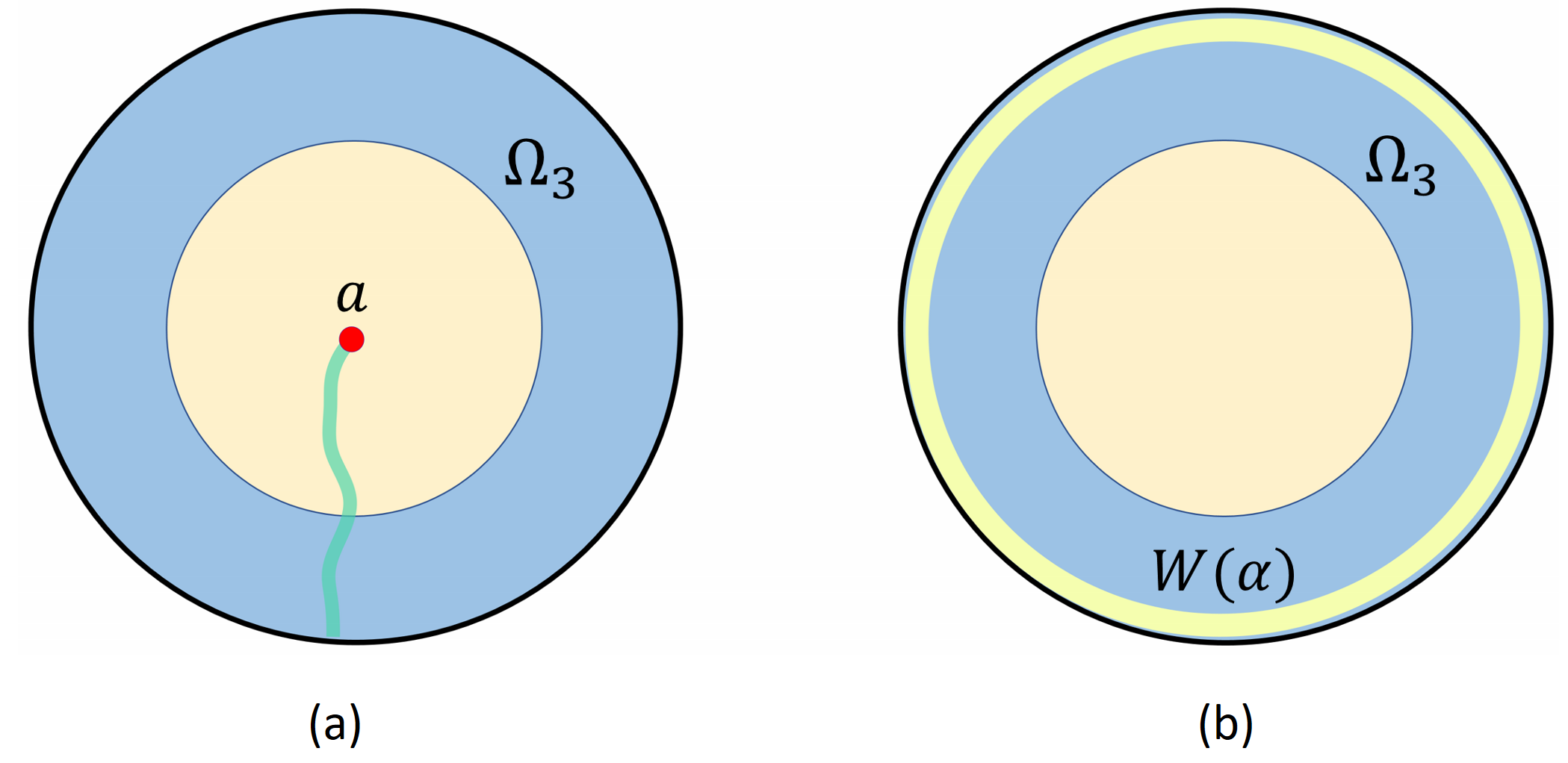}
	\caption{(a) A unitary string operator attaches to the boundary. It creates an anyon $a$ in the bulk. (b) The unitary operator $W(\alpha)$ is supported on the yellow loop along the boundary. It represent a braiding of $(\alpha,\bar{\alpha})$ pair around the boundary.}\label{W_alpha}
\end{figure}

Now consider a subsystem of $\Omega_3$ topology, see Fig. \ref{Omega_3 diagram}. It is an annulus with one edge identified with the boundary (recall that the relation to the boundary is part of the topological data). A natural motivation of considering $\Omega_3$ is that excitations in the bulk may be created by a string operator attached to the boundary, see Fig. \ref{W_alpha}(a). If so, it leaves footprints on $\Omega_3$. We will also see that the structure of $\Sigma(\Omega_3)$ is closely related to condensing of bulk anyons into the vacuum ($\alpha=1$) of the $K=\{1\}$ boundary and there is an infinite number of extremal points for non-Abelian $G$.

Define $\mathcal{H}^{\ast}(\Omega_3)=span\{\vert h,t\rangle\vert h,t\in G \}$ as the Hilbert space for the minimal diagram in Fig. \ref{Omega_3 diagram}, where  $\{\vert h,t\rangle\vert h,t\in G \}$ is an orthonormal basis. Define $\Sigma^{\ast}(\Omega_3)$ be a set of density matrices $\sigma^{\ast}_{\Omega_3}$ on $\mathcal{H}^\ast{(\Omega_3)}$ satisfying the following requirements:\\
1) $\sigma^{\ast}_{\Omega_3}=\sum_{h\in G} \sum_{\lambda} p^{\lambda}_{h}\,\vert \{h\};\lambda\rangle\langle \{h\};\lambda\vert$, where $\{p^{\lambda}_h \}$ is a probability distribution and $\vert \{h\};\lambda\rangle = \sum_{t\in G} c_{\lambda}(t)\vert h,t\rangle$ with complex coefficients $c_{\lambda}(t)$ satisfying $\sum_{t\in G}\vert c_{\lambda}(t)\vert^2=1$.\\
2) $B\sigma^{\ast}_{\Omega_3}=\sigma^{\ast}_{\Omega_3}$, where $B\vert h,t\rangle =\delta_{1,h}\vert h,t\rangle$.\\
3) $A_{1}^g \sigma^{\ast}_{\Omega_3} A^{\bar{g}}_1=\sigma^{\ast}_{\Omega_3}$ for $\forall g\in G$, where $A_1^{g}\vert h,t\rangle =\vert gh\bar{g}, gt\rangle$.

We find that the problem of finding $\Sigma^{\ast}(\Omega_3)$ maps exactly to a problem solved in proposition \ref{Prop. Omega_3 related}. The result is that $\Sigma^{\ast}(\Omega_3)$ has a set of extremal points $\sigma^{\ast(R,z)}_{\Omega_3}$:
\begin{equation}
\sigma^{\ast (R,z)}_{\Omega_3} = \frac{1}{n_R}   \sum_{j=1}^{n_R} \vert z (j,R)\rangle \langle z(j,R)\vert, 
\qquad\qquad \vert z(j,R)\rangle =\sum_{j'} z_{j'} \sqrt{\frac{n_R}{\vert G\vert}} \sum_{g\in G} \bar{\Gamma}^{jj'}_R (g) \vert 1,g\rangle.
\end{equation}
Here, the complex numbers $z_{j'}$ satisfy $\sum_{j'=1}^{n_R} \vert z_{j'}\vert^2 =1$. $R\in (G)_{ir}$. The parameter $\{z_{j'}\}$ has equivalence (redundancy) $\{z_{j'}\}\sim \{z_{j'} e^{i\theta}\}$ and $\sigma^{\ast (R,z)}_{\Omega_3}$ is really parameterized by points on the manifold $S^{2n_R-1}/S^1$. 

Let us use the notation $\langle z\vert z'\rangle \equiv \sum_{j'=1}^{n_R} \bar{z}_{j'} z'_{j'}$. One can show:
\begin{equation}
\langle z(j,R)\vert z'(j',R')\rangle =\delta_{R,R'}\delta_{j,j'}\langle z\vert z'\rangle \quad\Rightarrow\quad tr[\sigma^{\ast(R,z)}_{\Omega_3}\cdot \sigma^{\ast(R',z')}_{\Omega_3} ] = \delta_{R,R'}\frac{1}{n_R} \vert \langle z\vert z'\rangle\vert^2,
\qquad  S(\sigma^{\ast(R,z)}_{\Omega_3})= \ln {n_R}.
\end{equation} 
From the similarity between $\Sigma (\Omega_3)$ and $\Sigma^{\ast}({\Omega_3})$  we find that the extremal points of $\Sigma(\Omega_3)$ have the same parametrization, i.e. $\sigma^{(R,z)}_{\Omega_3}$, with the properties:
\begin{equation}
tr[\sigma^{(R,z)}_{\Omega_3}\cdot \sigma^{(R',z')}_{\Omega_3} ] = \delta_{R,R'}\frac{1}{n_R\cdot \vert G\vert^{n-1}} \vert \langle z\vert z'\rangle\vert^2, \quad \quad S(\sigma^{(R,z)}_{\Omega_3})= \ln n_R + (n-1)\ln \vert G\vert.
\end{equation}
The following structures of $\Sigma(\Omega_3)$ are topological invariants (note that $\sigma^1_{\Omega_3}\equiv tr_{\bar{\Omega}_3}\vert\psi\rangle\langle \psi\vert=\sigma^{(Id,z)}_{\Omega_3}$ ):
\begin{equation}
\frac{tr[\sigma^{(R,z)}_{\Omega_3}\cdot \sigma^{(R',z')}_{\Omega_3} ]}{tr[\sigma^{1}_{\Omega_3}\cdot \sigma^{1}_{\Omega_3} ]}=\delta_{R,R'}\frac{1}{d_{a}} \vert \langle z\vert z'\rangle\vert^2 ,
\qquad \quad S(\sigma^{(R,z)}_{\Omega_3})=S(\sigma^{1}_{\Omega_3}) + \ln d_a.
\end{equation}
Here we have used $d_{a}= n_R$ for $a=(\{1\},R)$.

A string operator attaches to the boundary which creates a single anyon $a$ in the bulk which could prepare extremal points of $\Sigma(\Omega_3)$. This process can be thought of as creating a pair $(a,{\alpha})$ with $a=(\{1\},R)$ and $\alpha=1$, a special case of the string operator in Fig. \ref{Omega_2 condense}:
\begin{equation}
\vert \varphi^{(\{1\},R)(u,v)}_{\xi'}\rangle\equiv \sqrt{\vert c\vert\cdot n_R}\, F_{\xi'}^{(\{1\}, R)(u,v)}\vert\psi\rangle,
\quad\quad\quad 
\vert \varphi^{(R,z)u}_{\xi'}\rangle = \sum_{j'=1}^{n_R} z_{j'} \vert \varphi^{(\{1\},R)(u,v=({1,j'}))}_{\xi'}\rangle.
\end{equation}
Calculation shows:
\begin{equation}
tr_{\bar{\Omega}_3} \vert \varphi^{(R,z)u}_{\xi'}\rangle \langle \varphi^{(R,z)u}_{\xi'}\vert =\sigma^{(R,z)}_{\Omega_3}.
\end{equation}

Let us consider braiding  an $(\alpha,\bar{\alpha})$ pair around the boundary, assuming there are no other excitations along the boundary. It corresponds to a unitary operator $W(\alpha)$ acting on a closed loop along the boundary (a closed string version of $I^{\alpha}_{\zeta}$), see Fig. \ref{W_alpha}(b). $W(\alpha)W(\beta)=W(\alpha\beta)$, $W(\alpha)^{\dagger}=W(\bar{\alpha})$.
One can check:
\begin{equation}
W(\alpha)\vert \psi\rangle =\vert \psi \rangle\qquad\qquad \forall \alpha\in G. \label{alpha braiding on GS}
\end{equation}
The braiding $W(\alpha)$ generates a structure preserving bijective mapping $W(\alpha)$: $\Sigma(\Omega_3)\to \Sigma(\Omega_3)$ such that $\sigma_{\Omega_3}\to W(\alpha) \sigma_{\Omega_3}W(\bar{\alpha})$. One can check the mapping on extremal points:
\begin{equation}
W(\alpha)\sigma^{(R,z)}_{\Omega_3} W(\bar{\alpha})= \sigma^{(R,\tilde{z})}_{\Omega_3}\qquad\qquad \textrm{with}\qquad \tilde{z}_b=\sum_{c=1}^{n_R}\bar{\Gamma}^{bc}_{R}(\alpha) z_c.
\end{equation}
The mapping does not mix different $R$, and on each manifold $S^{2n_R-1}/S^1$ it realize a group action of $G$.

Furthermore, $W(\alpha)$ provides a simple proof of the structure of $\Sigma(\Omega_2)_{bulk}$  claimed in Sec. \ref{Sec. Omega_2_bulk}. Consider a unitary operator $U_{bulk}$ supported on a bulk subsystem and  $\vert \varphi_{bulk}\rangle \equiv U_{bulk}\vert\psi\rangle$:
\begin{equation}
W(\alpha)\vert\psi\rangle =\vert\psi\rangle,\qquad [W(\alpha), U_{bulk}]=0\qquad\Rightarrow\qquad W({\alpha})\vert \varphi_{bulk}\rangle =\vert \varphi_{bulk}\rangle. \label{W_alpha_identity}
\end{equation}
Let us assume it prepares an element of $\Sigma(\Omega_2)$, i.e. $\rho_{\Omega_2}\equiv tr_{\bar{\Omega}_2}\vert\varphi_{bulk}\rangle\langle \varphi_{bulk}\vert \in \Sigma(\Omega_2)$ then, by definition, $\rho_{\Omega_2}\in \Sigma(\Omega_2)_{bulk}$. Any extremal point of $\Sigma(\Omega)_{bulk}$ can be written in this form since  purification exists. From Eq. (\ref{W_alpha_identity}) and with $W(\alpha)$ written as a product of unitary operators on $\Omega_2$ and $\bar{\Omega}_2$, i.e. $W(\alpha)=W_{\Omega_2}({\alpha})\otimes W_{\bar{\Omega}_2}(\alpha)$  one derives:
\begin{equation}
\rho_{\Omega_2}= W_{\Omega_2} (\alpha) \,\rho_{\Omega_2}W_{\Omega_2}{(\bar{\alpha})}\qquad\qquad \forall \alpha\in G.
\end{equation} 
Writing $\rho_{\Omega_2}=\sum_{h} p_h \sigma^{h}_{\Omega_2}$ and noticing that $W_{\Omega_2}(\alpha)\sigma^{h}_{\Omega_2} W_{{\Omega_2}}(\bar{\alpha})=\sigma^{\alpha h\bar{\alpha}}_{\Omega_2}$, one finds $p_h= p_{h'}$ for $h$ and $h'$ belonging to the same conjugacy class. Therefore, $\Sigma(\Omega_2)_{bulk}$ can only be a subset of the result claimed in Eq. (\ref{claim}). Finally, due to the  explicitly constructed extremal points, $\Sigma(\Omega_2)_{bulk}$ is no smaller than what is claimed in Eq. (\ref{claim}), so the result  in Eq. (\ref{claim}) is proved.

\section{Summary}

We have introduced the information convex $\Sigma(\Omega)$, a set of reduced density matrices that minimize energy in subsystem $\Omega$, to capture the topological invariants of a 2D topological order both in the bulk and on the gapped boundaries. Using quantum double models and their gapped boundaries as an example, we show how the information convex reveals and characterizes (i) bulk anyons (or bulk superselection sectors), (ii) boundary topological excitations (or boundary superselection sectors), (iii) the condensation rules from bulk anyons to boundary topological excitations. Recent progress in cold atoms provides a potential measurement for the information convex structure in interference experiments \cite{2015arXiv150901160I,2016Sci...353..794K}. As a powerful tool to study topological phases, the information convex can also be generalized to topological orders in higher spatial dimensions and fracton orders \cite{2016PhRvB..94w5157V}.

%

\section*{Acknowledgement}

We are grateful to Joshuah Heath, Isaac Kim, Shinsei Ryu, Ashvin Vishwanath, Alexei Kitaev, Xie Chen, Chao-Ming Jian, Tim Hsieh, Fiona Burnell and Xueda Wen for helpful discussions, and especially to Kohtaro Kato for kindly explaining his results, to Adam Kaufman for discussions on experimental measurements of entanglement entropy. B.S. also thanks Stuart Raby for discussions and encouragement. This work is supported by the National Science Foundation under Grant No. NSF DMR-1653769 (BS, YML).

\appendix

\section{Some useful group theory} \label{Appendix B}
We collect some results useful in the study of quantum double models in Sec. \ref{Quantum Double} which could be appreciated at the level of finite groups.
\subsection{Finite groups and representations - some basics}\label{Basic group theory}
We summarize some basic results about finite groups and representations. Our notation is adapted from Refs.   \cite{2008PhRvB..78k5421B,2013PhRvB..88k5133K,2017CMaPh.355..645C}.

Let $G$ be a finite group. $\vert G\vert$ is the number of group elements in $G$.  We use $(G)_{ir}$ to denote the set of irreducible representations of $G$ and use $(G)_{cj}$ to denote the set of conjugacy classes of $G$.  For a group element $g\in G$, we use $\bar{g}$ to denote its inverse element.

For a (unitary) representation $R$ of $G$, we denote its dimension using $n_R$ and $\Gamma_{R}(g)$ is a $n_R\times n_R$ dimensional unitary matrix associated with representation $R$ (in a chosen basis) and $\Gamma^{jj'}_{R}(g)$ with $j,j'=1,\cdots,n_R$ being the components of the matrix. $\bar{\Gamma}^{jj'}_{R}$ is the complex conjugate of $\Gamma^{jj'}_{R}$.

The following results are useful:
\begin{equation}
\sum_{R\in (G)_{ir}} n_{R}^2 =\vert G\vert,     \label{Irreducible square}
\end{equation}
\begin{equation}
\sum_{g\in G} \Gamma^{ab}_{R}(g) \bar{\Gamma}^{a'b'}_{R'}(g)= \delta_{R,R'}\delta_{a,a'}\delta_{b,b'}\frac{\vert G\vert }{n_{R}} \qquad  \textrm{for}\qquad R,R'\in (G)_{ir}.    \label{Rep Ortho}
\end{equation}

Let $c\in (G)_{cj}$ be a conjugacy class, i.e. $c\equiv\{gr_c \bar{g}\vert g\in G \}$, where  $r_c\in G$ is a representative of $c$. For a finite group, the number of conjugate classes $c\in (G)_{cj}$ equals the number of irreducible representations $R\in (G)_{ir}$.
Given a conjugacy class $c$ and a representative $r_c\in c$, a centralizer group $E(c)=\{ g\in G \,\vert \, gr_c=r_c g \}$ can be defined. Note that,  $E(c)$ depends on the choice of $r_c$ in general. Let $\vert c\vert$ be the number of elements in $c$, one can check $\vert c\vert\cdot \vert E(c)\vert =\vert G\vert$.

Let $P(c)=\{ p_i\}_{i=1}^{\vert c\vert}$ be a set of representatives of $G/E(c)$. It satisfies:\\
1) $p_{i} E(c) \bigcap p_{j} E(c)=0$ for $i\ne j$ and $\bigcup_{i=1}^{\vert c\vert}\, p_i E(c)= G $. It implies that there is a unique decomposition of a group element $g\in G$ as $g=p_i m$, with $m\in E(c)$.\\
2) $c=\{c_i\}_{i=1}^{\vert c\vert}$, where $c_i\equiv p_i r_c \bar{p}_i$. 

Let $K\subseteq G$ be a subgroup of $G$. Let $T\in K\backslash G/K$ be a double coset, i.e. $T\equiv\{k_1 r_T k_2 \vert k_1,k_2\in K \}=Kr_T K$, where $r_T\in G$ is a representative of $T$. Given $T$ and $r_T$, one can define $K^{r_T}\equiv K\bigcap r_T K\bar{r}_T$ which is a subgroup of $K$. Note that, $K^{r_T}$ depends on the choice of $r_T$ in general. 

Let $Q=\{q_i\}_{i=1}^{\vert Q\vert}$ be a set of representatives of $K/K^{r_T}$. It satisfies:\\
1) $\vert Q\vert= \vert K\vert/ \vert K^{r_T}\vert=\vert T\vert/\vert K\vert$. Here $\vert T\vert$ is the number of elements in $T$.\\
2) $q_i K^{r_T}\bigcap q_j K^{r_T}=0$ for $i\ne j$ and $\bigcup_{i=1}^{\vert Q\vert}=q_i K^{r_T}=K$. \\
3) $q_ir_T K\bigcap q_j r_T K=0$ for $i\ne j$ and $T=\bigcup_{i=1}^{\vert Q\vert} q_i r_T K$.  It implies that there is a unique decomposition of  $g\in T$ as $g=q_i r_T k$ with $q_i\in Q$ and $k\in K$.

\subsection{Invariant operators and invariant density matrices on the group Hilbert space $\mathcal{H}_{G}$}\label{Invariant operators}

\begin{definition}[The group Hilbert space]
	The group Hilbert space for a finite group $G$ is defined as $\mathcal{H}_G\equiv span \{\,\vert g\rangle\,\vert\, g\in G\,\}$. Here $\langle g\vert h\rangle =\delta_{g,h}$ with $g,h\in G$. In other words, $\{ \vert g\rangle, g\in G \}$ is an orthonormal basis and the dimension of $\mathcal{H}_{G}$ is $\dim \mathcal{H}_{G}= \vert G\vert$.
\end{definition}

Consider the following mappings that take an operator acting on $\mathcal{H}_G$ to another operator acting on $\mathcal{H}_G$:

1) $L_g$-mapping is defined by  $O\to L_g O L_g^\dagger$. Here $L_g$ is an unitary operator such that $L_g\vert h\rangle =\vert gh\rangle$. 

2) $\tilde{L}_g$-mapping is defined by $O\to \tilde{L}_g O \tilde{L}_g^\dagger$, where $\tilde{L}_g$ is an unitary operator such that $\tilde{L}_g\vert h\rangle =\vert h\bar{g}\rangle$. Here $\bar{g}$ is the inverse of $g$.

\begin{definition}[$L_K$-invariant and $\tilde{L}_K$ invariant operators]
	An operator acting on $\mathcal{H}_{G}$ is said to be:\\
	1) $L_{K}$-invariant if it is invariant under $L_k$-mapping for $\forall k\in K$. Here $K\subseteq G$ is a subgroup.\\
	2) $\tilde{L}_{K}$-invariant if it is invariant under $\tilde{L}_k$-mapping for $\forall k\in K$. Here, $K\subseteq G$ is a subgroup.
\end{definition}

Our next task is to find the general form of certain invariant operators. In order to do so, the following basis will be useful.

\begin{Proposition}
	$\mathcal{H}_{G}$ has an orthonormal basis  $\{\vert a,b;R\rangle \}$, with $a,b=1,\cdots,n_R$ and $R\in (G)_{ir}$. Here 		
	\begin{equation}
	\vert a,b;R\rangle \equiv \sqrt{\frac{n_R}{\vert G\vert}}\sum_{g\in G}\Gamma^{ab}_{R}(g)\vert g\rangle.
	\end{equation}
	
\end{Proposition}
\begin{proof}
	First, one checks  $\langle a,b;R\vert a',b';R'\rangle=\delta_{R,R'}\delta_{a,a'}\delta_{b,b'}$ using Eq. (\ref{Rep Ortho}).
	Then, one chould check that the dimension of the vector space spanned by $\{\vert a,b;R\rangle \}$ is  $\vert G\vert$ using Eq. (\ref{Irreducible square}).
\end{proof}

\begin{Proposition}
	The space of $L_G$-invariant operators acting on  $\mathcal{H}_G$ is the complex vector space $span\{\sum_{a=1}^{n_R}\vert a,b;R\rangle\langle a,b';R\vert \}$ with $b,b'=1,\cdots, n_R$ and $R\in (G)_{ir}$. The dimension of the vector space is $\vert G\vert$. \label{L invariant operators}
\end{Proposition}
\begin{proof}
	First, note that the space of all operators acting on $\mathcal{H}_G$ is a complex vector space spanned by $\{\vert a,b;R\rangle\langle a',b',R'\vert \}$. Second, show that the space of $L_G$-invariant operators is spanned by $\{\sum_{h\in G} L_h \vert a,b;R\rangle\langle a',b',R'\vert L^{\dagger}_{h}\}$. 
	Then, show that 
	\begin{equation}
	span\{\sum_{h\in G} L_h \vert a,b;R\rangle\langle a',b',R'\vert L^{\dagger}_{h} \}
	= span\{ \sum_{a=1}^{n_R}\vert a,b;R\rangle\langle a,b';R\vert \}.
	\end{equation}
	In this step, $L_{h}\vert a,b;R\rangle =\sum_{c=1}^{n_R}\Gamma^{ac}_{R}(\bar{h})\vert c,b;R\rangle$ and Eq. (\ref{Rep Ortho}) is used.  
\end{proof}

\begin{Proposition}\label{Prop. Omega_3 related}
	The set of  $L_{G}$ invariant density matrices is a convex set $\Sigma_G(L_G)$ with extremal points parameterized by a manifold $S^{2n_R-1}/S^1$ for each $R\in (G)_{ir}$.
	The manifold  $S^{2n_R-1}/S^1$  is parameterized by complex numbers $\{z_b \}$ under equivalence $\{z_b\}\sim \{z_b e^{i\theta} \}$.
	Here $b=1,\cdots, n_R$   and $\sum_{b=1}^{n_R}\vert z_b\vert^2=1$ . The corresponding density matrix is
	\begin{equation}
	\rho_{R}(z)\equiv \frac{1}{n_R}\sum_{a=1}^{n_R}\vert{z}(a;R)\rangle \langle z(a;R)\vert\quad \textrm{with}\quad \vert z(a;R)\rangle\equiv \sum_{b=1}^{n_R} z_b\vert a,b;R\rangle, \qquad R\in (G)_{ir}.
	\end{equation}
\end{Proposition}

\begin{Proposition}
	The space of $L_G$-invariant and $\tilde{L}_G$-invariant operators acting on  $\mathcal{H}_G$ is the complex vector space $span\{\sum_{a,b=1}^{n_R}\vert a,b;R\rangle\langle a,b;R\vert \}$ with $R\in (G)_{ir}$. \label{LR invariant operators}
\end{Proposition}
\begin{proof}
	First, notice the result in proposition \ref{L invariant operators} that the space of $L_G$-invariant operators is spanned by $\{\sum_{a=1}^{n_R}\vert a,b;R\rangle\langle a,b';R\vert \}$ with $b,b'=1,\cdots, n_R$ and $R\in (G)_{ir}$. Second, show that the space of $L_G$-invariant and $\tilde{L}_G$-invariant operators is spanned by $\{\sum_{h\in G}\sum_{a=1}^{n_R} \tilde{L}_{h}\vert a,b;R\rangle\langle a,b';R\vert \tilde{L}_{h}^{\dagger} \}$. Finally, show that 
	\begin{equation}
	span\{	\sum_{h\in G}\tilde{L}_{h}\bigg( \sum_{a=1}^{n_R} \vert a,b;R\rangle \langle a,b';R\vert \bigg) \tilde{L}_{h}  \}
	= span\{  \sum_{a,b=1}^{n_R} \vert a,b;R\rangle\langle a,b; R\vert \}.
	\end{equation}
	In this step, $\tilde{L}_{h}\vert a,b;R\rangle =\sum_{c=1}^{n_R} \bar{\Gamma}^{bc}_R (\bar{h})\vert a,c; R\rangle $ and Eq. (\ref{Rep Ortho}) is used. 	
\end{proof}

\begin{Proposition}
	The set of  $L_{G}$-invariant and $\tilde{L}_G$-invariant density matrices is a convex set $\Sigma_{G}(L_G,\tilde{L}_G)$ with extremal points
	\begin{equation}
	\rho_{R}\equiv \frac{1}{n_R^2} \sum_{a,b=1}^{n_R} \vert a,b;R\rangle\langle a,b;R\vert,\qquad\qquad R\in (G)_{ir}.
	\end{equation}
\end{Proposition}

\section{The $S_3$ quantum double with $K=\{1\}$ boundary}\label{Appendix E}
In this appendix, we provide some additional details for the $S_3$ quantum double with $K=\{1\}$ boundary, i.e., explain our notations and point to some  references.

$S_3$ is the simplest non-Abelian finite group and the  $S_3$ quantum double model is discussed as examples by many references. For example,  \cite{2011CMaPh.306..663B} contains  detailed anyon types, fusion rules and $S$-matrix of the $S_3$ quantum double. On the other hand, our notation for bulk anyons $\{1,A,J^w,J^x,J^y,J^z,K^a, K^b \}$ is  similar with \cite{2013PhRvB..88k5133K}. The condensation rules for the $S_3$ quantum double with $K=\{1\}$ boundary is discussed \cite{2017CMaPh.355..645C} in a different physical context. The following is some details of our notation.

\subsection{Bulk superselection sectors}

Given that the bulk superselection sectors (or bulk anyon types) are labeled by $a=(c,R)$ with $c\in (G)_{cj}$ and $R\in (E(c))_{ir}$,
the 8 superselection sectors for the  $S_3$ quantum double model $\{1,A,J^w,J^x,J^y,J^z,K^a, K^b \}$ can be worked out. Here, $S_3=\{1,r,r^2,s,sr,sr^2\}$ with $r^3=s^2=1$ and $sr=r^2s$. See the following table for more details.
\begin{table}[h]
	\centering
	
	\begin{tabular}{|c| c| c | c | c | c | c | c | c |}
		\hline
		\multicolumn{1}{|c|}{$c\in (S_3)_{cj}$}& \multicolumn{3}{c|}{$c_1=\{1 \}$}& \multicolumn{3}{c|}{$c_r=\{r,r^2 \}$}& \multicolumn{2}{c|}{$c_s=\{s,sr,sr^2 \}$}\\
		\hline
		\multicolumn{1}{|c|} {$r_c$} & \multicolumn{3}{c|}{$1$}& \multicolumn{3}{c|}{$r$}& \multicolumn{2}{c|}{$s$}\\
		\hline
		\multicolumn{1}{|c|} {$E(c)$} & \multicolumn{3}{c|}{$S_3$}& \multicolumn{3}{c|}{$Z_3=\{1,r,r^2\}$}& \multicolumn{2}{c|}{$Z_2=\{1,s \}$}\\
		\hline
		\multicolumn{1}{|c|} {$P(c)=S_3/E(c)$} & \multicolumn{3}{c|}{$\{1\}$}& \multicolumn{3}{c|}{$\{1,s\}$}& \multicolumn{2}{c|}{$\{1,r,r^2\}$}\\
		\hline
		$R\in (E(c))_{ir}$ & $Id$ &$A$ & $B$ & $Id$ &$\omega$ &$\omega^2$ & $Id$ & $A$\\
		\hline
		$n_R$ & 1&1 &2 &1 &1 &1 &1 &1\\
		\hline
		$a$ & \,$1$\,& \,$A$\, & $J^w$ & \,$J^x$\, & \,$J^y$\, & \,$J^z$\, & \,\,\,\,\,$K^a$\,\,\,\,\, & $K^b$\\
		\hline
		$d_a=\vert c\vert\cdot n_R$ & 1&1 &2 &2 &2 &2 &3 &3\\
		\hline
	\end{tabular}
\end{table}

Here are the representations.\\
For $R\in (S_3)_{ir}$ the corresponding unitary matrices $\Gamma_R (g)$ with $g\in S_3$ are:\\
1) $\Gamma_{Id}(g)=1$ for $\forall g\in S_3$.\\
2) $\Gamma_{A}(r)=1$ and $\Gamma_A(s)=-1$. Other $\Gamma_A(g)$ for $g\in S_3$ can be obtained using $\Gamma_{A}(r)$ and $\Gamma_A(s)$.\\
3) $\Gamma_{B}(r)=\left(\begin{array}{rr}
\omega & 0 \\ 0 &\omega^2
\end{array} \right)$ and $\Gamma_B(s)=\left(\begin{array}{rr}
0 & 1 \\ 1 & 0
\end{array} \right)$. Here $\omega\equiv e^{i\frac{2\pi}{3}}$. Other $\Gamma_B(g)$ for $g\in S_3$ can be obtained using $\Gamma_{B}(r)$ and $\Gamma_B(s)$.\\
For $R\in (Z_3)_{ir}$, $Z_3=\{1,r,r^2\}$, the corresponding unitary matrices $\Gamma_R (g)$ with $g\in Z_3$ are:\\
1) $\Gamma_{Id}(g)=1$ for $\forall g\in Z_3$.\\
2) $\Gamma_{\omega}(1)=1$, $\Gamma_{\omega}(r)=\omega$, $\Gamma_{\omega}(r^2)=\omega^2$. Here $\omega\equiv e^{i\frac{2\pi}{3}}$.  \\
3) $\Gamma_{\omega^2}(1)=1$, $\Gamma_{\omega^2}(r)=\omega^2$, $\Gamma_{\omega^2}(r^2)=\omega$. Here $\omega\equiv e^{i\frac{2\pi}{3}}$.\\
For $R\in (Z_2)_{ir}$, $Z_2=\{1,s\}$, the corresponding unitary matrices $\Gamma_R (g)$ with $g\in Z_2$ are:\\
1) $\Gamma_{Id}(g)=1$ for $\forall g\in Z_2$.\\
2) $\Gamma_{A}(1)=1$, $\Gamma_{A}(s)=-1$.

\subsection{The fusion rules of $S_3$ quantum double}
The following fusion rules of the $S_3$ quantum double model can be found in \cite{2013PhRvB..88k5133K}.
\begin{eqnarray}
\left\{
\begin{array}{lll}
A\times A&=&1\\
A\times K^{a/b}&=&K^{b/a}\\
A\times J^{\alpha}&=&J^{\alpha}
\end{array}
\right.\qquad
\left\{
\begin{array}{lll}
J^{\alpha}\times J^{\alpha}&=&1+A+J^{\alpha}\\
J^{\alpha}\times J^{\beta}&=&J^{\gamma}+J^{\delta}
\end{array}
\right.\qquad
\left\{
\begin{array}{lll}
J^{\alpha}\times K^{\beta} &=& K^a + K^b\\
K^{\alpha}\times K^{\alpha} &=& 1 + J^w +J^x +J^y +J^z\\
K^a\times K^b &=& A+J^w +J^x +J^y +J^z
\end{array}
\right. .
\end{eqnarray}
Here $\alpha$, $\beta$, $\gamma$ and $\delta$ are running indices. $\alpha$, $\beta$, $\gamma$ and $\delta$ with values different from each other in the equation which has all of them, i.e. in $J^{\alpha}\times J^{\beta}=J^{\gamma}+J^{\delta}$. The fusion rules (in a different notation) together with the calculation method for a general quantum double model can be found in \cite{2011CMaPh.306..663B}.

\subsection{$K=\{1\}$ boundary superselection sectors and condensation rules}
According to Sec. \ref{Sec. K={1}}, the boundary superselection sectors (boundary topological excitation types) of a $K=\{1\}$ boundary are labeled by the group element $\alpha\in G$. Therefore, for $G=S_3$ we have 6 types $\{1,r,r^2,s,sr,sr^2\}$. $\alpha\in S_3$ correspond to $(T,R)=(\{\alpha\}, Id)$.
\begin{table}[h]
	\centering
	
	\begin{tabular}{|c| c| c | c | c | c | c | }
		\hline
		{$T\in \{1\}\backslash S_3/\{1\}$}& $\{1\}$ &$\{r\}$ &$\{r^2\}$ &$\{s\}$ & $\{sr\}$& $\{sr^2\}$ \\
		\hline
		{$r_T$}& $1$ &$r$ &$r^2$ &$s$ & $sr$& $sr^2$ \\
		\hline
		{$K^{r_T}$} & $\{1\}$ &$\{1\}$ &$\{1\}$ &$\{1 \}$ & $\{1\}$ & $\{\{1\}$ \\
		\hline
		{$Q=K/K^{r_T}$} & $\{1\}$ &$\{1\}$ &$\{1\}$ &$\{1 \}$ & $\{1\}$ & $\{\{1\}$ \\
		\hline
		$\{s_i\}_{i=1}^{\vert Q\vert}$ & $\{1\}$ &$\{r\}$ &$\{r^2\}$ &$\{s\}$ & $\{sr\}$& $\{sr^2\}$ \\
		\hline
		$R\in (K^{r_T})_{ir}$ & $Id$ &$Id$ & $Id$  & $Id$ &$Id$  &$Id$ \\
		\hline
		$n_R$ & 1 &1 &1 &1 &1 &1\\
		\hline
		$\alpha$ & $1$ & $r$ & $r^2$ & $s$ & $sr$ & $sr^2$\\
		\hline
		$d_{\alpha}$ & 1&1 &1 &1 &1 &1 \\
		\hline
	\end{tabular}
\end{table}

The following condensation rules are discussed \cite{2017CMaPh.355..645C} in the physical context of confined boundary excitations. 

\begin{table}[h]
	\centering
	
	\begin{tabular}{|c|| c| c | c | c|c | c|c | c|c | c|c|c | c|c|c |}
		\hline
		\multicolumn{1}{|c||}{$c\in (S_3)_{cj}$}& \multicolumn{3}{c|}{$c_1=\{1 \}$}& \multicolumn{6}{c|}{$c_r=\{r,r^2 \}$}& \multicolumn{6}{c|}{$c_s=\{s,sr,sr^2 \}$}\\
		\hline
		$a$ & \,$1$\,& \,$A$\, & $J^w$ &    \multicolumn{2}{c|}{\,\,\,\,$J^x$\,\,\,\,} & \multicolumn{2}{c|}{\,\,\,\,$J^y$\,\,\,\,} & \multicolumn{2}{c|}{\,\,\,\,$J^z$\,\,\,\,} & \multicolumn{3}{c|}{\,\,\,\,$K^a$\,\,\,\,}&\multicolumn{3}{c|}{\,\,\,\,$K^b$\,\,\,\,}\\
		\hline
		$\alpha$ & \,$1$\,& \,$1$\, & $2\cdot 1$ & \,$r$\, & $ r^2$ & \,$r$\, & $ r^2$ & \,$r$\, & $ r^2$ & \,$s$\,& $sr$ & $sr^2$ & \,$s$\,& $sr$ & $sr^2$\\
		\hline
	\end{tabular}
\end{table}

The same condensation rules also apply to our case, i.e. condense a bulk anyon into a deconfined boundary topological excitation. This is suggested by the construction of ribbon operators and the results of $\Sigma(\Omega_2)$, $\Sigma(\Omega_2)_{bulk}$ and $\Sigma(\Omega_3)$, see Sec. \ref{Sec. K={1}}. 

\emph{Note added.} Months after the completion of this work, more evidence has been collected. It is now clear that information convex $\Sigma(\Omega)$ could coherently encode fusion multiplicities $N_{ab}^c$, $N_{\alpha\beta}^{\gamma}$ and condensation multiplicities $N_{a}^{\alpha}$ for  suitably chosen topologies of $\Omega$.  Here $N_{ab}^c$ is the fusion multiplicity for bulk anyons i.e. $a\times b= \sum_c N_{ab}^c \,c$. $N_{\alpha\beta}^{\gamma}$ is the fusion multiplicity for boundary topological excitations $\alpha \times \beta = \sum_{\gamma } N_{\alpha\beta}^{\gamma} \,\gamma$. $N_{a}^{\alpha}$ is the condensation multiplicity satisfying $a= \sum_{\alpha} N_a^{\alpha} \,\alpha$. Furthermore, the following are shown in Ref. \cite{2018arXiv181001986S}.
\begin{itemize}
	\item  This coherent encoding of multiplicities and strong subadditivity provides a new derivation of the topological contributions to the von Neumann entropy, i.e. the $\ln d_a$ from bulk anyons and the $\ln d_{\alpha}$ from boundary topological excitations.
	\item The result in Eq.(\ref{eq:condense}), is a special case of a more general result
	\begin{equation}
	\sigma_{\Omega_2}^a = \sum_{\alpha} \frac{N_{a}^{\alpha} d_{\alpha}}{d_a} \sigma_{\Omega_2}^{\alpha},
	\end{equation}
	where $	\sigma_{\Omega_2}^a $ is the reduced density matrix on $\Omega_2$ from the quantum state shown in Fig. \ref{Convex_2}(c).
\end{itemize}

\bibliography{ref_combined}
\bibliographystyle{apsrev}

\end{document}